\documentclass{article}
\usepackage{arxiv}
\usepackage[T1]{fontenc}
\usepackage{amsmath,amssymb,amsthm}
\usepackage{booktabs}
\usepackage{microtype}
\usepackage{array}
\usepackage{url}
\ifdefined\kanjiskip
  \PassOptionsToPackage{dvipdfmx}{graphicx}
  \PassOptionsToPackage{dvipdfmx}{xcolor}
\fi
\usepackage{tikz}
\usetikzlibrary{shapes.geometric,shapes.misc,decorations.pathreplacing,positioning,fit}
\tikzset{flbox/.style={draw,rounded corners=1.5pt,align=left,inner sep=2pt},flterm/.style={flbox,rounded corners=3.5mm,fill=black!8,inner xsep=2.2mm},fldec/.style={flbox,double,double distance=0.7pt},flarr/.style={->,>=stealth,semithick}}

\newtheorem{theorem}{Theorem}

\newtheorem{corollary}[theorem]{Corollary}
\theoremstyle{definition}
\newtheorem{definition}[theorem]{Definition}
\theoremstyle{remark}
\newtheorem{remark}[theorem]{Remark}

\newcommand{\Code}{\mathit{Code}}
\newcommand{\pos}{\mathit{pos}}
\newcommand{\rmL}{\mathrm{L}}
\newcommand{\rmR}{\mathrm{R}}
\newcommand{\rmZ}{\mathrm{Z}}

\title{On Small Universal Reversible Turing Machines:\\
 Improved Conversions, the Two-State Case,\\
 and the Three-State Two-Symbol Case}
\renewcommand{\shorttitle}{On Small Universal Reversible Turing Machines}
\author{Tetsuo Yokoyama\\Nanzan University}
\date{October 2026}

\begin{document}
\maketitle

\begin{abstract}
A reversible Turing machine (RTM) is a Turing machine in which every instantaneous description has at most one predecessor.
Writing URTM$(m,n)$ for a universal RTM with $m$ states and $n$ symbols, we construct URTM$(3,600)$, URTM$(4,38)$, URTM$(5,36)$, URTM$(6,28)$, URTM$(7,26)$, URTM$(8,21)$, URTM$(9,8)$, URTM$(10,7)$, and URTM$(85,2)$; each of them has fewer states or fewer symbols than each of the nine URTMs in Morita's book (2017).
Most of them come from improved versions of Morita's conversions to three states, to four states, and to two symbols; for example, the improved conversion to three states uses $O(mn^2)$ symbols instead of $O(m^2n^3)$.
URTM$(10,7)$ comes from merging two states of Morita's URTM$(13,7)$, recording the difference in tags on the tape cells, and then merging symbols, three times; one more such step gives URTM$(9,8)$.
We also show that the halting problem of RTMs with at most two states is decidable for finite instantaneous descriptions, so that no RTM with at most two states is universal under any notion of universality that reduces the halting problem of Turing machines to it by a computable encoding into finite instantaneous descriptions; in particular, there is no URTM with at most two states in our formalization of Morita's definition. We also show that no three-state, two-symbol RTM is universal in that formalization.
\end{abstract}

\section{Introduction}
\label{sec:intro}

Reversible computing concerns computations in which each step uniquely determines its predecessor.
Bennett showed that irreversible computations can be simulated reversibly without loss of computational power~\cite{Bennett73}.
A reversible Turing machine (RTM) is one of the basic models of reversible computing, and a universal RTM is called a URTM.
As for ordinary universal Turing machines (UTMs)~\cite{Rogozhin96,KudlekRogozhin02,NearyWoods09,WoodsNeary09}, a natural question is how small a URTM can be.
We measure the size by the numbers of states and symbols, and write URTM$(m,n)$ for a URTM with $m$ states and $n$ symbols.

Morita's book~\cite{Morita17} lists nine URTMs.
Six of them, URTM$(10,8)$, $(13,7)$, $(15,6)$, $(17,5)$, $(24,4)$, and $(32,3)$, simulate cyclic tag systems with halting condition directly; URTM$(17,5)$ is due to Morita and Yamaguchi.
The other three, URTM$(4,168)$, $(138,2)$, and $(3,36654)$, are obtained from them by general conversions that reduce the number of states or of symbols.
The book remarks that these three leave room for reduction, and it states that the existence of a two-state URTM is open.
In a later survey, after recalling that two-state TMs are universal in the irreversible case, Morita writes that ``it is unknown whether the class of two-state RTMs is universal''~\cite{Morita23}.

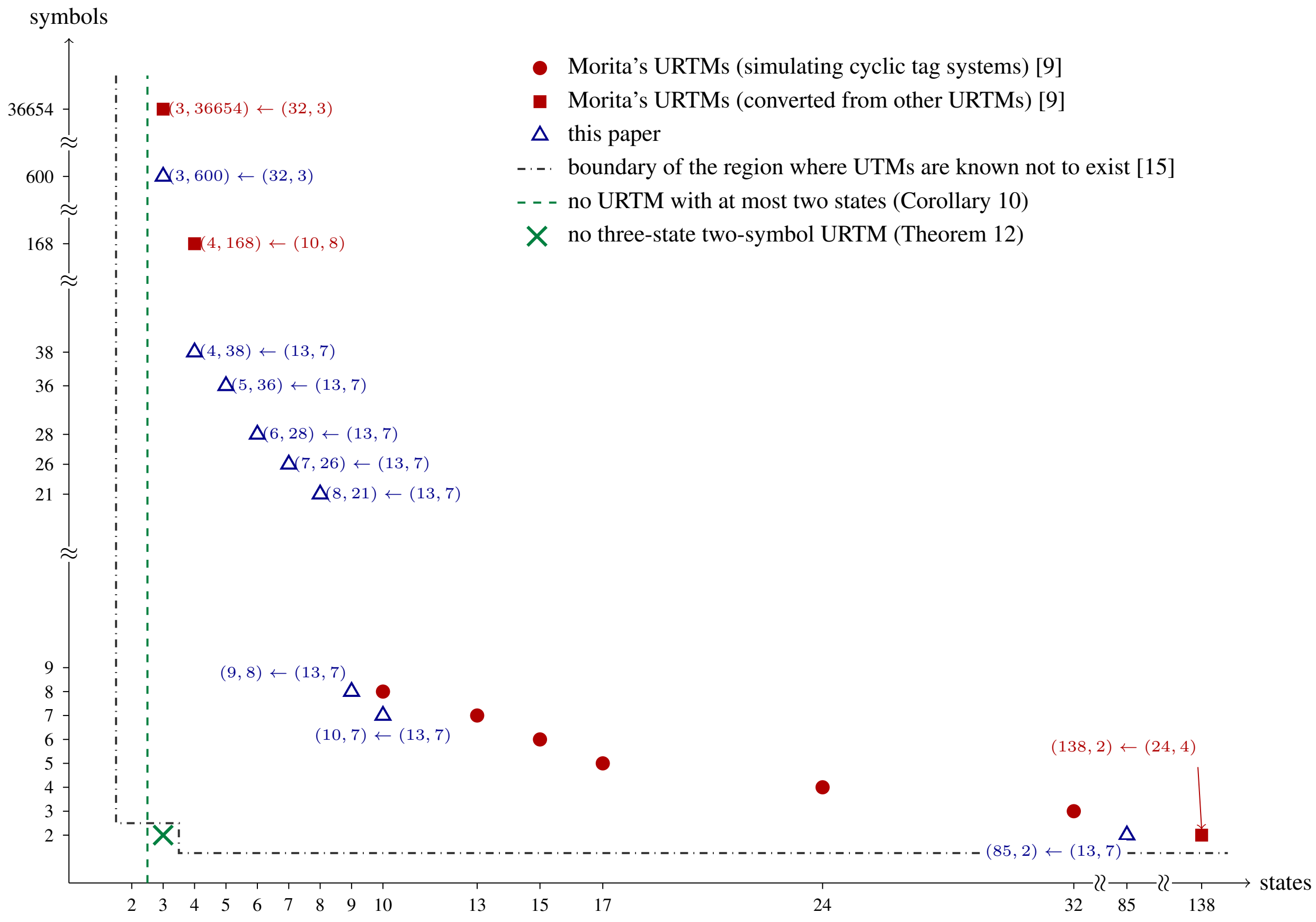
\begin{figure}[tb]
\centering
\resizebox{\textwidth}{!}{%
\begin{tikzpicture}[font=\scriptsize,
  wave/.style={fill=white, inner sep=0.6pt, font=\small}]
\definecolor{cM}{RGB}{176,0,0}\definecolor{cU}{RGB}{0,0,140}\definecolor{cO}{RGB}{0,128,64}
\draw[->] (0,0) -- (15.8,0) node[anchor=west,font=\small] {states};
\draw[->] (0,0) -- (0,11.3) node[anchor=south,font=\small] {symbols};
\foreach \x/\l in {0.840/2,1.260/3,1.680/4,2.100/5,2.520/6,2.940/7,3.360/8,3.780/9,4.200/10,5.460/13,6.300/15,7.140/17,10.080/24,13.440/32,14.150/85,15.150/138}
  \draw (\x,0) -- (\x,-0.08) node[below] {\l};
\foreach \y/\l in {0.640/2,0.960/3,1.280/4,1.600/5,1.920/6,2.240/7,2.560/8,2.880/9,5.200/21,5.600/26,6.000/28,6.650/36,7.100/38,8.550/168,9.450/600,10.350/36654}
  \draw (0,\y) -- (-0.08,\y) node[left] {\l};
\node[wave,rotate=90] at (13.800,0) {$\approx$};
\node[wave,rotate=90] at (14.650,0) {$\approx$};
\node[wave] at (0,4.400) {$\approx$};
\node[wave] at (0,8.050) {$\approx$};
\node[wave] at (0,9.000) {$\approx$};
\node[wave] at (0,9.900) {$\approx$};
\draw[dashed, cO, thick] (1.050,0) -- (1.050,10.800);
\draw[black!80, dash dot, thick] (0.630,10.800) -- (0.630,0.800) -- (1.470,0.800) -- (1.470,0.400) -- (15.500,0.400);
\node[cross out,draw=cO,very thick,inner sep=0pt,minimum size=6pt] at (1.260,0.640) {};
\foreach \x/\y in {4.200/2.560,5.460/2.240,6.300/1.920,7.140/1.600,10.080/1.280,13.440/0.960}
  \node[circle,draw=cM,fill=cM,inner sep=0pt,minimum size=5pt] at (\x,\y) {};
\foreach \x/\y in {15.150/0.640,1.680/8.550,1.260/10.350}
  \node[rectangle,draw=cM,fill=cM,inner sep=0pt,minimum size=4.5pt] at (\x,\y) {};
\foreach \x/\y in {1.260/9.450,1.680/7.100,2.100/6.650,2.520/6.000,2.940/5.600,3.360/5.200,3.780/2.560,4.200/2.240,14.150/0.640}
  \node[regular polygon,regular polygon sides=3,draw=cU,fill=white,thick,inner sep=0pt,minimum size=6.5pt] at (\x,\y) {};
\node[anchor=west,font=\tiny,text=cM,inner sep=1.5pt] at (1.260,10.350) {$(3,36654)\leftarrow(32,3)$};
\node[anchor=west,font=\tiny,text=cM,inner sep=1.5pt] at (1.680,8.550) {$(4,168)\leftarrow(10,8)$};
\node[anchor=south east,font=\tiny,text=cM,fill=white,inner sep=1.5pt] at (15.150,1.650) {$(138,2)\leftarrow(24,4)$};
\draw[->,cM,thin] (15.100,1.550) -- (15.150,0.720);
\node[anchor=west,font=\tiny,text=cU,inner sep=1.5pt] at (1.260,9.450) {$(3,600)\leftarrow(32,3)$};
\node[anchor=west,font=\tiny,text=cU,inner sep=1.5pt] at (1.680,7.100) {$(4,38)\leftarrow(13,7)$};
\node[anchor=west,font=\tiny,text=cU,inner sep=1.5pt] at (2.100,6.650) {$(5,36)\leftarrow(13,7)$};
\node[anchor=west,font=\tiny,text=cU,inner sep=1.5pt] at (2.520,6.000) {$(6,28)\leftarrow(13,7)$};
\node[anchor=west,font=\tiny,text=cU,inner sep=1.5pt] at (2.940,5.600) {$(7,26)\leftarrow(13,7)$};
\node[anchor=west,font=\tiny,text=cU,inner sep=1.5pt] at (3.360,5.200) {$(8,21)\leftarrow(13,7)$};
\node[anchor=south east,font=\tiny,text=cU,inner sep=1.5pt] at (3.780,2.640) {$(9,8)\leftarrow(13,7)$};
\node[anchor=north,font=\tiny,text=cU,inner sep=1.5pt] at (4.200,2.130) {$(10,7)\leftarrow(13,7)$};
\node[anchor=north east,font=\tiny,text=cU,fill=white,inner sep=1.5pt] at (14.150,0.580) {$(85,2)\leftarrow(13,7)$};
\node[circle,draw=cM,fill=cM,inner sep=0pt,minimum size=5pt] at (6.300,10.900) {};
\node[anchor=west,font=\footnotesize] at (6.55,10.90) {Morita's URTMs (simulating cyclic tag systems)~\cite{Morita17}};
\node[rectangle,draw=cM,fill=cM,inner sep=0pt,minimum size=4.5pt] at (6.300,10.450) {};
\node[anchor=west,font=\footnotesize] at (6.55,10.45) {Morita's URTMs (converted from other URTMs)~\cite{Morita17}};
\node[regular polygon,regular polygon sides=3,draw=cU,fill=white,thick,inner sep=0pt,minimum size=6.5pt] at (6.300,10.000) {};
\node[anchor=west,font=\footnotesize] at (6.55,10.00) {this paper};
\draw[black!80, dash dot, thick] (6.000,9.550) -- (6.520,9.550);
\node[anchor=west,font=\footnotesize] at (6.55,9.55) {boundary of the region where UTMs are known not to exist~\cite{WoodsNeary09}};
\draw[dashed,cO,thick] (6.000,9.100) -- (6.520,9.100);
\node[anchor=west,font=\footnotesize] at (6.55,9.10) {no URTM with at most two states (Corollary~\ref{cor:twostate})};
\node[cross out,draw=cO,very thick,inner sep=0pt,minimum size=6pt] at (6.260,8.650) {};
\node[anchor=west,font=\footnotesize] at (6.55,8.65) {no three-state two-symbol URTM (Theorem~\ref{thm:threetwo})};
\end{tikzpicture}}
\caption{Numbers of states and symbols of URTMs}
\label{fig:compare}
\end{figure}

\paragraph{Contributions.}
\begin{itemize}
\item \emph{New upper bounds} (Section~\ref{sec:constr}).
We give nine URTMs, $(3,600)$, $(4,38)$, $(5,36)$, $(6,28)$, $(7,26)$, $(8,21)$, $(9,8)$, $(10,7)$, and $(85,2)$, none of which is dominated by Morita's nine points; that is, compared with each of the nine points, each of them has fewer states or fewer symbols (Figure~\ref{fig:compare}).
The three-state machine improves the conversion of Morita's Theorem~5.6~\cite{Morita17}, the machines with four to eight states come from a block decomposition of the conversion of Morita's Theorem~5.5 followed by symbol merging, the ten-state machine merges two states of URTM$(13,7)$, keeps them apart by tags on the cells, and then merges symbols, three times, and the nine-state machine does this once more, and the two-symbol machine optimizes the conversion of Morita's Theorem~5.4.
The block decomposition gives a sequence of machines with four to nine states in which more states mean fewer symbols (Table~\ref{tab:blocks}); no conversion in the book yields five to nine states.
\item \emph{Universality in the sense of Morita's definition} (Section~\ref{sec:univ}).
Morita's Definition~7.1~\cite{Morita17} asks for total recursive encoding and decoding functions, but the book does not give them precisely for its URTMs, noting that they are not difficult to define from the constructions but their descriptions would be lengthy.
We give computable encodings and decodings for all 13 non-dominated points and prove universality in this sense.
\item \emph{Nonexistence for two states} (Section~\ref{sec:twostate}).
The halting problem of RTMs with at most two states is decidable for finite instantaneous descriptions, uniformly in the machine.
Hence no URTM with at most two states exists, for any number of symbols.
The key fact is that, in a two-state RTM, the state equals the direction of the last move, so a half-line of the tape can tell the rest of the tape only that the head has come back.
\item \emph{Nonexistence for three states and two symbols} (Section~\ref{sec:threetwo}).
No three-state, two-symbol RTM is universal.
\item \emph{Formalization} (Section~\ref{sec:formal}).
All theorems above are formalized and proved in Lean~4~\cite{Lean4} with Mathlib~\cite{Mathlib20}, using only the standard axioms.
The formalized machine model reproduces the halting step counts of the published test runs of all nine of Morita's machines, except one published count (Section~\ref{sec:formal}).
\end{itemize}
For each number of states, the constructions give an upper bound on the minimum number of symbols of a URTM. We do not claim that any of the new points is minimal.

Figure~\ref{fig:compare} shows Morita's nine points, our nine points, and the nonexistence results.
The dash-dotted line in Figure~\ref{fig:compare} bounds the region where UTMs are known not to exist~\cite{WoodsNeary09}: one-state machines, two-state and three-state two-symbol machines (under a definition that leaves one entry of the table empty for halting), and one-symbol machines.
These results concern TMs that need not be reversible, and their treatment of halting and of zero-shifts differs from ours, so we do not use them for URTMs in our sense.
The green lines mark our own nonexistence results (Sections~\ref{sec:twostate} and~\ref{sec:threetwo}), which are stated for the same notion of universality as Theorem~\ref{thm:exists}.

\paragraph{Related work.}
Small UTMs have been studied for a long time; Woods and Neary survey the results~\cite{WoodsNeary09}.
The non-dominated small UTMs listed in Morita's book include UTM$(2,18)$ and UTM$(4,6)$ of Rogozhin~\cite{Rogozhin96}, UTM$(3,9)$ of Kudlek and Rogozhin~\cite{KudlekRogozhin02}, and UTM$(5,5)$, $(6,4)$, $(9,3)$, and $(15,2)$ of Neary and Woods~\cite{NearyWoods09}.
These machines need not be reversible, and their conventions on halting and on stationary moves differ from those used here.
The URTMs of Morita and his coauthors are collected in~\cite[Ch.~7]{Morita17}, together with the conversions that we improve.
Turing machines and universal Turing machines have been formalized in Isabelle/HOL~\cite{XuZhangUrban13}, Matita~\cite{AspertiRicciotti15}, and Coq~\cite{ForsterKunzeWuttke20}, and Mathlib contains a universal partial recursive function. %
These formalizations treat general constructions rather than small machines, and none of them is about reversible computation.
Closest to our two-state result is Dudenhefner's decision procedure, certified in Coq, for the halting problem of reversible two-counter machines with a particular instruction set; it contrasts with Morita's result that reversible two-counter machines with another instruction set are universal~\cite{Dudenhefner22}.
To our knowledge, small URTMs and their universality in the sense of Morita's Definition~7.1 have not been formalized before.

\section{Preliminaries}
\label{sec:prelim}

We follow Morita's book~\cite{Morita17} and state the definitions in the form used in the formalization.

\begin{definition}[Machines and runs]
A \emph{quintuple} $r=[p,s,t,D,q]$ consists of a source state $p$, a read symbol $s$, a written symbol $t$, a direction $D\in\{\rmL,\rmZ,\rmR\}$, and a target state $q$, where $p,s,t,q\in\mathbb{N}$.
The directions L, Z, and R move the head by $-1$, $0$, and $+1$; a quintuple with direction Z is a \emph{zero-shift}.
A Turing machine (TM) is a finite list $T$ of quintuples; only membership in $T$ matters. We also call a TM a \emph{table}.
No initial or halting states are specified; halting is defined by the absence of an applicable quintuple.
An \emph{instantaneous description} (ID) is a triple $(\mathit{st},\pos,\mathit{tape})$ with $\mathit{st}\in\mathbb{N}$, $\pos\in\mathbb{Z}$, and $\mathit{tape}:\mathbb{Z}\to\mathbb{N}$; the symbol 0 is the blank.
A quintuple $r$ applies to an ID $\alpha$ if $r.p=\alpha.\mathit{st}$ and $r.s=\alpha.\mathit{tape}(\alpha.\pos)$; the result writes $r.t$ at $\alpha.\pos$, moves the head by the displacement of $r.D$, and enters $r.q$.
An ID is \emph{halting} for $T$ if no quintuple of $T$ applies to it.
We write $\mathrm{Halts}_T(\alpha)$ if $T$ reaches a halting ID from $\alpha$ in finitely many (possibly zero) steps.
\end{definition}

\begin{definition}[Determinism and reversibility]
$T$ is \emph{deterministic} if any two quintuples of $T$ with the same source state and read symbol are equal.
$T$ is \emph{reversible} if any two distinct quintuples of $T$ with the same target state have the same direction and different written symbols.
\end{definition}

This is Morita's Definition~5.4~\cite{Morita17}, specialized to the quintuple form.
By reversibility, all quintuples entering a state move in the same direction (the unique shift-direction property, \cite[Lemma~5.5]{Morita17}).
For an RTM $T$ we write $Q_+$ and $Q_-$ for the sets of states entered by right moves and by left moves, and $m_+=|Q_+|$, $m_-=|Q_-|$. A state entered only by zero-shifts, or not entered by any quintuple, belongs to neither set.

\begin{definition}[Size]
$\mathrm{NStates}(T)$ is the number of distinct source states of $T$, and $\mathrm{NSyms}(T)$ is the number of elements of $\{0\}\cup\{r.s,r.t\mid r\in T\}$.
A point $(m,n)$ \emph{dominates} $(m',n')$ if $m\le m'$, $n\le n'$, and $(m,n)\ne(m',n')$.
\end{definition}

Halting states, which have no outgoing quintuples, are not counted, as in Morita's book (Section~\ref{sec:counting}).

\paragraph{Codes of finite IDs.}
An ID is \emph{finite} if it has finitely many non-blank cells.
Let $\Code=\mathbb{N}\times\mathrm{List}\,\mathbb{N}\times\mathrm{List}\,\mathbb{N}$.
For $p\in\mathbb{Z}$ and $(q,\mathit{ls},\mathit{rs})\in\Code$, the ID $\mathrm{toID}(p,(q,\mathit{ls},\mathit{rs}))$ has state $q$ and head position $p$, carries $\mathit{rs}$ on the cells $p,p+1,\dots$ and $\mathit{ls}$ on the cells $p-1,p-2,\dots$, and is blank elsewhere.
For an ID $\alpha$, $\alpha.\pos$ is its head position.
$\mathrm{IsIDOf}(T,\alpha)$ holds if $\alpha$ is finite, its state occurs in $T$, and every cell holds the blank or a symbol occurring in $T$.
The enumeration $\mathrm{tm}(i)$, $i\in\mathbb{N}$, decodes $i$ into a list of 5-tuples of natural numbers by the standard bijection in Mathlib, reads each 5-tuple as a quintuple whose direction is L, Z, or R according as the fourth component is 0, 1, or at least 2, and replaces a non-deterministic result by the empty list.

\begin{definition}[Universality, after {\cite[Def.~7.1]{Morita17}}]
\label{def:univ}
A table $U$ satisfies $\mathrm{IsUTMStrict}(U)$ if there are an encoding $c:\mathbb{N}\times\Code\to\Code$ and a decoding $d:\Code\to\mathrm{Option}\,\Code$, both computable, such that for all $i\in\mathbb{N}$ and all $a\in\Code$ with $\mathrm{IsIDOf}(\mathrm{tm}(i),\mathrm{toID}(0,a))$:
\begin{enumerate}
\item $\mathrm{IsIDOf}(U,\mathrm{toID}(0,c(i,a)))$;
\item if $\mathrm{tm}(i)$ reaches a halting ID $\beta$ from $\mathrm{toID}(0,a)$, then $U$ reaches a halting ID $\gamma$ from $\mathrm{toID}(0,c(i,a))$, $\gamma$ has a code (some $g$ with $\mathrm{toID}(\gamma.\pos,g)=\gamma$), and every such code $g$ satisfies $d(g)=\mathrm{some}(b)$ for some $b$ with $\mathrm{toID}(\beta.\pos,b)=\beta$;
\item if $\mathrm{tm}(i)$ does not halt from $\mathrm{toID}(0,a)$, then $U$ does not halt from $\mathrm{toID}(0,c(i,a))$.
\end{enumerate}
The weaker predicate $\mathrm{IsUTM}(U)$ drops condition~1 and requires the decoding in condition~2 only for some code of $\gamma$.
\end{definition}

$\mathrm{IsUTMStrict}$ implies $\mathrm{IsUTM}$.
We use the strong form for existence and the weak form for nonexistence whenever possible, so both kinds of results hold under either form, except the nonexistence of three-state, two-symbol URTMs (Theorem~\ref{thm:threetwo}), which is stated for the strong form.
The requirement in condition~2 that \emph{every} code of $\gamma$ decodes correctly matters: a code can be padded with trailing blanks without changing the ID it denotes, so a condition on \emph{some} code alone would let the correctness of decoding depend on how the code is written (the number of trailing blanks), which is weak.
Morita's IDs are finite strings with the state inserted just before the symbol under the head, without a head position, whereas our IDs are two-way infinite tapes; we prove in Lean that, for deterministic tables, $\mathrm{IsUTMStrict}$ implies the formulation on string IDs. In that formulation an ID is a state with two finite strings, the head reading the first symbol of the second one, and two IDs are identified if they differ only by blanks at either end; a table $U$ satisfies it if there are computable $c$ and $d$ such that, for every $i$ and every ID $\alpha$ of $\mathrm{tm}(i)$, if $\mathrm{tm}(i)$ halts from $\alpha$ in $\beta$, then $U$ halts from $c(i,\alpha)$ in some $\gamma$ and $d(\gamma)$ is identified with $\beta$, and if $\mathrm{tm}(i)$ does not halt from $\alpha$, neither does $U$ from $c(i,\alpha)$.

The proofs of universality go through the following cyclic tag systems with halting condition (CTSHs).

\begin{definition}[Cyclic tag system with halting condition {\cite[Def.~7.5]{Morita17}}]
A CTSH is $\hat C=(k,(\mathrm{halt},w_1,\dots,w_{k-1}))$ with $k\ge 1$ and $w_j\in\{Y,N\}^*$.
An ID is a pair $(v,j)$ of a string $v\in\{Y,N\}^*$ and a phase $j\in\{0,\dots,k-1\}$.
The transitions are $(Yv,j)\Rightarrow(vw_j,j+1\bmod k)$ if $j\ne 0$, and $(Nv,j)\Rightarrow(v,j+1\bmod k)$ for every $j$.
The IDs $(Yv,0)$ and $(\lambda,j)$ are halting, where $\lambda$ is the empty string.
\end{definition}

Morita's six original machines simulate an arbitrary CTSH started from $(v,0)$, and CTSHs can simulate arbitrary TMs through 2-tag systems~\cite{CockeMinsky64}, \cite[Sect.~7.2]{Morita17}.

\section{Constructions}
\label{sec:constr}

All our machines are obtained from Morita's six original machines by conversions.
Table~\ref{tab:conversions} summarizes how each conversion changes the numbers of states and symbols; its notation is defined in the respective sections, and an indented row improves the conversion of the row above it.
Throughout, $T$ is the RTM to which a conversion is applied and $T'$ is the converted RTM; $m,n$ are the numbers of states and symbols of $T$, and $m',n'$ those of $T'$.
States of $T$ are written in lower case, such as $p,q,p_i$, and states of $T'$ in upper case, such as $\rmL_0,H_+,A_-$.
When we distinguish the machines obtained by different conversions, we write $T'_1,T'_2,\dots$.

\begin{table}[tb]
\centering
\caption{Conversions and their effect on the numbers of states and symbols}
\label{tab:conversions}
\small
\resizebox{\textwidth}{!}{%
\begin{tabular}{llllc}
\toprule
Conversion & States & Symbols & Example & Source \\
\midrule
\multicolumn{5}{l}{Reducing states (symbols increase)} \\
\quad To four states & 4 & $(2m+1)n$ & $(10,8)\to(4,168)$ & \cite[Thm.~5.5]{Morita17} \\
\qquad Disjoint ranges by direction & 4 & $(m+1)n$ & $(10,8)\to(4,88)$ & Sect.~\ref{sec:four} \\
\qquad Reverse counting & 4 & $\le(\max(m_+,m_-)+1)n$ & $(10,8)\to(4,48)$ & Sect.~\ref{sec:four} \\
\qquad Blocks and merging of roles & 4--9 & $n+\sum_{i\in V}|\Sigma_i|$ & $(15,6)\to(7,32)$ & Sect.~\ref{sec:blocks} \\
\quad To three states & 3 & $N(2M-1)$ & $(32,3)\to(3,36654)$ & \cite[Thm.~5.6]{Morita17} \\
\qquad One symbol per cell, shared counters & 3 & $O(mn^2)$ & $(32,3)\to(3,600)$ & Sect.~\ref{sec:three} \\
\multicolumn{5}{l}{Reducing symbols (states increase)} \\
\quad To two symbols & $\le mk+P_r+P_w$ & 2 & $(24,4)\to(138,2)$ & \cite[Thm.~5.4]{Morita17} \\
\qquad Code optimization, state merging & per table & 2 & $(13,7)\to(85,2)$ & Sect.~\ref{sec:two} \\
\multicolumn{5}{l}{Same number of states} \\
\quad Symbol merging & unchanged & number of colors & $(4,43)\to(4,38)$ & Sect.~\ref{sec:merge} \\
\multicolumn{5}{l}{One state fewer (symbols increase)} \\
\quad Merging two states by cell tags & $m-1$ & per table & $(13,7)\to(12,8)$ & Sect.~\ref{sec:tags} \\
\bottomrule
\end{tabular}}
\end{table}

\subsection{Four states: disjoint ranges by direction and reverse counting}
\label{sec:four}

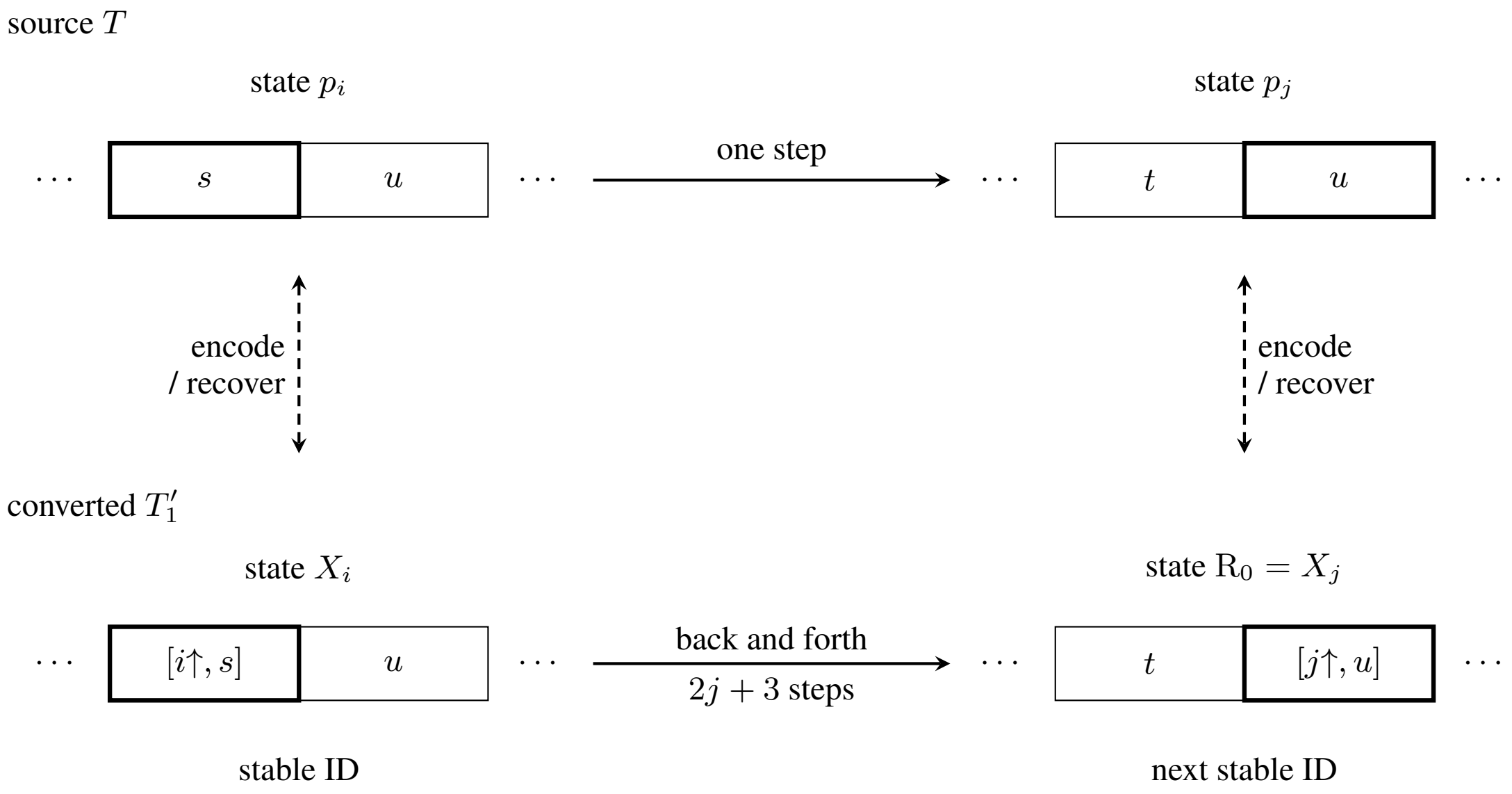
\begin{figure}[tb]
\centering
\begin{tikzpicture}[
  x=1mm,y=1mm,>=stealth,font=\small,
  cell/.style={draw,minimum width=18mm,minimum height=7mm,inner sep=0pt},
  head/.style={cell,line width=1.2pt},
  note/.style={font=\footnotesize},
  correspondence/.style={<->,densely dashed,thick}
]
\node[anchor=west] at (-20,43) {source $T$};
\node at (9,37) {state $p_i$};
\node at (99,37) {state $p_j$};
\node[head] at (0,28) {$s$};
\node[cell] at (18,28) {$u$};
\node at (-14,28) {$\cdots$};
\node at (32,28) {$\cdots$};
\node[cell] at (90,28) {$t$};
\node[head] at (108,28) {$u$};
\node at (76,28) {$\cdots$};
\node at (122,28) {$\cdots$};
\draw[->,thick] (37,28) -- node[above,note] {one step} (71,28);

\draw[correspondence] (9,19) -- node[left,note,align=right] {encode\\/ recover} (9,2);
\draw[correspondence] (99,19) -- node[right,note,align=left] {encode\\/ recover} (99,2);

\node[anchor=west] at (-20,-3) {converted $T'_1$};
\node at (9,-9) {state $X_i$};
\node at (99,-9) {state $\rmR_0=X_j$};
\node[head] at (0,-18) {$[i{\uparrow},s]$};
\node[cell] at (18,-18) {$u$};
\node at (-14,-18) {$\cdots$};
\node at (32,-18) {$\cdots$};
\node[cell] at (90,-18) {$t$};
\node[head] at (108,-18) {$[j{\uparrow},u]$};
\node at (76,-18) {$\cdots$};
\node at (122,-18) {$\cdots$};
\draw[->,thick] (37,-18) -- node[above,note] {back and forth}
  node[below,note] {$2j+3$ steps} (71,-18);
\node[note] at (9,-28) {stable ID};
\node[note] at (99,-28) {next stable ID};
\end{tikzpicture}
\caption{Correspondence between stable IDs and IDs of the source machine $T$}
\label{fig:stable-id-correspondence}
\end{figure}

Morita's conversion to four states~\cite[Sect.~5.3.4]{Morita17} keeps the number of the current state of $T$ on the tape, as a counter attached to the symbol under the head; we write $T'_1$ for the machine it produces.
To execute a quintuple $[p,s,t,D,q]$ of $T$, $T'_1$ writes the counter of $q$ into the current cell, places a counter with the initial value in the target cell, and then moves back and forth between the two cells, decrementing the counter of the source cell and incrementing that of the target cell, until the whole number has been carried over.
To distinguish the source cell from the target cell, each counter carries an up mark or a down mark.
We number the states of $T$ as $p_0,\dots,p_{m-1}$ and write $[i{\uparrow},s]$ and $[i{\downarrow},s]$ for the symbol $s$ with a counter of value $i$ and an up or a down mark.
We call a symbol of $T'_1$ made of a symbol of $T$ with a number and a mark a \emph{counter symbol}, and a symbol of $T$ itself a \emph{plain symbol}.
There are $m$ values $0,\dots,m-1$ and two marks, so $T'_1$ has $(2m+1)n$ symbols: the $n$ plain symbols and $2mn$ marked counter symbols.
From URTM$(10,8)$ it gives URTM$(4,168)$: with $m=10$ and $n=8$, the number of symbols is $n'=(2m+1)n=(2\cdot 10+1)\cdot 8=168$.
$T'_1$ has the four states $\mathrm{L}_0,\mathrm{L}_1,\mathrm{R}_0,\mathrm{R}_1$ ($m'=4$).
For a right-moving quintuple $[p_i,s,t,\rmR,p_j]$ of $T$, $T'_1$ uses the rules (named as in Morita's book)
\begin{center}
\small
\begin{tabular}{lllllll}
\toprule
& state of $T'_1$ & read & write & move & next & values of $k$ \\
\midrule
$\tilde\delta_1,\tilde\delta_2$ & $X_i$ & $[i{\uparrow},s]$ & $[j{\downarrow},t]$ & $\rmR$ & $\rmR_0$ & \\
$\tilde\delta_3$ & $\rmR_0$ & $s'$ & $[0{\uparrow},s']$ & $\rmL$ & $\rmL_0$ & \\
$\tilde\delta_4$ & $\rmL_0$ & $[k{\downarrow},x]$ & $[(k-1){\downarrow},x]$ & $\rmR$ & $\rmR_1$ & $1,\dots,m-1$ ($m-1$ values) \\
$\tilde\delta_5$ & $\rmR_1$ & $[k{\uparrow},s']$ & $[(k+1){\uparrow},s']$ & $\rmL$ & $\rmL_0$ & $0,\dots,m-2$ ($m-1$ values) \\
$\tilde\delta_6$ & $\rmL_0$ & $[0{\downarrow},x]$ & $x$ & $\rmR$ & $\rmR_0$ & \\
\bottomrule
\end{tabular}
\end{center}
where $X_i$ is $\rmL_0$ if $p_i\notin Q_+$ (the rule is then $\tilde\delta_1$) and $\rmR_0$ if $p_i\in Q_+$ (the rule is then $\tilde\delta_2$); the rules $\tilde\delta_3$ to $\tilde\delta_6$ do not depend on the quintuple and are given for all symbols $s'$ and $x$ of $T$.
For a left-moving quintuple, the first rule moves left into $\rmL_0$, and in the other rules L and R are exchanged both in the directions and in the names of the states ($\rmL_1$ is used only in these left-moving rules).
Figure~\ref{fig:four} shows how it executes a right-moving quintuple; each row gives the state and the two cells $c$ and $c+1$ ($u$ is the symbol in cell $c+1$), and the thick frame marks the head.
The execution takes $2j+3$ steps.
Figure~\ref{fig:four-flow} shows the same execution as a flowchart; each box is headed by a rule and the state of $T'_1$ that executes it, and the test in $\rmL_0$ decides whether the back-and-forth phase continues.
We call an ID \emph{stable} if it has exactly one counter symbol on the whole tape, this symbol has an up mark and is under the head, and $T'_1$ is in $\rmL_0$ or $\rmR_0$.
An ID of $T$ in state $p_i$ with $s$ under the head corresponds to the stable ID of $T'_1$ obtained by replacing that symbol by $[i{\uparrow},s]$ and the state by $X_i$, and a step of $T$ corresponds to the execution of $T'_1$ from a stable ID to the next stable ID~\cite[proof of Thm.~5.5]{Morita17}. The first and the last rows of Fig.~\ref{fig:four} are stable IDs. Figure~\ref{fig:stable-id-correspondence} shows the correspondence with the IDs of the source machine $T$.
Figure~\ref{fig:stable-id-correspondence} also shows the execution of a right-moving quintuple $[p_i,s,t,\rmR,p_j]$: the thick frame marks the head, the left and right cells are $c$ and $c+1$, and the other cells do not change. To go from the lower row back to the upper row, replace the counter symbol by its plain symbol and take the number of the counter as the state number of $T$; the head position is kept.
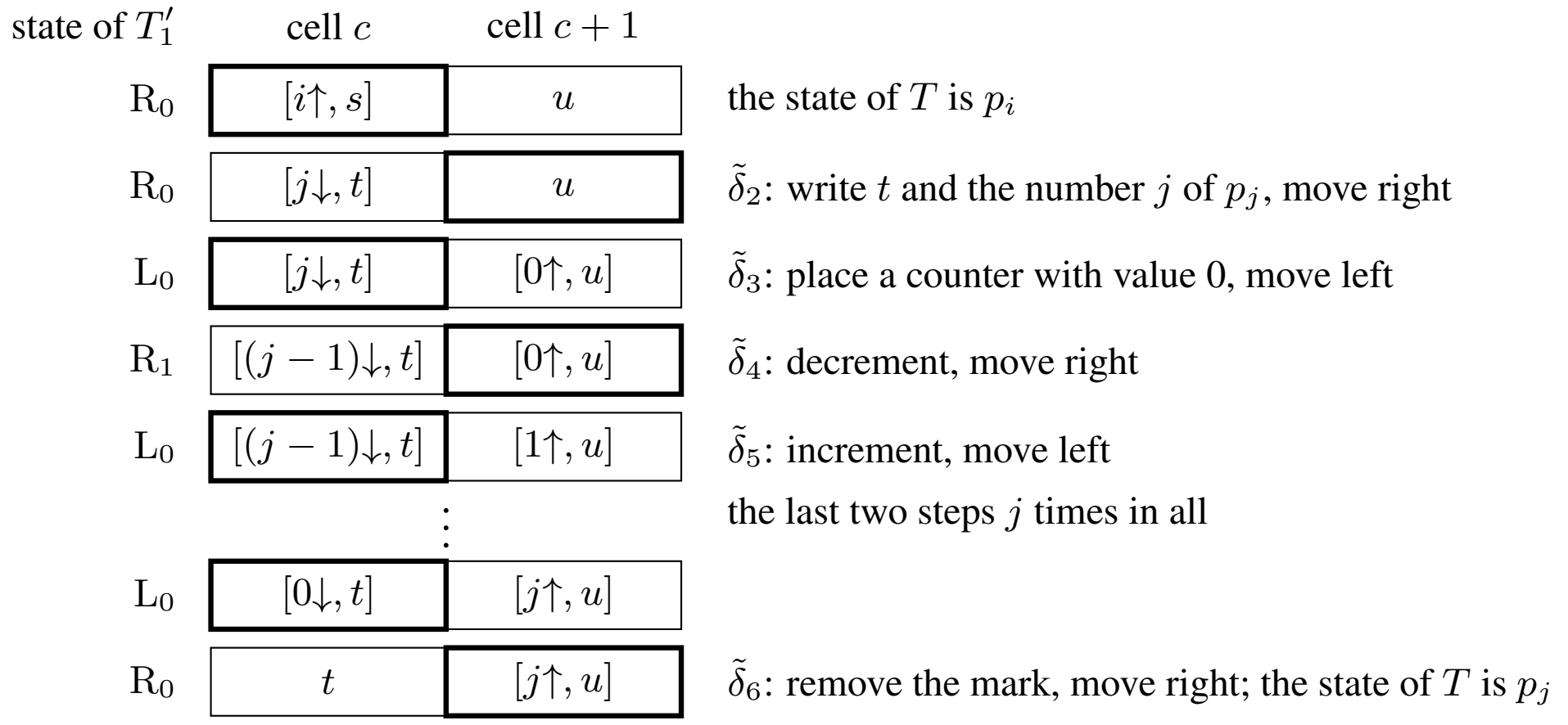
\begin{figure}[tb]
\centering
\begin{tikzpicture}[x=1mm,y=1mm,cell/.style={draw,minimum width=19mm,minimum height=5.5mm,inner sep=1pt,font=\small}]
\def\tmrow#1#2#3#4#5#6{%
  \node[anchor=east,font=\small] at (-1,#1) {$#2$};
  \node[cell] at (10,#1) {$#3$};
  \node[cell] at (29,#1) {$#4$};
  \ifnum#5=0 \draw[very thick] (0.5,{#1-2.75}) rectangle (19.5,{#1+2.75});\else \draw[very thick] (19.5,{#1-2.75}) rectangle (38.5,{#1+2.75});\fi
  \node[anchor=west,font=\footnotesize] at (41,#1) {#6};}
\node[font=\footnotesize] at (10,6) {cell $c$};
\node[font=\footnotesize] at (29,6) {cell $c+1$};
\node[font=\footnotesize,anchor=east] at (-1,6) {state of $T'_1$};
\tmrow{0}{\mathrm{R}_0}{[i{\uparrow},s]}{u}{0}{the state of $T$ is $p_i$}
\tmrow{-7}{\mathrm{R}_0}{[j{\downarrow},t]}{u}{1}{$\tilde\delta_2$: write $t$ and the number $j$ of $p_j$, move right}
\tmrow{-14}{\mathrm{L}_0}{[j{\downarrow},t]}{[0{\uparrow},u]}{0}{$\tilde\delta_3$: place a counter with value 0, move left}
\tmrow{-21}{\mathrm{R}_1}{[(j-1){\downarrow},t]}{[0{\uparrow},u]}{1}{$\tilde\delta_4$: decrement, move right}
\tmrow{-28}{\mathrm{L}_0}{[(j-1){\downarrow},t]}{[1{\uparrow},u]}{0}{$\tilde\delta_5$: increment, move left}
\node at (19.5,-33.5) {$\vdots$};
\node[anchor=west,font=\footnotesize] at (41,-33.5) {the last two steps $j$ times in all};
\tmrow{-40}{\mathrm{L}_0}{[0{\downarrow},t]}{[j{\uparrow},u]}{0}{}
\tmrow{-47}{\mathrm{R}_0}{t}{[j{\uparrow},u]}{1}{$\tilde\delta_6$: remove the mark, move right; the state of $T$ is $p_j$}
\end{tikzpicture}
\caption{$T'_1$ (Morita's four-state conversion) executing a right-moving quintuple $[p_i,s,t,\mathrm{R},p_j]$ of $T$}
\label{fig:four}
\end{figure}
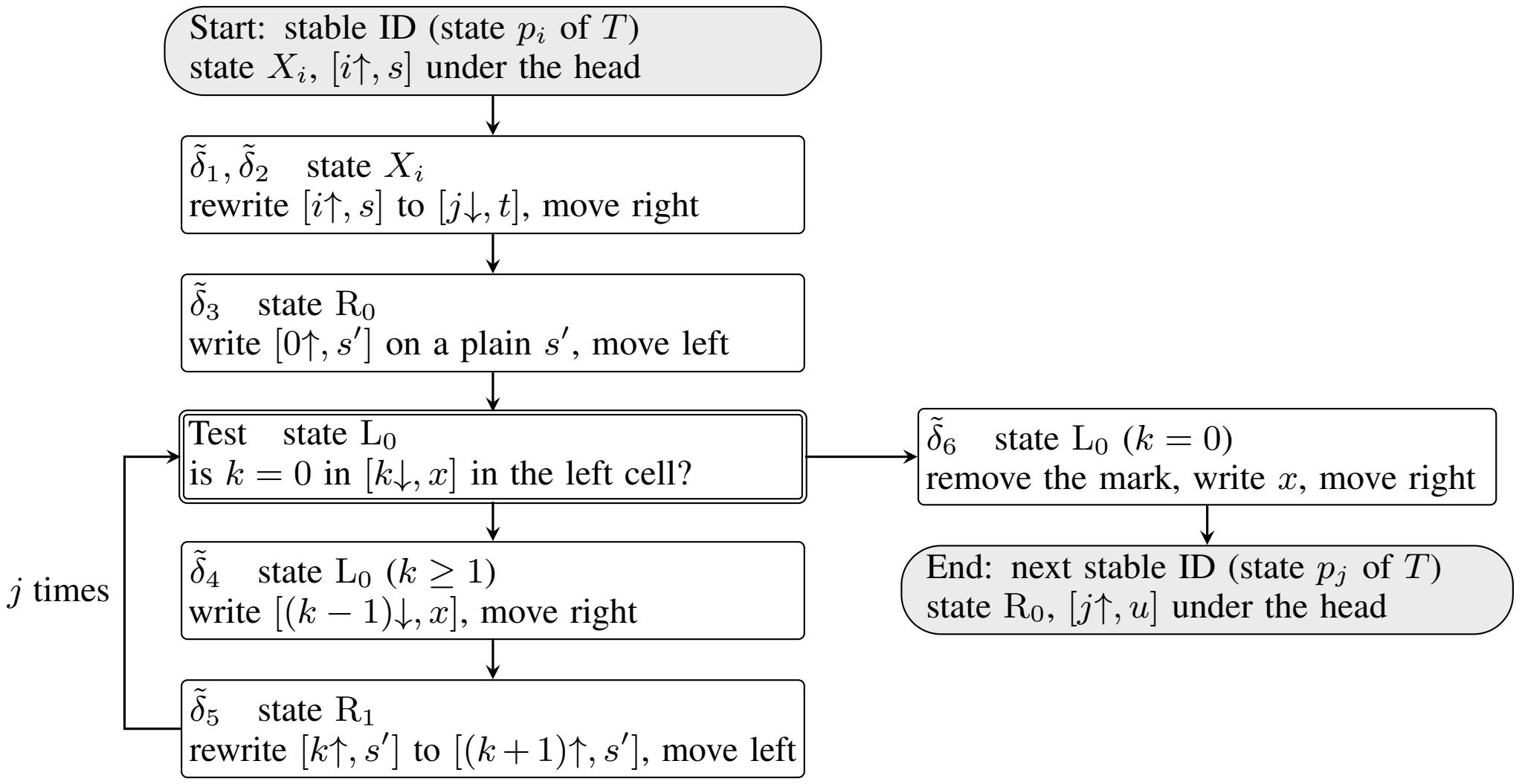
\begin{figure}[tb]
\centering
\begin{tikzpicture}[x=1mm,y=1mm,font=\footnotesize,node distance=3.5mm]
\node[flterm,text width=54mm] (S) at (0,0) {Start: stable ID (state $p_i$ of $T$)\\ state $X_i$, $[i{\uparrow},s]$ under the head};
\node[flbox,text width=54mm] (A) [below=of S] {$\tilde\delta_1,\tilde\delta_2$\quad state $X_i$\\ rewrite $[i{\uparrow},s]$ to $[j{\downarrow},t]$, move right};
\node[flbox,text width=54mm] (B) [below=of A] {$\tilde\delta_3$\quad state $\rmR_0$\\ write $[0{\uparrow},s']$ on a plain $s'$, move left};
\node[fldec,text width=54mm] (Q) [below=of B] {Test\quad state $\rmL_0$\\ is $k=0$ in $[k{\downarrow},x]$ in the left cell?};
\node[flbox,text width=54mm] (C) [below=of Q] {$\tilde\delta_4$\quad state $\rmL_0$ ($k\ge 1$)\\ write $[(k-1){\downarrow},x]$, move right};
\node[flbox,text width=54mm] (D) [below=of C] {$\tilde\delta_5$\quad state $\rmR_1$\\ rewrite $[k{\uparrow},s']$ to $[(k+1){\uparrow},s']$, move left};
\node[flbox,text width=50mm] (F) [right=10mm of Q] {$\tilde\delta_6$\quad state $\rmL_0$ ($k=0$)\\ remove the mark, write $x$, move right};
\node[flterm,text width=50mm] (E) [below=of F] {End: next stable ID (state $p_j$ of $T$)\\ state $\rmR_0$, $[j{\uparrow},u]$ under the head};
\draw[flarr] (S) -- (A); \draw[flarr] (A) -- (B); \draw[flarr] (B) -- (Q);
\draw[flarr] (Q) -- (C); \draw[flarr] (C) -- (D);
\draw[flarr] (D.west) -- ++(-5,0) |- node[pos=0.25,left] {$j$ times} (Q.west);
\draw[flarr] (Q) -- (F); \draw[flarr] (F) -- (E);
\end{tikzpicture}
\caption{Flowchart of $T'_1$ executing a right-moving quintuple}
\label{fig:four-flow}
\end{figure}

\begin{figure}[tb]
\centering
\begin{tikzpicture}[x=1mm,y=1mm,font=\scriptsize,node distance=3mm]
\node[flterm,text width=38mm] (rS) at (0,0) {Start: stable ID\\ $H_e$ reads $[\nu(p),s]$ ($T$ in $p$)};
\node[flbox,text width=38mm] (rA) [below=of rS] {$\theta_1$\quad $H_e$\\ write $[\nu(q),t]$, move right};
\node[flbox,text width=38mm] (rB) [below=of rA] {$\theta_2$\quad $H_+$\\ write $[\ell_+,s']$ on $s'$, move left};
\node[fldec,text width=38mm] (rQ) [below=of rB] {Test\quad $A_-$\\ reads the source $[i,x]$};
\node[flbox,text width=38mm] (rC) [below=of rQ] {$\theta_3$\quad $A_-$ ($i>\ell_+$)\\ write $[i-1,x]$, move right};
\node[flbox,text width=38mm] (rD) [below=of rC] {$\theta_4$\quad $A_+$\\ rewrite $[j,s']$ to $[j+1,s']$, move left};
\node[flbox,text width=24mm] (rF) [right=4mm of rQ] {$\theta_5$\quad $A_-$ ($i=\ell_+$)\\ remove the counter, move right};
\node[flterm,text width=24mm] (rE) [below=of rF] {End: next stable ID\\ $H_+$ reads $[\nu(q),u]$};
\node[anchor=south west,font=\footnotesize,text width=72mm,align=left] at ([shift={(-4,1.5)}]rS.north west) {Right-moving quintuple ($q\in Q_+$)};
\draw[flarr] (rS) -- (rA); \draw[flarr] (rA) -- (rB); \draw[flarr] (rB) -- (rQ);
\draw[flarr] (rQ) -- (rC); \draw[flarr] (rC) -- (rD);
\draw[flarr] (rD.west) -- ++(-3.5,0) |- (rQ.west);
\draw[flarr] (rQ) -- (rF); \draw[flarr] (rF) -- (rE);
\node[flterm,text width=38mm] (lS) at (82,0) {Start: stable ID\\ $H_e$ reads $[\nu(p),s]$ ($T$ in $p$)};
\node[flbox,text width=38mm] (lA) [below=of lS] {$\theta_1$\quad $H_e$\\ write $[\nu(q),t]$, move left};
\node[flbox,text width=38mm] (lB) [below=of lA] {$\theta_2$\quad $H_-$\\ write $[\ell_-,s']$ on $s'$, move right};
\node[fldec,text width=38mm] (lQ) [below=of lB] {Test\quad $A_+$\\ reads the source $[i,x]$};
\node[flbox,text width=38mm] (lC) [below=of lQ] {$\theta_3$\quad $A_+$ ($i>\ell_-$)\\ write $[i-1,x]$, move left};
\node[flbox,text width=38mm] (lD) [below=of lC] {$\theta_4$\quad $A_-$\\ rewrite $[j,s']$ to $[j+1,s']$, move right};
\node[flbox,text width=24mm] (lF) [right=4mm of lQ] {$\theta_5$\quad $A_+$ ($i=\ell_-$)\\ remove the counter, move left};
\node[flterm,text width=24mm] (lE) [below=of lF] {End: next stable ID\\ $H_-$ reads $[\nu(q),u]$};
\node[anchor=south west,font=\footnotesize,text width=72mm,align=left] at ([shift={(-4,1.5)}]lS.north west) {Left-moving quintuple ($q\in Q_-$)};
\draw[flarr] (lS) -- (lA); \draw[flarr] (lA) -- (lB); \draw[flarr] (lB) -- (lQ);
\draw[flarr] (lQ) -- (lC); \draw[flarr] (lC) -- (lD);
\draw[flarr] (lD.west) -- ++(-3.5,0) |- (lQ.west);
\draw[flarr] (lQ) -- (lF); \draw[flarr] (lF) -- (lE);
\end{tikzpicture}
\caption{Flowcharts of $T'_2$ executing a right-moving and a left-moving quintuple}
\label{fig:fourdisj-flow}
\end{figure}
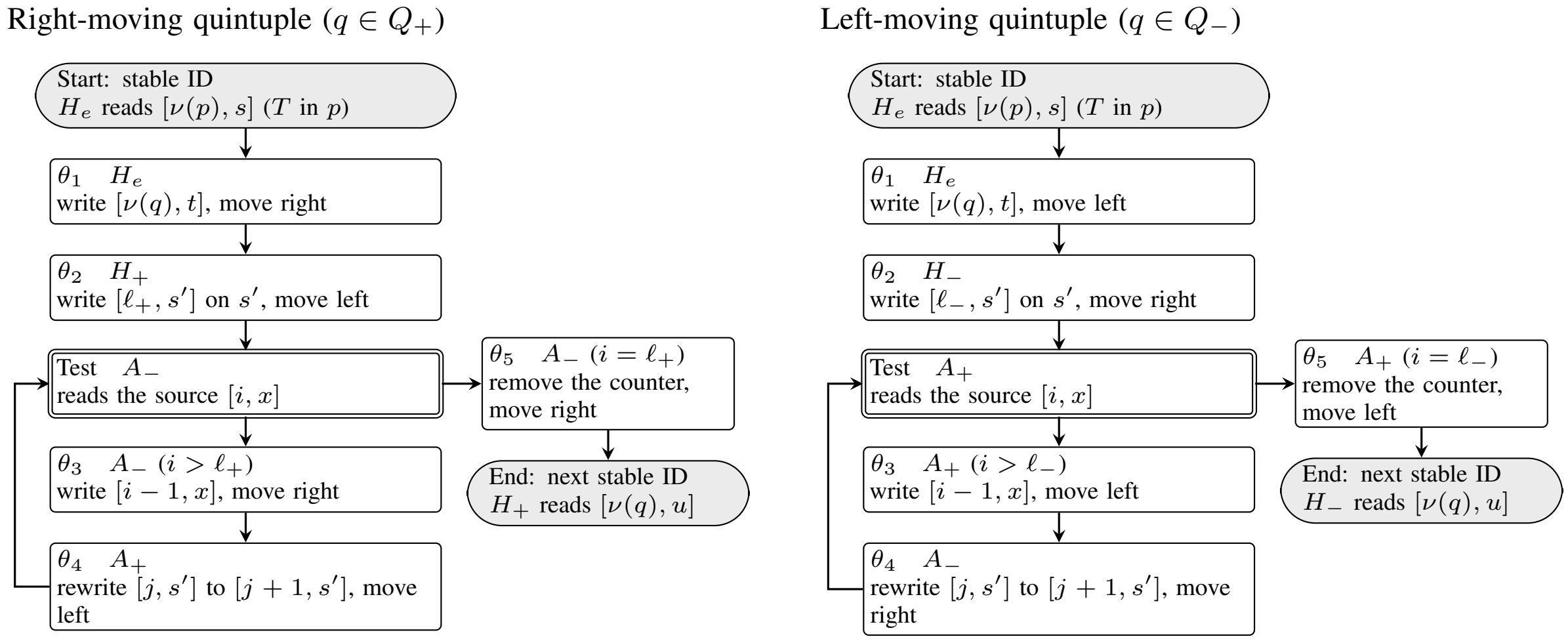

\paragraph{Disjoint ranges by direction.}
In $T'_1$, $\rmL_0$ reads both the counter of a stable ID (with an up mark) and the counter on the source side during the back-and-forth phase (with a down mark), and so does $\rmR_0$; the marks tell these two apart. If the state of stable IDs is separated from the back-and-forth states, and the two roles of a back-and-forth state are told apart by the counter value, the marks should not be needed.
In the four-state conversions (Sections~\ref{sec:four} and~\ref{sec:blocks}), a state that is not entered by any quintuple is put into $Q_+$. In Morita's tables to which we apply the conversions, the only zero-shifts are the quintuples into halting states, and the conversions map them separately to halting rules (the halting quintuples of Section~\ref{sec:univ}). Hence every state of $T$ is in exactly one of $Q_+$ and $Q_-$, and $m_++m_-=m$.
Number the states of $Q_+$ by $1,\dots,m_+$ and those of $Q_-$ by $m_++1,\dots,m$.
We write $T'_2$ for the machine this conversion produces. $T'_2$ has the four states $H_+,H_-,A_+,A_-$: $H$ is the state of stable IDs, $A$ is a back-and-forth state, and the index is the direction by which the state is entered. Its symbols are the plain symbols $s$ and the counter symbols $[i,s]$.
Here too, a stable ID has exactly one counter symbol on the whole tape, under the head, and $T'_2$ in an $H$-state; it represents the ID of $T$ obtained by replacing $[i,s]$ by $s$ and taking the state numbered $i$. While a quintuple is executed, there are two counter symbols, one in the source cell and one in the target cell (Fig.~\ref{fig:fourdisj}).
In the table below, $\nu$ is the numbering, $[\ell_D,h_D]$ is the range of numbers of $Q_D$, $\bar D$ is the direction opposite to $D$, and $s'$ and $x$ are arbitrary symbols of $T$; in subscripts, R is identified with $+$ and L with $-$.
For a quintuple $[p,s,t,D,q]$ of $T$ with $p\in Q_e$, $T'_2$ uses the rules
\begin{center}
\small
\begin{tabular}{lllllll}
\toprule
& state of $T'_2$ & read & write & move & next & range of values \\
\midrule
$\theta_1$ & $H_e$ & $[\nu(p),s]$ & $[\nu(q),t]$ & $D$ & $H_D$ & \\
$\theta_2$ & $H_D$ & $s'$ & $[\ell_D,s']$ & $\bar D$ & $A_{\bar D}$ & \\
$\theta_3$ & $A_{\bar D}$ & $[i,x]$ & $[i-1,x]$ & $D$ & $A_D$ & $i=\ell_D+1,\dots,h_D$ ($h_D-\ell_D$ values) \\
$\theta_4$ & $A_D$ & $[j,s']$ & $[j+1,s']$ & $\bar D$ & $A_{\bar D}$ & $j=\ell_D,\dots,h_D-1$ ($h_D-\ell_D$ values) \\
$\theta_5$ & $A_{\bar D}$ & $[\ell_D,x]$ & $x$ & $D$ & $H_D$ & \\
\bottomrule
\end{tabular}
\end{center}
Each row is a rule scheme, which gives one rule for each quintuple of $T$, each choice of the symbols $s'$ and $x$, and each value. We call these schemes $\theta_1,\dots,\theta_5$; they correspond to Morita's $\tilde\delta_1$ and $\tilde\delta_2$, $\tilde\delta_3$, $\tilde\delta_4$, $\tilde\delta_5$, and $\tilde\delta_6$, respectively.
The sum of the two counters is invariant during the back-and-forth phase, and $\theta_5$ fires exactly when the whole number has been carried, so a stable ID is restored after $3+2(\nu(q)-\ell_D)$ steps.
Figure~\ref{fig:fourdisj} shows the execution of a right-moving quintuple in the same form as Figure~\ref{fig:four}, with $i=\nu(p)$ and $j=\nu(q)$.
The state $A_+$ acts as the target side of a right move and as the source side of a left move, but the counter values it reads in the two roles lie in disjoint ranges, so the table is deterministic; reversibility follows by the same separation together with the reversibility of $T$.
Figure~\ref{fig:fourdisj-flow} shows the flowcharts of a right-moving and a left-moving quintuple side by side; the test and $\theta_3$, $\theta_4$, $\theta_5$ are the rules of $A_+$ and $A_-$, so the two directions use the same two states and tell the roles apart by the range of the value read.
In Morita's conversion, the up and down marks distinguish the source cell from the target cell. Here the subscript of the state (the direction by which the cell was entered) and the range of the counter value make this distinction.
Hence the marks are no longer needed, and the number of symbols becomes $(m+1)n$.
From URTM$(10,8)$ this gives URTM$(4,88)$: with $m=10$ and $n=8$, the number of symbols is $n'=(m+1)n=(10+1)\cdot 8=88$.
\begin{figure}[tb]
\centering
\begin{tikzpicture}[x=1mm,y=1mm,cell/.style={draw,minimum width=19mm,minimum height=5.5mm,inner sep=1pt,font=\small}]
\def\tmrow#1#2#3#4#5#6{%
  \node[anchor=east,font=\small] at (-1,#1) {$#2$};
  \node[cell] at (10,#1) {$#3$};
  \node[cell] at (29,#1) {$#4$};
  \ifnum#5=0 \draw[very thick] (0.5,{#1-2.75}) rectangle (19.5,{#1+2.75});\else \draw[very thick] (19.5,{#1-2.75}) rectangle (38.5,{#1+2.75});\fi
  \node[anchor=west,font=\footnotesize] at (41,#1) {#6};}
\node[font=\footnotesize] at (10,6) {cell $c$};
\node[font=\footnotesize] at (29,6) {cell $c+1$};
\node[font=\footnotesize,anchor=east] at (-1,6) {state of $T'_2$};
\tmrow{0}{H_e}{[i,s]}{u}{0}{the state of $T$ is $p$}
\tmrow{-7}{H_+}{[j,t]}{u}{1}{$\theta_1$: write $t$ and the number $j$ of $q$, move right}
\tmrow{-14}{A_-}{[j,t]}{[\ell_+,u]}{0}{$\theta_2$: place a counter with value $\ell_+$, move left}
\tmrow{-21}{A_+}{[j-1,t]}{[\ell_+,u]}{1}{$\theta_3$: decrement, move right}
\tmrow{-28}{A_-}{[j-1,t]}{[\ell_++1,u]}{0}{$\theta_4$: increment, move left}
\node at (19.5,-33.5) {$\vdots$};
\node[anchor=west,font=\footnotesize] at (41,-33.5) {the last two steps $j-\ell_+$ times in all};
\tmrow{-40}{A_-}{[\ell_+,t]}{[j,u]}{0}{}
\tmrow{-47}{H_+}{t}{[j,u]}{1}{$\theta_5$: remove the counter, move right: the state of $T$ is $q$}
\end{tikzpicture}
\caption{$T'_2$ (disjoint ranges by direction) executing a right-moving quintuple $[p,s,t,\mathrm{R},q]$ of $T$}
\label{fig:fourdisj}
\end{figure}

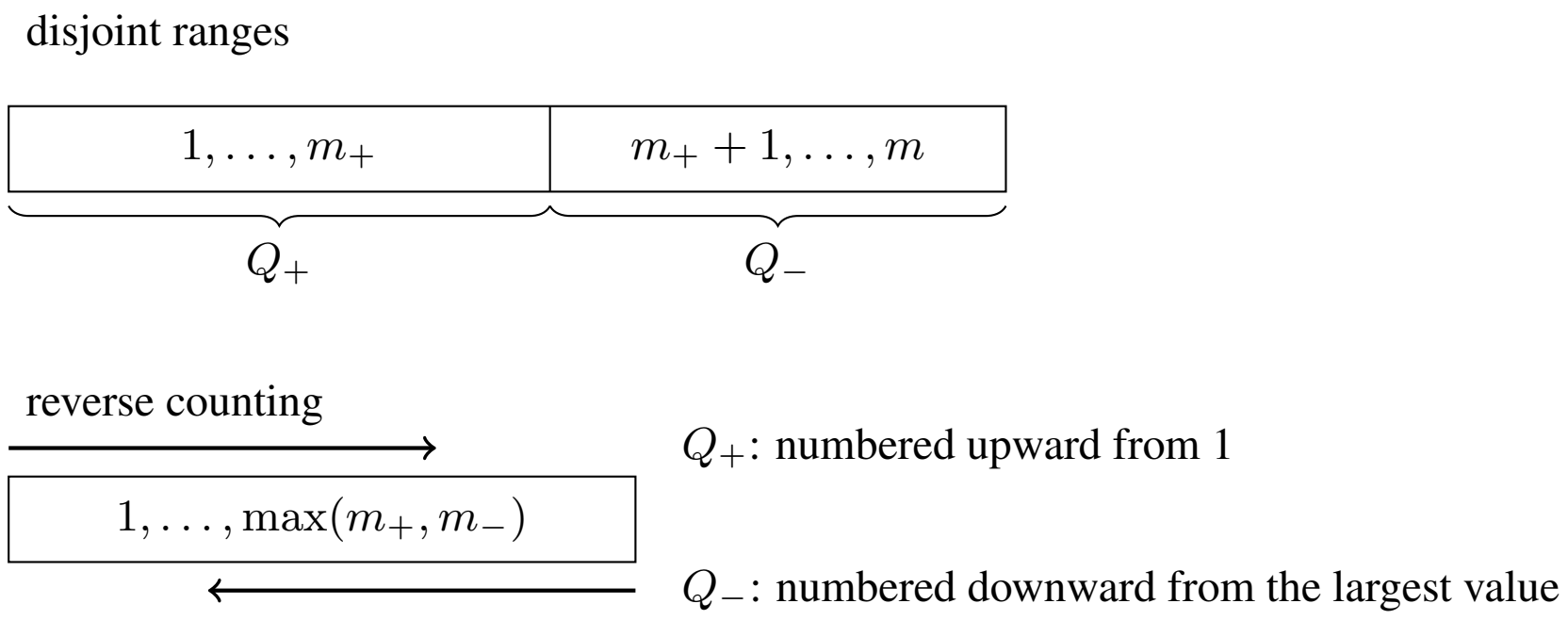
\begin{figure}[tb]
\centering
\begin{tikzpicture}[x=1mm,y=1mm,font=\small]
\node[anchor=west] at (0,8) {disjoint ranges};
\draw (0,-3) rectangle (70,3); \draw (38,-3) -- (38,3);
\node at (19,0) {$1,\dots,m_+$}; \node at (54,0) {$m_++1,\dots,m$};
\draw[decorate,decoration={brace,mirror,amplitude=4pt}] (0,-4) -- (38,-4) node[midway,below=4pt] {$Q_+$};
\draw[decorate,decoration={brace,mirror,amplitude=4pt}] (38,-4) -- (70,-4) node[midway,below=4pt] {$Q_-$};
\node[anchor=west] at (0,-18) {reverse counting};
\draw (0,-29) rectangle (44,-23);
\node at (22,-26) {$1,\dots,\max(m_+,m_-)$};
\draw[->,thick] (0,-21) -- (30,-21);
\node[anchor=west] at (46,-21) {$Q_+$: numbered upward from 1};
\draw[->,thick] (44,-31) -- (14,-31);
\node[anchor=west] at (46,-31) {$Q_-$: numbered downward from the largest value};
\end{tikzpicture}
\caption{Numbers of the states of $T$ put on counter symbols by $T'_2$ and $T'_3$}
\label{fig:ranges}
\end{figure}

\begin{figure}[tb]
\centering
\begin{tikzpicture}[x=1mm,y=1mm,font=\scriptsize,node distance=3mm]
\node[flterm,text width=38mm] (rS) at (0,0) {Start: stable ID\\ $H_e$ reads $[\nu(p),s]$ ($T$ in $p$)};
\node[flbox,text width=38mm] (rA) [below=of rS] {$\theta_1$\quad $H_e$\\ write $[\nu(q),t]$, move right};
\node[flbox,text width=38mm] (rB) [below=of rA] {$\theta_2$\quad $H_+$\\ write $[\ell,s']$ on $s'$, move left};
\node[fldec,text width=38mm] (rQ) [below=of rB] {Test\quad $A_-$\\ reads the source $[i,x]$};
\node[flbox,text width=38mm,line width=1.1pt] (rC) [below=of rQ] {$\theta_3$\quad $A_-$ ($i\ne\ell$), shared 1\\ rewrite $[i,x]$ to $[i-1,x]$, move right};
\node[flbox,text width=38mm,line width=1.1pt] (rD) [below=of rC] {$\theta_4$\quad $A_+$, shared 2\\ rewrite $[i,x]$ to $[i+1,x]$, move left};
\node[flbox,text width=24mm] (rF) [right=4mm of rQ] {$\theta_5$\quad $A_-$ ($i=\ell$)\\ remove the counter, move right};
\node[flterm,text width=24mm] (rE) [below=of rF] {End: next stable ID\\ $H_+$ reads $[\nu(q),u]$};
\node[anchor=south west,font=\footnotesize,text width=72mm,align=left] at ([shift={(-4,1.5)}]rS.north west) {Right-moving quintuple ($q\in Q_+$, sentinel $\ell$ the minimum)};
\draw[flarr] (rS) -- (rA); \draw[flarr] (rA) -- (rB); \draw[flarr] (rB) -- (rQ);
\draw[flarr] (rQ) -- (rC); \draw[flarr] (rC) -- (rD);
\draw[flarr] (rD.west) -- ++(-3.5,0) |- (rQ.west);
\draw[flarr] (rQ) -- (rF); \draw[flarr] (rF) -- (rE);
\node[flterm,text width=38mm] (lS) at (82,0) {Start: stable ID\\ $H_e$ reads $[\nu(p),s]$ ($T$ in $p$)};
\node[flbox,text width=38mm] (lA) [below=of lS] {$\theta_1$\quad $H_e$\\ write $[\nu(q),t]$, move left};
\node[flbox,text width=38mm] (lB) [below=of lA] {$\theta_2$\quad $H_-$\\ write $[h,s']$ on $s'$, move right};
\node[fldec,text width=38mm] (lQ) [below=of lB] {Test\quad $A_+$\\ reads the source $[i,x]$};
\node[flbox,text width=38mm,line width=1.1pt] (lC) [below=of lQ] {$\theta_3$\quad $A_+$ ($i\ne h$), shared 2\\ rewrite $[i,x]$ to $[i+1,x]$, move left};
\node[flbox,text width=38mm,line width=1.1pt] (lD) [below=of lC] {$\theta_4$\quad $A_-$, shared 1\\ rewrite $[i,x]$ to $[i-1,x]$, move right};
\node[flbox,text width=24mm] (lF) [right=4mm of lQ] {$\theta_5$\quad $A_+$ ($i=h$)\\ remove the counter, move left};
\node[flterm,text width=24mm] (lE) [below=of lF] {End: next stable ID\\ $H_-$ reads $[\nu(q),u]$};
\node[anchor=south west,font=\footnotesize,text width=72mm,align=left] at ([shift={(-4,1.5)}]lS.north west) {Left-moving quintuple ($q\in Q_-$, sentinel $h$ the maximum)};
\draw[flarr] (lS) -- (lA); \draw[flarr] (lA) -- (lB); \draw[flarr] (lB) -- (lQ);
\draw[flarr] (lQ) -- (lC); \draw[flarr] (lC) -- (lD);
\draw[flarr] (lD.west) -- ++(-3.5,0) |- (lQ.west);
\draw[flarr] (lQ) -- (lF); \draw[flarr] (lF) -- (lE);
\end{tikzpicture}
\caption{Flowcharts of $T'_3$ executing a right-moving and a left-moving quintuple}
\label{fig:reverse-flow}
\end{figure}
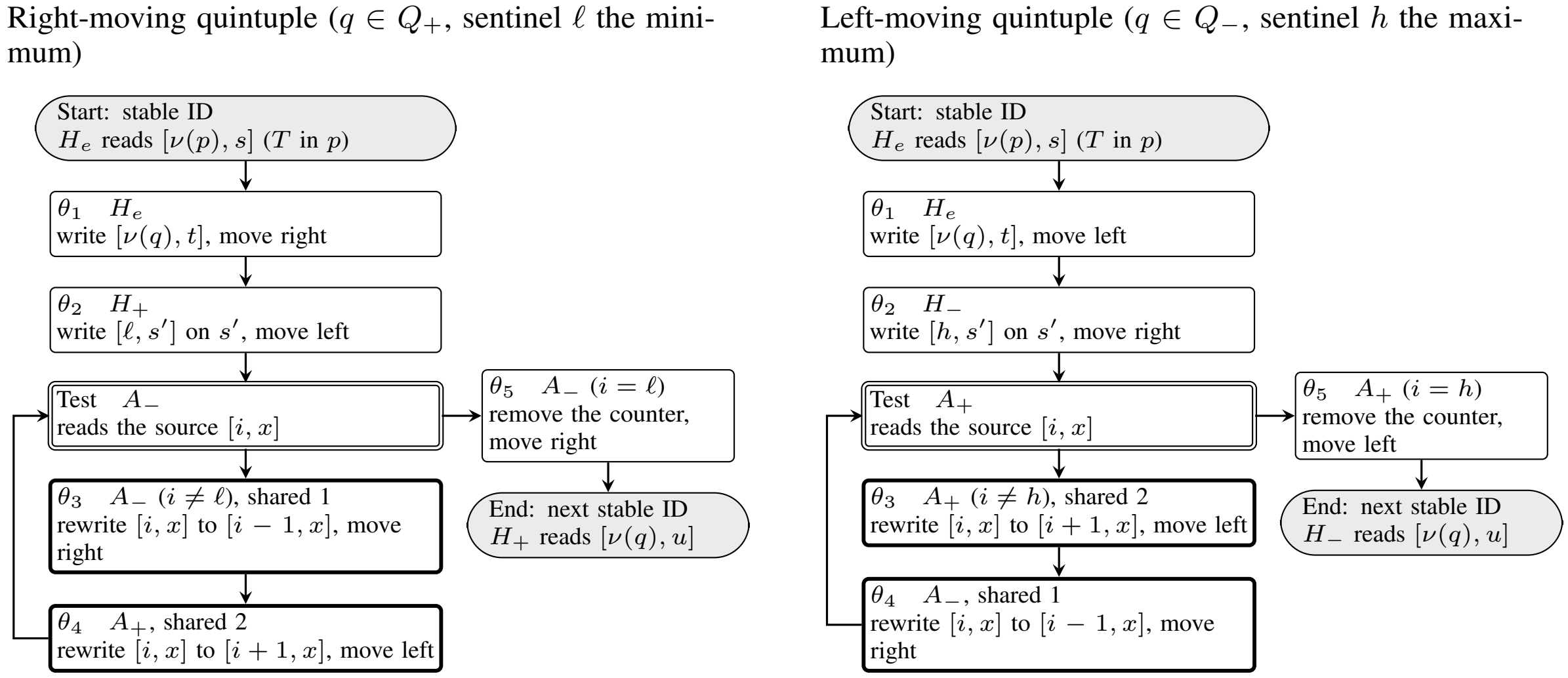

\paragraph{Reverse counting.}
$T'_2$ uses separate ranges of values for $Q_+$ and $Q_-$, but a single carry uses only one of them, so we would like both directions to share one range.
If the counters are counted in opposite directions for the two directions of motion (the sentinel, that is, the initial value that $\theta_2$ places and $\theta_5$ detects, is the smallest value for $Q_+$ and the largest for $Q_-$), then the decrement rule $\theta_3$ of one direction and the increment rule $\theta_4$ of the other become literally the same quintuple. For example, if $Q_-$ counts downward, $\theta_3$ of $Q_+$ and $\theta_4$ of $Q_-$ both take the form $[A_-,[i,x],[i-1,x],\rmR,A_+]$; since they are the same rule, the two roles need not be told apart by ranges of values.
The two directions can then share one range of values, and the number of symbols is at most $(\max(m_+,m_-)+1)n$. We write $T'_3$ for the machine obtained in this way.
Figure~\ref{fig:ranges} compares the two numberings.
Figure~\ref{fig:reverse-flow} shows the flowcharts of $T'_3$; the boxes with thick borders are the rules that are literally the same quintuple for the two directions.
From URTM$(10,8)$ this gives URTM$(4,48)$: with $m_+=m_-=5$ and $n=8$, the number of symbols is $n'=(\max(m_+,m_-)+1)n=(5+1)\cdot 8=48$.
We prove that it is universal in the sense of Definition~\ref{def:univ}, but it is dominated by URTM$(4,38)$.

\subsection{Block decomposition and merging of roles}
\label{sec:blocks}

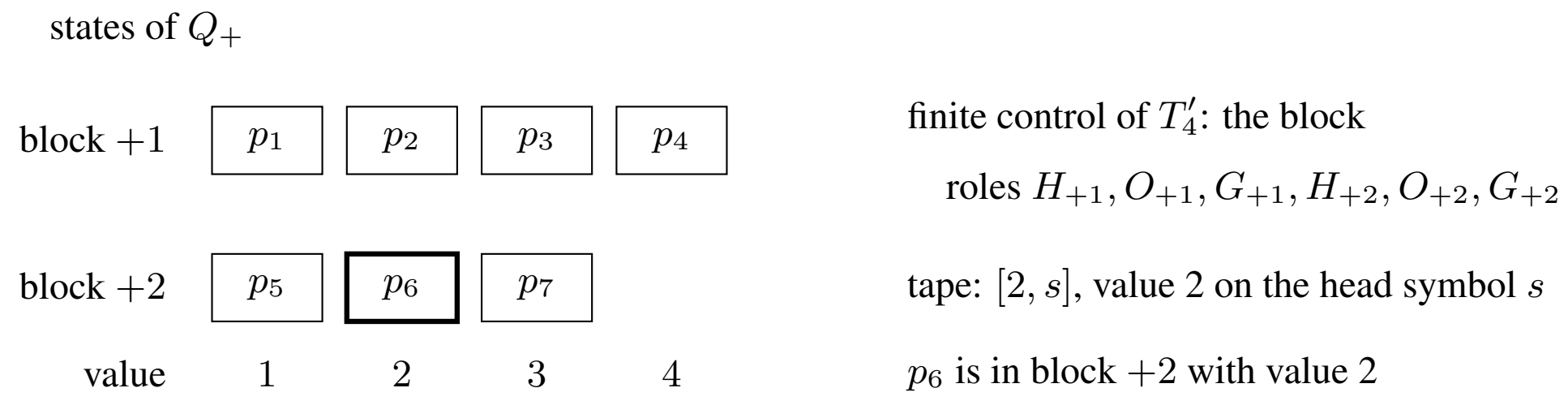
\begin{figure}[tb]
\centering
\begin{tikzpicture}[x=1mm,y=1mm,font=\small,st/.style={draw,minimum width=9mm,minimum height=5.5mm,inner sep=1pt}]
\node[anchor=west] at (-14,9) {states of $Q_+$};
\node[anchor=east] at (-2,0) {block $+1$};
\node[anchor=east] at (-2,-12) {block $+2$};
\foreach \k in {1,...,4} \node[st] at ({(\k-1)*11+5},0) {$p_{\k}$};
\foreach \k/\v in {5/1,6/2,7/3} \node[st] at ({(\v-1)*11+5},-12) {$p_{\k}$};
\foreach \v in {1,...,4} \node[font=\footnotesize] at ({(\v-1)*11+5},-19) {$\v$};
\node[anchor=east,font=\footnotesize] at (-2,-19) {value};
\draw[very thick] (11.5,-14.75) rectangle (20.5,-9.25);
\node[anchor=west] at (56,2) {finite control of $T'_4$: the block};
\node[anchor=west] at (56,-4) {\quad roles $H_{+1},O_{+1},G_{+1},H_{+2},O_{+2},G_{+2}$};
\node[anchor=west] at (56,-12) {tape: $[2,s]$, value 2 on the head symbol $s$};
\node[anchor=west,font=\footnotesize] at (56,-19) {$p_6$ is in block $+2$ with value 2};
\end{tikzpicture}
\caption{Splitting the numbers of the states of $Q_+$ of $T$ into blocks and values for $T'_4$ (an example with seven states)}
\label{fig:blocks}
\end{figure}

In $T'_3$ the number of values on the tape still grows with the number of states of $T$. Moving part of the number into the finite control should increase the states but decrease the values on the tape.
Split the number of a state into a pair (block, value within the block), and write $T'_4$ for the machine obtained in this way.
The finite control of $T'_4$ holds the block and the tape holds only the value, so a counter symbol $[i,s]$ does not record the block, and different blocks reuse the same values.
Let there be $b_+$ blocks for $Q_+$ and $b_-$ blocks for $Q_-$; we write $+k$ for the $k$th block of $Q_+$ and $-k$ for the $k$th block of $Q_-$.
Each block $\beta$ has three roles: $H_\beta$ (stable), $O_\beta$ (the source side of a carry), and $G_\beta$ (the target side of a carry).
Figure~\ref{fig:blocks} shows the splitting for seven states in two blocks.
For a quintuple $[p,s,t,D,q]$ of $T$ with $p\in Q_e$, let $\beta=\mathrm{blk}(q)$ be the block of $q$; then $T'_4$ uses the rules below. A role is a state before merging; the state column gives roles, and the merging of roles described below turns several roles into one state.
\begin{center}
\small
\begin{tabular}{lllllll}
\toprule
& role & read & write & move & next & range of values \\
\midrule
$\theta_1$ & $H_{\mathrm{blk}(p)}$ & $[\mathrm{val}(p),s]$ & $[\mathrm{val}(q),t]$ & $D$ & $H_\beta$ & \\
$\theta_2$ & $H_\beta$ & $s'$ & $[\sigma_\beta,s']$ & $\bar D$ & $O_\beta$ & \\
$\theta_3$ & $O_\beta$ & $[i,x]$ & $[i-\varepsilon,x]$ & $D$ & $G_\beta$ & after $\sigma_\beta$ up to $\mathrm{val}(q)$ ($\lvert\mathrm{val}(q)-\sigma_\beta\rvert$ values) \\
$\theta_4$ & $G_\beta$ & $[j,s']$ & $[j+\varepsilon,s']$ & $\bar D$ & $O_\beta$ & from $\sigma_\beta$ up to before $\mathrm{val}(q)$ (as many) \\
$\theta_5$ & $O_\beta$ & $[\sigma_\beta,x]$ & $x$ & $D$ & $H_\beta$ & \\
\bottomrule
\end{tabular}
\end{center}
Here $\mathrm{val}(q)$ is the value of $q$ within its block, $\sigma_\beta$ is the sentinel of block $\beta$ (the value that $\theta_2$ places and $\theta_5$ detects), and $\varepsilon$ is $+1$ if $\mathrm{val}(q)\ge\sigma_\beta$ and $-1$ otherwise. The back-and-forth phase ends after $3+2\lvert\mathrm{val}(q)-\sigma_\beta\rvert$ steps.
With one block in each direction, $\mathrm{val}=\nu$, $\sigma_{+1}=\ell_+$, and $\sigma_{-1}=\ell_-$, reading the roles $H_{+1},O_{+1},G_{+1}$ as $H_+,A_-,A_+$ and $H_{-1},O_{-1},G_{-1}$ as $H_-,A_+,A_-$ gives the table of $T'_2$ in Section~\ref{sec:four}.
The roles still have to know their block in the back-and-forth phase, because $\theta_5$ must enter the correct $H$-state; this is what costs states.

\begin{figure}[tb]
\centering
\begin{tikzpicture}[x=1mm,y=1mm,font=\footnotesize,node distance=3.5mm]
\node[flterm,text width=46mm] (S) at (0,0) {Start: stable ID (state $p$ of $T$)\\ $H$ of the block of $p$ reads $[i,s]$};
\node[flbox,text width=46mm] (A) [below=of S] {$\theta_1$\quad $H$ of the block of $p$\\ write $[6,t]$, move right (value 6 of $a$)};
\node[fldec,text width=90mm] (Q) [below=of A] {Test\quad state $A_+$\\ the symbol under the head decides the role};
\node[flbox,text width=44mm,dashed] (C2) [below=7mm of Q] {Counter with value 1--4\\ $A_+$ acts as $G_{+1}$ or $O_{-1}$ of a carry (not on this run)};
\node[flbox,text width=44mm] (C1) [left=4mm of C2] {Plain $u$: $\theta_2$ ($A_+$ acts as $H_a$)\\ write $[6,u]$, move left; next $H_-$};
\node[flterm,text width=44mm] (C3) [right=4mm of C2] {Counter with value 6: end\\ stable ID of $a$ (state $a$ of $T$)};
\node[flbox,text width=44mm] (C4) [below=of C1] {$\theta_5$\quad $H_-$ ($O_a$) reads the sentinel $[6,t]$\\ remove the counter, move right; next $A_+$};
\draw[flarr] (S) -- (A); \draw[flarr] (A) -- (Q);
\path (Q.south) ++(0,-3.5) coordinate (b);
\draw[semithick] (Q.south) -- (b);
\draw[flarr] (b) -| (C1.north); \draw[flarr,dashed] (b) -- (C2.north); \draw[flarr] (b) -| (C3.north);
\draw[flarr] (C1) -- (C4);
\draw[flarr] (C4.west) -- ++(-4,0) |- (Q.west);
\end{tikzpicture}
\caption{Flowchart of URTM$(4,46)$ entering the state $a$ of a single-state block}
\label{fig:single-flow}
\end{figure}
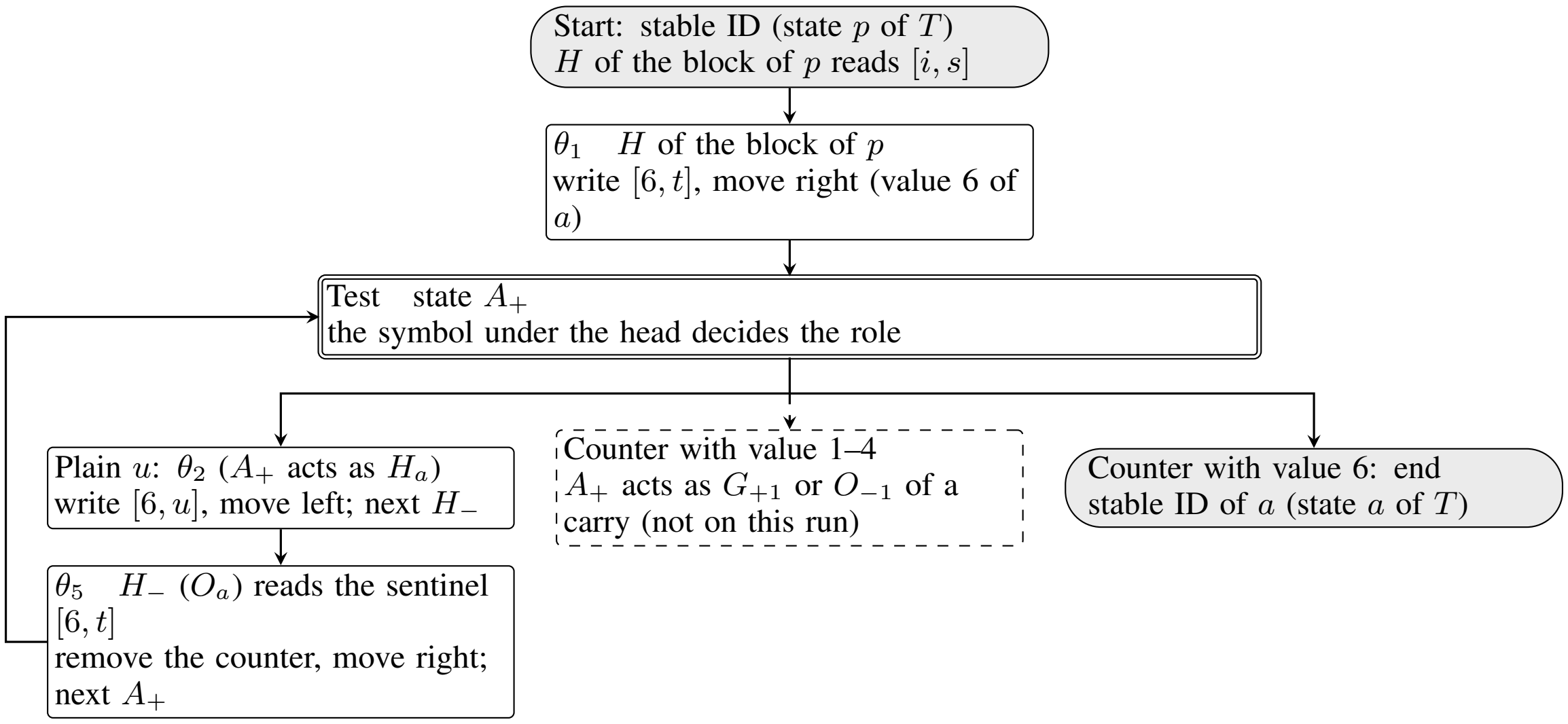

With three roles per block as separate states, $3(b_++b_-)$ states are needed. Roles that are entered by moves in the same direction and read disjoint values, however, should be able to share one state.
A \emph{merging of roles} groups the roles into classes and turns the roles of each class into a single state.
Roles in the same class must be entered by moves in the same direction, and the merged table must remain deterministic and reversible.
For example, $T'_2$ and $T'_3$ of Section~\ref{sec:four} are the case where all states of $Q_+$ form one block $+1$ and all states of $Q_-$ form one block $-1$ ($b_+=b_-=1$). Since there is only one block in each direction, all that the finite control holds about the block is the direction, and the number of a state is put on the counter symbol as its value.
There are six roles $H_{+1},O_{+1},G_{+1},H_{-1},O_{-1},G_{-1}$, so without merging there would be six states.
In Fig.~\ref{fig:fourdisj}, $A_-$, moving back and forth in cell $c$, plays the role $O_{+1}$, and $A_+$, in cell $c+1$, plays the role $G_{+1}$.
Merging $G_{+1}$ and $O_{-1}$, both entered by moving right, into one state $A_+$, and $O_{+1}$ and $G_{-1}$, both entered by moving left, into $A_-$ gives four states (Table~\ref{tab:roles}).
$T'_2$ tells apart the two roles merged into one state by the ranges of values; $T'_3$ uses reverse counting to make the rules of the two roles the same quintuples, so that they also share one range.
URTM$(4,46)$ has two blocks in each direction and hence twelve roles, which are again merged into four classes; blocks $\pm1$ are the main blocks and blocks $\pm2$ are blocks with a single state, described next.
\begin{table}[tb]
\centering
\caption{Examples of merging of roles in $T'_3$ and $T'_4$}
\label{tab:roles}
\small
\begin{tabular}{llll}
\toprule
State & Entered & $T'_3=$ URTM$(4,48)$ & $T'_4=$ URTM$(4,46)$ \\
 & & (one block per direction) & (two blocks per direction) \\
\midrule
$H_+$ & right & $H_{+1}$ & $H_{+1},\ G_{+2},\ O_{-2}$ \\
$A_+$ & right & $G_{+1},\ O_{-1}$ & $G_{+1},\ O_{-1},\ H_{+2}$ \\
$H_-$ & left & $H_{-1}$ & $H_{-1},\ O_{+2},\ G_{-2}$ \\
$A_-$ & left & $O_{+1},\ G_{-1}$ & $O_{+1},\ G_{-1},\ H_{-2}$ \\
\bottomrule
\end{tabular}
\end{table}
A further merging that turned out to be useful uses a block with a single state $a$; we call the block that holds all the other states of the same direction the \emph{main block}.
We give $a$ a value outside the range of the main block and make the sentinel of its block the same value; then nothing is carried on entering $a$, and there is no back-and-forth phase (three steps, by $\theta_1$, $\theta_2$, and $\theta_5$).
We then write $H_a,O_a,G_a$ for the roles of the block of $a$ and merge them into states of the main blocks: $H_a$ into the back-and-forth state entered by moves in the same direction as $a$ ($A_+$ if $a\in Q_+$), and $O_a$ into the $H$-state entered by moves in the opposite direction ($H_-$ if $a\in Q_+$). $G_a$ has no rules, so it can be put into any class entered by moves in its direction.
Since the back-and-forth state otherwise reads only values in the main range, reading a value outside this range signals that the machine is in $a$; thus $a$ costs no state of its own.
Figure~\ref{fig:single} shows the execution in URTM$(4,46)$ that enters some $a\in Q_+$; the states are named as in Section~\ref{sec:four}, with $A_+$ the back-and-forth state and $H_-$ the $H$-state of the main blocks.
Figure~\ref{fig:single-flow} shows the same execution as a flowchart: the symbol read by $A_+$ decides whether it acts as the $H$ role of $a$, as a role of the main block, or as the stable state of $a$.
\begin{figure}[tb]
\centering
\begin{tikzpicture}[x=1mm,y=1mm,cell/.style={draw,minimum width=19mm,minimum height=5.5mm,inner sep=1pt,font=\small}]
\def\tmrow#1#2#3#4#5#6{%
  \node[anchor=east,font=\small] at (-1,#1) {$#2$};
  \node[cell] at (10,#1) {$#3$};
  \node[cell] at (29,#1) {$#4$};
  \ifnum#5=0 \draw[very thick] (0.5,{#1-2.75}) rectangle (19.5,{#1+2.75});\else \draw[very thick] (19.5,{#1-2.75}) rectangle (38.5,{#1+2.75});\fi
  \node[anchor=west,font=\footnotesize] at (41,#1) {#6};}
\node[font=\footnotesize] at (10,6) {cell $c$};
\node[font=\footnotesize] at (29,6) {cell $c+1$};
\node[font=\footnotesize,anchor=east] at (-1,6) {state of $T'_4$};
\tmrow{0}{H_{\mathrm{blk}(p)}}{[i,s]}{u}{0}{the state of $T$ is $p$ (value $i$)}
\tmrow{-7}{A_+}{[6,t]}{u}{1}{$\theta_1$: write $t$ and the value 6 of $a$, move right; $H_a$ is $A_+$}
\tmrow{-14}{H_-}{[6,t]}{[6,u]}{0}{$\theta_2$: place the sentinel 6, move left; $O_a$ is $H_-$}
\tmrow{-21}{A_+}{t}{[6,u]}{1}{$\theta_5$: sentinel read, remove the counter, move right; $T$ is in $a$}
\end{tikzpicture}
\caption{$T'_4$ executing a right-moving quintuple $[p,s,t,\mathrm{R},a]$ of $T$ into the state $a$ of a single-state block ($T$ is URTM$(10,8)$ and $T'_4$ is URTM$(4,46)$)}
\label{fig:single}
\end{figure}

Finally, only the counter symbols that can actually occur are kept.
If $V$ is the set of counter values used and $\Sigma_i$ is the set of symbols of $T$ that can carry the value $i$, the number of symbols is
\[
  n+\sum_{i\in V}|\Sigma_i|\ \le\ (|V|+1)n .
\]
The choice of the blocks, of the values, and of the merging is a combinatorial optimization problem.
We enumerated partitions of the roles and searched each partition using constraint programming. To explore beyond the enumerated cases, we also used simulated annealing that changes the blocks, counter values, and partitions of the roles. We do not claim optimality of the resulting constructions in general.
This gives URTM$(4,46)$ from $(10,8)$, for example, with $V=\{1,2,3,4,6\}$: the main blocks use the values 1 to 4, and two states placed in blocks of their own share the value 6 (the values were chosen by the optimization).
Table~\ref{tab:blocks} lists the six machines with four to nine states obtained in this way. The number of states is the number of classes of roles and need not equal the number of blocks; for example, $(7,32)$ has blocks $(2,2)$ and seven states.
They are all dominated by the machines of Section~\ref{sec:merge}, but they are universal in their own right, and they are the inputs to symbol merging.
Morita's book~\cite{Morita17} contains no conversion that produces five to nine states.

\begin{table}[tb]
\centering
\caption{Machines obtained by block decomposition and merging of roles}
\label{tab:blocks}
\small
\begin{tabular}{llcllrr}
\toprule
Point & Original & Blocks $(b_+,b_-)$ & $V$ & $|\Sigma_i|$ for $i\in V$ & $n+\sum|\Sigma_i|$ & $(|V|+1)n$ \\
\midrule
$(4,46)$ & $(10,8)$ & $(2,2)$ & $\{1,2,3,4,6\}$ & 8, 8, 8, 8, 6 & 46 & 48 \\
$(5,45)$ & $(15,6)$ & $(3,2)$ & $\{1,\dots,8\}$ & 3, 5, 5, 5, 5, 5, 5, 6 & 45 & 54 \\
$(6,37)$ & $(10,8)$ & $(3,3)$ & $\{1,2,5,6\}$ & 6, 7, 8, 8 & 37 & 40 \\
$(7,32)$ & $(15,6)$ & $(2,2)$ & $\{1,\dots,5\}$ & 6, 5, 5, 5, 5 & 32 & 36 \\
$(8,28)$ & $(17,5)$ & $(4,4)$ & $\{2,4,5,6,7\}$ & 3, 5, 5, 5, 5 & 28 & 30 \\
$(9,27)$ & $(15,6)$ & $(5,4)$ & $\{1,2,3,4\}$ & 6, 5, 5, 5 & 27 & 30 \\
\bottomrule
\end{tabular}
\end{table}

\begin{remark}
Three states cannot be reached in this framework, which merges neither two $H$-roles nor the $H$- and $G$-roles of the same block. The former is outside the framework. The latter cannot be merged if the block contains two states that read a common symbol $x$: if $q$ is the one closer to the sentinel, the $H$-role reads $[\mathrm{val}(q),x]$ by $\theta_1$ and the $G$-role reads the same symbol by $\theta_4$, with different actions.
If $Q_-$ has at least two states, then either some block of $Q_-$ has two or more states, and its $H$- and $G$-roles fall into different classes, or there are two single-state blocks with two $H$-roles.
In either case the roles entered by left moves need at least two classes.
The same holds for $Q_+$, so at least four states are needed for all six of Morita's machines.
This remark is an argument in this paper and is not formalized.
\end{remark}

\subsection{Symbol merging}
\label{sec:merge}

In a converted machine $T'$, each counter symbol is read by few states and written into few states. The more states a table has, the fewer states use each symbol and the more pairs of symbols are never read by the same state and never written into the same state, so there should be more room for merging. In Table~\ref{tab:merge}, too, the fraction removed by merging roughly grows with the number of states; this is an observation, not a theorem.
UTM$(3,9)$ of Kudlek and Rogozhin~\cite{KudlekRogozhin02} can be read as UTM$(3,10)$ of Rogozhin~\cite{Rogozhin96} with the blank made the same symbol as $\overleftarrow{1}$ and with the encodings of the initial word and of the halting symbol changed accordingly.
We apply systematically this idea of reducing symbols by making several of them one symbol.
Two symbols $x$ and $y$ \emph{conflict} if some state reads both, some state is entered by quintuples writing $x$ and $y$, or both can occur on the tape of a halting ID reached from an encoded input.
The \emph{conflict graph} has the symbols as vertices and an edge between any two conflicting symbols.
Merging pairwise non-conflicting symbols is a coloring of the conflict graph (an assignment of colors in which adjacent vertices get different colors), and symbols of the same color become one symbol.
The merged table is deterministic and reversible if the unmerged one is, and it follows the run of the unmerged one step by step.

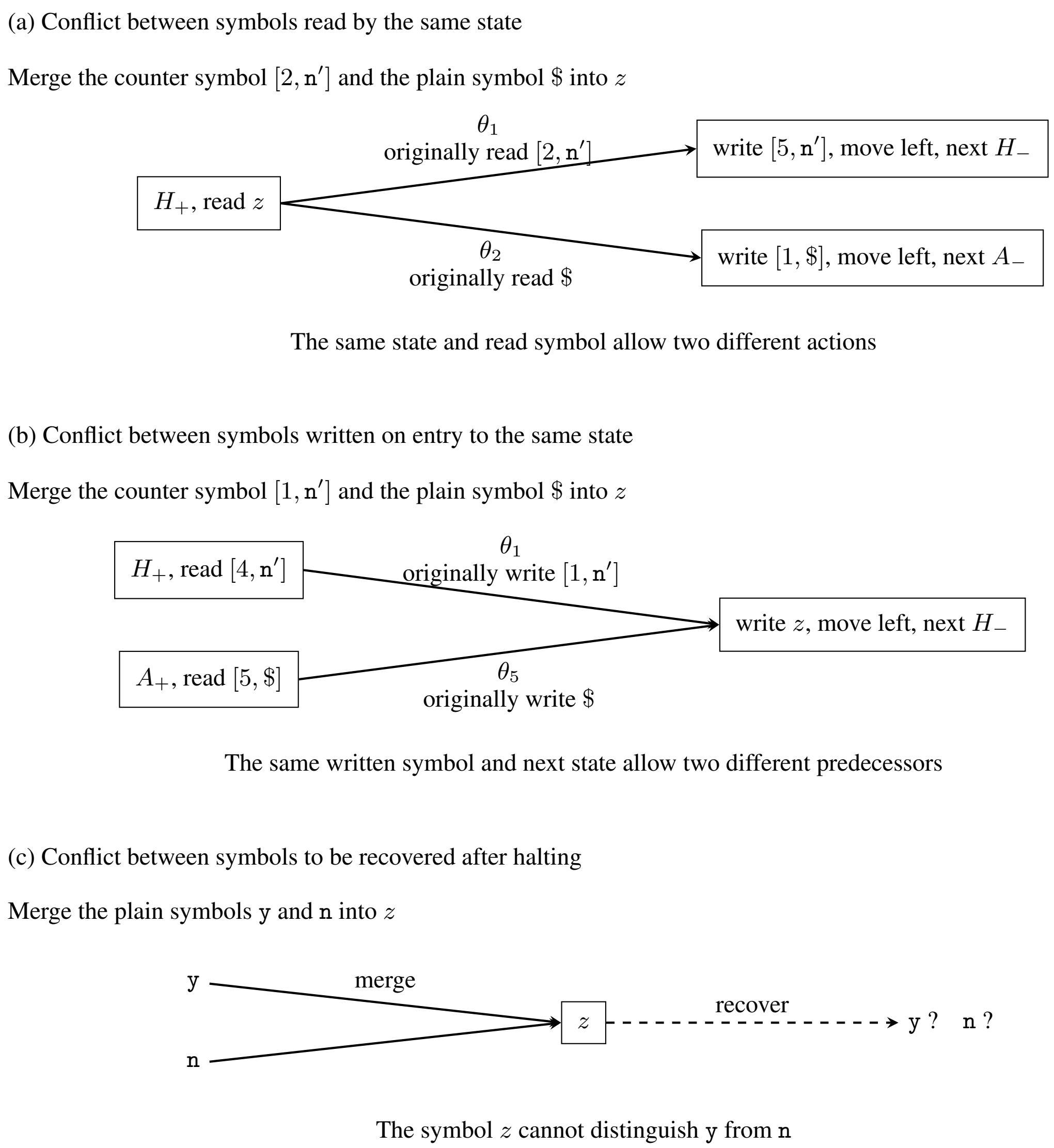
\begin{figure}[tb]
\centering
\begin{tikzpicture}[
  x=1mm,y=1mm,>=stealth,font=\small,
  box/.style={draw,align=center,inner sep=2mm},
  note/.style={font=\footnotesize}
]
\node[anchor=west] at (0,0) {(a) Conflict between symbols read by the same state};
\node[anchor=west,note] at (0,-7) {Merge the counter symbol $[2,\mathtt{n}']$ and the plain symbol $\mathtt{\$}$ into $z$};
\node[box] (read) at (27,-23) {$H_+$, read $z$};
\node[box] (action1) at (112,-16) {write $[5,\mathtt{n}']$, move left, next $H_-$};
\node[box] (action2) at (112,-30) {write $[1,\mathtt{\$}]$, move left, next $A_-$};
\draw[->,thick] (read.east) -- node[above,note,align=center] {$\theta_1$\\originally read $[2,\mathtt{n}']$} (action1.west);
\draw[->,thick] (read.east) -- node[below,note,align=center] {$\theta_2$\\originally read $\mathtt{\$}$} (action2.west);
\node[note] at (75,-41) {The same state and read symbol allow two different actions};

\node[anchor=west] at (0,-53) {(b) Conflict between symbols written on entry to the same state};
\node[anchor=west,note] at (0,-60) {Merge the counter symbol $[1,\mathtt{n}']$ and the plain symbol $\mathtt{\$}$ into $z$};
\node[box] (pred1) at (27,-70) {$H_+$, read $[4,\mathtt{n}']$};
\node[box] (pred2) at (27,-84) {$A_+$, read $[5,\mathtt{\$}]$};
\node[box] (written) at (112,-77) {write $z$, move left, next $H_-$};
\draw[->,thick] (pred1.east) -- node[above,note,align=center] {$\theta_1$\\originally write $[1,\mathtt{n}']$} (written.west);
\draw[->,thick] (pred2.east) -- node[below,note,align=center] {$\theta_5$\\originally write $\mathtt{\$}$} (written.west);
\node[note] at (75,-95) {The same written symbol and next state allow two different predecessors};

\node[anchor=west] at (0,-107) {(c) Conflict between symbols to be recovered after halting};
\node[anchor=west,note] at (0,-114) {Merge the plain symbols $\mathtt{y}$ and $\mathtt{n}$ into $z$};
\node (y) at (25,-123) {$\mathtt{y}$};
\node (n) at (25,-133) {$\mathtt{n}$};
\node[box] (z) at (75,-128) {$z$};
\node (recover) at (122,-128) {$\mathtt{y}$ ?\quad $\mathtt{n}$ ?};
\draw[->,thick] (y.east) -- node[above,note] {merge} (z.west);
\draw[->,thick] (n.east) -- (z.west);
\draw[->,thick,dashed] (z.east) -- node[above,note] {recover} (recover.west);
\node[note] at (75,-142) {The symbol $z$ cannot distinguish $\mathtt{y}$ from $\mathtt{n}$};
\end{tikzpicture}
\caption{Examples of conflicts under symbol merging}
\label{fig:symbol-conflicts}
\end{figure}

For example, consider $T'_3$ obtained from URTM$(10,8)$ (with the symbols $\mathtt{b},\mathtt{y},\mathtt{n},\mathtt{n}',\mathtt{Y},\mathtt{N},\mathtt{*},\mathtt{\$}$), that is, URTM$(4,48)$.
The state $H_+$ reads $[2,\mathtt{n}']$ by $\theta_1$ and the plain symbol $\mathtt{\$}$ by $\theta_2$, so these two conflict (merged, the rule of $H_+$ would not be determined).
The state $H_-$ is entered by the $\theta_1$ rule $[H_+,[4,\mathtt{n}'],[1,\mathtt{n}'],\rmL,H_-]$, which writes $[1,\mathtt{n}']$, and by the $\theta_5$ rule $[A_+,[5,\mathtt{\$}],\mathtt{\$},\rmL,H_-]$, which writes $\mathtt{\$}$, so these two also conflict (merged, the previous ID of $H_-$ would not be determined).
The plain symbols $\mathtt{y}$ and $\mathtt{n}$ conflict because both can occur on a halting tape.
Of the 1128 pairs of the 48 symbols, only $\mathtt{n}'$ and $[3,\mathtt{n}]$ do not conflict. The symbol $\mathtt{n}'$ is read only by $H_+$ and written only by quintuples entering $H_-$; the symbol $[3,\mathtt{n}]$ is read only by $A_+$ and $A_-$ and written only by quintuples entering $H_+$, $A_+$, and $A_-$; and counter symbols do not occur on a halting tape.
With so few states and uniform back-and-forth rules, almost all pairs conflict.
Figure~\ref{fig:symbol-conflicts} illustrates these conflicts. Each panel is a separate example merging only the indicated pair of symbols into $z$; arrows show the merged rules or maps on symbols.

The condition for a conflict can be relaxed as follows. If two quintuples from the same state read $x$ and $y$, have the same direction and the same target state, and write two symbols that are also merged into one, then after merging the two quintuples become the same quintuple. The same holds for two quintuples that enter the same state writing $x$ and $y$ if they have the same source state and the same direction and read two symbols that are also merged into one. Such a pair need not be treated as a conflict, since the two equal quintuples are kept as one. The counts in the example above are for the condition before relaxing.

\begin{theorem}[Symbol merging]
\label{thm:merge}
Let $(c,d)$ witness $\mathrm{IsUTMStrict}(U)$, and assume that every halting ID reached from $\mathrm{toID}(0,c(i,a))$ is in a state that is not a source state and has all its tape symbols in a fixed finite set $F$.
Let $h,\rho:\mathbb{N}\to\mathbb{N}$ be primitive recursive with $h(0)=\rho(0)=0$ and $\rho(h(x))=x$ for every $x\in F$.
If the table $h(U)$, obtained by replacing the read and written symbols of every quintuple by their images under $h$, is deterministic, then $\mathrm{IsUTMStrict}(h(U))$.
\end{theorem}

Here $h$ is the map of the coloring, which sends each symbol to its color, that is, to the merged symbol, and $\rho$ recovers the unmerged symbols on halting tapes; the hypothesis on the halting states prevents a run from continuing from a halting ID when merging makes the rule of another symbol applicable, and it holds because, in runs from encoded inputs, our block tables halt only by entering a halting state.
Applied to the fixed tables of Table~\ref{tab:blocks}, merging removes one to three symbols.
Much more is gained by choosing the blocks and values again, by simulated annealing, with the number of colors after merging as the objective.
The best results come from the block decompositions of URTM$(10,7)$ for four and five states, of URTM$(9,8)$ for six and seven states, and of URTM$(13,7)$ for eight states (Table~\ref{tab:merge}); URTM$(10,7)$ and URTM$(9,8)$ are obtained from URTM$(13,7)$ in Section~\ref{sec:tags}. The search allows the relaxed condition; only the table of $(4,38)$ uses it (one pair of quintuples becomes the same quintuple).
Their tables are not printed here; they are provided as text files in the directory tables of the repository \url{https://github.com/yokoyama-lab/small-machine-zoo}, together with a description that gives the classes of merged symbols and the roles merged into each state.
For each table of Table~\ref{tab:merge}, we prove in Lean that, even under the relaxed condition, no map of the symbols of the table before merging gives a deterministic and reversible table with fewer symbols than the merged table.
The proof picks as many symbols of the table before merging as the merged table has, and shows that every map giving a deterministic and reversible table sends them to pairwise distinct symbols.
We do not claim minimality over the choices of the original machine, the blocks, and the values, or among all URTMs with the same number of states.

\begin{table}[tb]
\centering
\caption{Machines obtained by symbol merging}
\label{tab:merge}
\small
\begin{tabular}{llcrr}
\toprule
Point & Original & Blocks $(b_+,b_-)$ & Symbols before merging & Symbols after merging \\
\midrule
$(4,38)$ & $(10,7)$ & $(2,2)$ & 43 & 38 \\
$(5,36)$ & $(10,7)$ & $(3,2)$ & 43 & 36 \\
$(6,28)$ & $(9,8)$ & $(3,3)$ & 46 & 28 \\
$(7,26)$ & $(9,8)$ & $(4,3)$ & 46 & 26 \\
$(8,21)$ & $(13,7)$ & $(4,4)$ & 40 & 21 \\
\bottomrule
\end{tabular}
\end{table}

\subsection{Merging two states using tags on cells}
\label{sec:tags}

\begin{figure}[tb]
\centering
\begin{tikzpicture}[x=1mm,y=1mm,cell/.style={draw,minimum width=17mm,minimum height=5.5mm,inner sep=1pt,font=\small}]
\def\tagrow#1#2#3#4#5{%
  \node[anchor=east,font=\small] at (-1,#1) {#2};
  \node[cell] at (9,#1) {$#3$};
  \node[cell] at (26,#1) {$#4$};
  \draw[very thick] (17.5,{#1-2.75}) rectangle (34.5,{#1+2.75});
  \node[anchor=west,font=\footnotesize,text width=74mm,align=left] at (37,#1) {#5};}
\node[font=\footnotesize] at (9,6) {cell $c$};
\node[font=\footnotesize] at (26,6) {cell $c+1$};
\node[font=\footnotesize,anchor=east] at (-1,6) {machine, state};
\tagrow{0}{$T$, $p$}{t}{u}{$T$ is in $p$ and next applies its quintuple $[p,u,\dots]$}
\tagrow{-7}{$T'$, $r$}{(t,k)}{(u,1)}{$T'$ is in $r=\mu(p)$; the tag $1$ selects $[r,(u,1),\dots]$, the image of $[p,u,\dots]$}
\tagrow{-18}{$T$, $q$}{t'}{u}{$T$ is in $q$ and next applies its quintuple $[q,u,\dots]$}
\tagrow{-25}{$T'$, $r$}{(t',k')}{(u,2)}{$T'$ is in $r=\mu(q)$; the tag $2$ selects $[r,(u,2),\dots]$, the image of $[q,u,\dots]$}
\end{tikzpicture}
\caption{Merging two states $p$ and $q$ of $T$ into one state $r$ of $T'$ using tags on cells}
\label{fig:tags}
\end{figure}
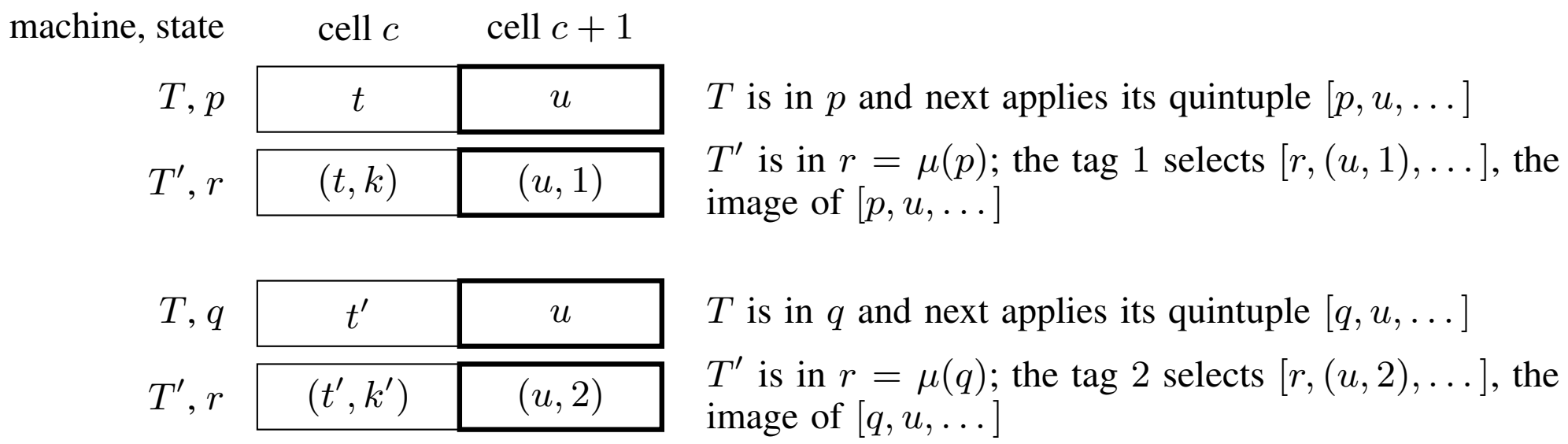

This conversion removes one state of an RTM $T$ and adds tags to its symbols.
Let $p$ and $q$ be two states of $T$ such that every quintuple of $T$ that enters $p$ or $q$ moves in the same direction.
The converted machine $T'$ is defined as follows.
\begin{itemize}
\item States: those of $T$, with $p$ and $q$ replaced by one state $r$. The map $\mu$ sends $p$ and $q$ to $r$ and fixes the other states.
\item Symbols: pairs $(s,k)$ of a symbol $s$ of $T$ and a tag $k$. The blank is the blank of $T$ with tag $0$.
\item Quintuples: when $T$ applies $[p_1,s,t,d,q_1]$, the machine $T'$ applies $[\mu(p_1),(s,k),(t,k'),d,\mu(q_1)]$, for tags $k$ and $k'$ chosen by the construction. Thus $T'$ runs in lockstep with $T$.
\item Constraints on the tags: when $T'$ is in $r$, the tag of the scanned symbol tells whether $T$ is in $p$ or $q$ (Figure~\ref{fig:tags}). Two quintuples of $T'$ that enter the same state write different symbols. The direction condition of reversibility holds because $p$ and $q$ are entered by moves in the same direction.
\item Encoding and decoding: the initial tape carries no tags, so $T'$ uses the encoding of $T$. Erasing the tags from a halting ID of $T'$ gives the halting ID of $T$.
\end{itemize}
In Figure~\ref{fig:tags}, $T'$ scans the cell $c+1$ in state $r$ after a move to the right. The cell holds the same symbol $u$ of $T$ in both cases, and its tag ($1$ or $2$) selects the quintuple to apply.

The tag of a cell can depend only on the sequence of states in which $T$ has read the cell so far.
We collect these sequences from runs of $T$ on random CTSHs and assign tags to the nodes of their prefix tree with an answer set solver. We check each resulting table by running it in lockstep with $T$ on further CTSHs.

\begin{figure}[tb]
\centering
\begin{tikzpicture}[x=1mm,y=1mm,
  mach/.style={draw,minimum width=19mm,minimum height=6mm,inner sep=1pt,font=\small},
  mid/.style={mach,dashed},
  fin/.style={mach,very thick,fill=black!10},
  stp/.style={anchor=west,font=\scriptsize,inner sep=1pt},
  arr/.style={-latex,thick}]
\node[mach] (a) at (0,0) {$(13,7)$};
\node[mid] (b) at (0,-12) {$(12,8)$};
\node[mid] (c) at (0,-24) {$(12,7)$};
\node[mid] (d) at (0,-36) {$(11,8)$};
\node[fin] (e) at (0,-48) {$(11,7)$};
\node[fin] (f) at (0,-60) {$(10,7)$};
\node[fin] (g) at (0,-72) {$(9,8)$};
\draw[arr] (a) -- node[stp,xshift=2mm] {tags: copy-n-2 into copy-end} (b);
\draw[arr] (b) -- node[stp,xshift=2mm] {symbols: $\mathtt{N}$ with $\hat{\mathtt{n}}$} (c);
\draw[arr] (c) -- node[stp,xshift=2mm] {tags: copy-y-2 into case-y-2} (d);
\draw[arr] (d) -- node[stp,xshift=2mm] {symbols: $\mathtt{N}$ with $\hat{\mathtt{y}}$} (e);
\draw[arr] (e) -- node[stp,xshift=2mm] {tags and symbols together: cycle-end into case-y-2} (f);
\draw[arr] (f) -- node[stp,xshift=2mm] {tags and symbols together: copy-n-3 into case-y-1} (g);
\node[fin] (h) at (100,0) {$(8,21)$};
\node[fin,minimum width=27mm] (i) at (100,-60) {$(4,38)$, $(5,36)$};
\node[fin,minimum width=27mm] (j) at (100,-72) {$(6,28)$, $(7,26)$};
\draw[arr] (a) -- node[above,font=\scriptsize] {blocks, symbols (Table~\ref{tab:merge})} (h);
\draw[arr] (f) -- node[above,font=\scriptsize,pos=0.8] {blocks, symbols} (i);
\draw[arr] (g) -- node[above,font=\scriptsize,pos=0.8] {blocks, symbols} (j);
\end{tikzpicture}
\caption{Machines obtained from URTM$(13,7)$}
\label{fig:tagchain}
\end{figure}
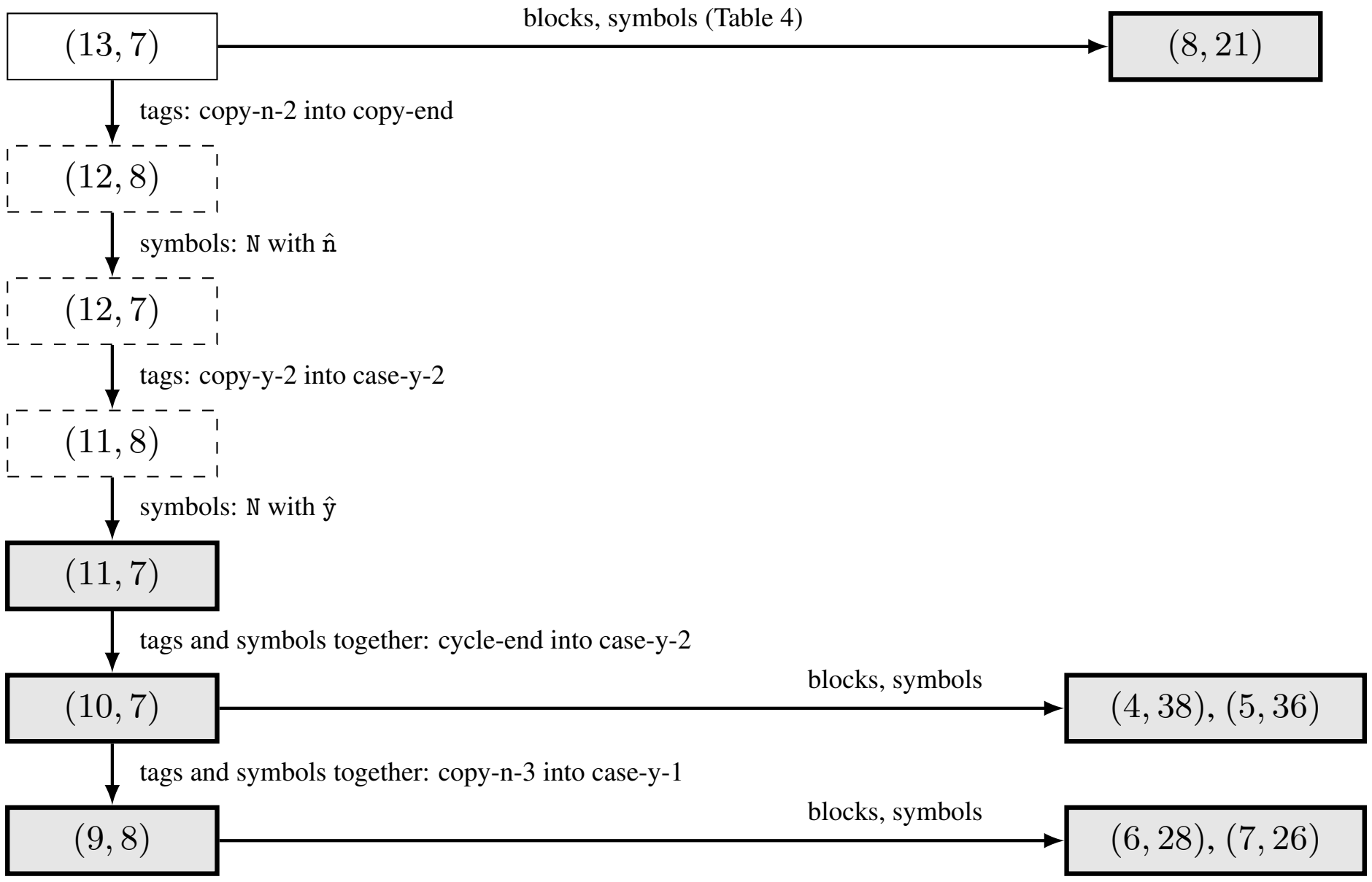

We apply this conversion and then symbol merging (Section~\ref{sec:merge}).
Figure~\ref{fig:tagchain} shows the applications to URTM$(13,7)$.
State names are those of Morita's description file~\cite{MoritaFiles}, with hyphens for underscores.
The symbols of URTM$(13,7)$ are $\mathtt{b},\mathtt{y},\mathtt{n},\mathtt{Y},\mathtt{N},\ast,\mathtt{\$}$, and $\hat{x}$ denotes the symbol $x$ with the tag added at that stage.
In the last two steps, a constraint solver finds the tags and the merge of symbols together. It searches for a table with the given number of symbols that runs in lockstep with the machine it is obtained from; the intermediate table with tags need not be deterministic and reversible.

Boxes in Figure~\ref{fig:tagchain} give (states, symbols).
Thick boxes are the tables whose universality we prove; dashed boxes are intermediate tables, for which we do not prove it.
We prove universality of URTM$(11,7)$, URTM$(10,7)$, and URTM$(9,8)$ by following each step of the proof for the machine it is obtained from.
We do not claim minimality.
\begin{itemize}
\item URTM$(11,7)$, from URTM$(13,7)$.
  No tags remain. URTM$(11,7)$ has the 59 quintuples of URTM$(13,7)$, with copy-y-2 renamed to case-y-2 and copy-n-2 renamed to copy-end. In 8 of them, the read or the written symbol is changed to $\mathtt{N}$.
  These 8 quintuples belong to three pairs of sweeps, each going over a part of the tape and back. The sweep of copy-y-1 (to the right) writes $\mathtt{N}$ over each $\mathtt{y}$ and over the blank where URTM$(13,7)$ appends $\mathtt{y}$, and the sweep of copy-y-3 (to the left) writes $\mathtt{y}$ over each $\mathtt{N}$. The sweeps of copy-n-1 and copy-n-3 do the same with $\mathtt{n}$. The sweep of case-y-1 (to the left) writes $\mathtt{N}$ over each $\mathtt{y}$, and the sweep of case-y-2 (to the right) writes $\mathtt{y}$ back.
  After each step of the CTSH, the tape equals that of URTM$(13,7)$.
  The halting tapes differ only when URTM$(11,7)$ halts in the state $\mathsf{halt}$. It then halts during the sweep of case-y-1, so the symbols $\mathtt{y}$ between $\mathtt{\$}$ and the mark $\mathtt{Y}$ are $\mathtt{N}$.
  The decoding of URTM$(13,7)$ does not read these cells (it counts the separators $\ast$ and reads the string after the mark $\mathtt{Y}$).
  Encoding and decoding: those of URTM$(13,7)$.
\item URTM$(10,7)$, from URTM$(11,7)$.
  Of the 59 quintuples of URTM$(11,7)$, 4 have two images because they are used on cells of different colors, and each of the others has one image. The 63 images include 6 pairs of identical quintuples, so there are 57 quintuples.
  Each of the 7 symbols stands for a set of symbols with tags and is named after a symbol of URTM$(13,7)$.
  After each step of the CTSH, the tape is that of URTM$(11,7)$, except that the separators $\ast$ after the phase marker are written as $\mathtt{Y}$. During each sweep, every occurrence of a given symbol that the head has passed is recolored in the same way.
  Encoding: that of URTM$(13,7)$.
  Decoding: that of URTM$(13,7)$, except that we count the cells holding $\ast$ or $\mathtt{Y}$ as separators. When URTM$(10,7)$ halts in the state $\mathsf{halt}$, we subtract one from this count for the mark. After removing trailing blanks, we read the string as the longest suffix consisting of $\mathtt{y}$ and $\mathtt{n}$. On a halting tape, this is the part after the last $\mathtt{\$}$.
\item URTM$(9,8)$, from URTM$(10,7)$.
  Of the 57 quintuples of URTM$(10,7)$, 4 have two images because they are used on cells of different colors, and each of the others has one image. The 61 images include 7 pairs of identical quintuples, so there are 54 quintuples.
  URTM$(9,8)$ runs in lockstep with URTM$(10,7)$, and their tapes differ only in the colors of the cells. After each step of the CTSH, the tape equals that of URTM$(10,7)$. During each sweep, every occurrence of a given symbol that the head has passed is recolored in the same way.
  Encoding and decoding: those of URTM$(10,7)$.
\end{itemize}

The same conversion merges copy-y-2 into case-y-1 in URTM$(10,8)$ and gives URTM$(9,14)$. Symbol merging then gives URTM$(9,9)$.
Both are dominated by URTM$(9,8)$, and we do not use them.

\subsection{Three states}
\label{sec:three}

\begin{figure}[tb]
\centering
\begin{tikzpicture}[x=1mm,y=1mm,font=\footnotesize]
\node[flbox,text width=46mm] (QS) at (0,0) {Carrying states $q(s)$ ($n$)\\ remember the symbol of $T$ carried from the previous pair and rewrite pairs while moving right (repeating $\hat\delta_2,\hat\delta_3$)};
\node[flbox,text width=38mm] (QP) [right=14mm of QS] {Mark-carrying states $q(p',u)$ ($m_+n$)\\ after a right move of $T$, remember the target $p'\in Q_+$ and $u$, and write the mark on the next pair};
\node[flbox,text width=28mm] (QM) [below=16mm of QS] {Turning state (1)\\ after adding a pair at the right end};
\node[flbox,text width=32mm] (QL) [left=24mm of QM] {Returning state (1)\\ move left without rewriting ($\hat\delta_{10},\hat\delta_{11}$)};
\node[flbox,text width=38mm] (HS) [below=16mm of QP] {Post-halting states $h(s)$ ($n$)\\ carry symbols to the right end ($\hat\delta_{12}$)};
\draw[flarr] (QL.north) |- node[pos=0.75,above] {$\hat\delta_1$ start a sweep} (QS.west);
\draw[flarr] ([yshift=2mm]QS.east) -- node[above] {$\hat\delta_5$} ([yshift=2mm]QP.west);
\draw[flarr] ([yshift=-2mm]QP.west) -- node[below] {$\hat\delta_6$} ([yshift=-2mm]QS.east);
\draw[flarr] (QS.south -| QM.north) -- node[left] {$\hat\delta_8$ right end} (QM.north);
\draw[flarr] (QM.west) -- node[above] {$\hat\delta_9$ turn} (QL.east);
\draw[flarr] ([xshift=20mm]QS.south) |- node[pos=0.25,right] {$\hat\delta_4$ halt} (HS.west);
\draw[flarr] (QP.south) -- node[right] {$\hat\delta_7$ halt} (HS.north);
\end{tikzpicture}
\caption{Kinds of states of $\hat T$ in Morita's Lemma~5.9~\cite{Morita17} and the transitions between them}
\label{fig:hat-flow}
\end{figure}
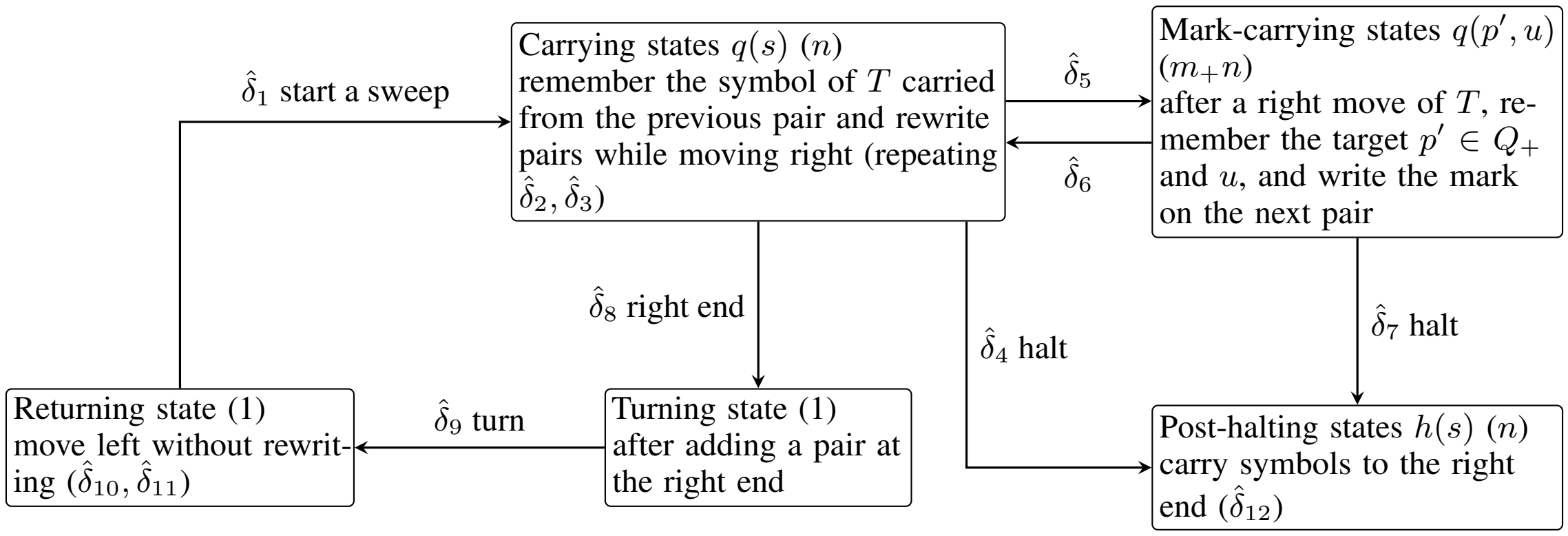

Morita's conversion to three states~\cite[Sect.~5.3.5]{Morita17} has two stages (Fig.~\ref{fig:three-pipeline}).
\begin{figure}[tb]
\centering
\begin{tikzpicture}[x=1mm,y=1mm,font=\footnotesize]
\node[flbox,text width=66mm] (T) at (0,0) {Original machine $T$: $m$ states, $n$ symbols\\ moves both ways and rewrites the tape};
\node[flbox,text width=66mm] (H) at (0,-40) {One-way-updating machine $\hat T$: $M=m_+n+2n+2$ states, $N=mn^2+n^2+1$ symbols\\ rewrites only while sweeping right and returns to the left end in one state without rewriting; one sweep is one step of $T$};
\node[flbox,text width=66mm] (D) at (0,-80) {Three-state machine $T^\dagger$ (states $\rmL$, $\rmR_0$, $\rmR_1$)\\ $N(2M-1)$ symbols};
\draw[flarr] (T) -- node[right,text width=60mm,align=left,xshift=1mm] {Morita's Lemma~5.9~\cite{Morita17}: pack two cells of $T$ into one pair, mark the state on a pair, and shift the content one cell to the right in every sweep} (H);
\draw[flarr] (H) -- node[right,text width=60mm,align=left,xshift=1mm] {Morita's Lemma~5.10~\cite{Morita17}: number the states of $\hat T$ by $1,\dots,M-1$ and carry the number to the next cell as up and down counters} (D);
\node[anchor=north west,align=left,font=\scriptsize] at (-34,-90) {For $(32,3)$: $m=32$, $n=3$, $m_+=18$, so $M=62$, $N=298$, and $298\cdot 123=36654$ symbols};
\end{tikzpicture}
\caption{The two stages of Morita's conversion to three states}
\label{fig:three-pipeline}
\end{figure}
Morita's Lemma~5.9~\cite{Morita17} turns $T$ into a \emph{one-way-updating} RTM $\hat T$, which rewrites the tape only while sweeping to the right and returns to the left end in a single state without rewriting.
To distinguish the blank of $T$ from the blank of $\hat T$ and to extend the used region one cell at a time, its tape packs two cells of $T$ into one cell.
$\hat T$ has $N=mn^2+n^2+1$ symbols and $M=m_+n+2n+2$ states.
The symbols are the blank of $\hat T$, the $n^2$ pairs $[s,t]$ holding the contents of two cells of $T$, and the $mn^2$ pairs $[[p,s],t]$ whose left component is the head cell of $T$ and which are marked with the state $p$ of $T$; thus $N=1+n^2+mn^2$.
The states are of four kinds, and $M=1+(n+1)+m_+n+n$.
\begin{itemize}
\item One state returns to the left end without rewriting.
\item During the sweep to the right, $n$ states remember the symbol of $T$ carried from the previous cell, and one state turns back after adding a pair at the right end of the used region. Through this carrying, the symbols of $T$ shift one place to the right in every sweep.
\item After a right-moving quintuple of $T$ is executed, $m_+n$ states remember the target state $q\in Q_+$ and the carried symbol and go to write the mark on the pair in the next cell. The target of a left move can be written as a mark on the pair in the current cell, so these states are needed only for the states in $Q_+$.
\item After $T$ halts, $n$ states carry symbols to the right end.
\end{itemize}
Figure~\ref{fig:hat-flow} shows the transitions between these states. The labels $\hat\delta_1,\dots,\hat\delta_{12}$ are the names that Morita gives to the rules in the proof of Lemma~5.9~\cite{Morita17}; we use them only to relate the figure to that proof.
Morita's Lemma~5.10~\cite{Morita17} then represents the state $q_i$ of $\hat T$ by a counter $[i,x]$ with an up or a down mark on the symbol under the head and uses three states $\rmL$, $\rmR_0$, $\rmR_1$, which gives $N(2M-1)$ symbols.
Figure~\ref{fig:three-carry} shows how it executes a right-moving quintuple of $\hat T$; as in Fig.~\ref{fig:four} of Section~\ref{sec:four}, the number $j$ of the state is carried to the right by counting down the down-marked counter in cell $c$ while counting up the up-marked counter in cell $c+1$, and the execution takes $2j+1$ steps.
For URTM$(32,3)$ we have $m=32$, $n=3$, and $m_+=18$, so $N=32\cdot 3^2+3^2+1=298$ and $M=18\cdot 3+2\cdot 3+2=62$; the result has $N(2M-1)=298\cdot 123=36654$ symbols, which gives URTM$(3,36654)$.
Of these symbols, 85\% are counters that combine a marked pair, one of the $m_+n$ states of $\hat T$ entered after a right move, and an up or down mark.
In a one-way-updating machine every carry goes to the right, so each state always reads a fixed side and the marks should not be needed; moreover, most counters are never read, and packing two cells into one creates the $n^2$ pairs.
We reduce the count in four steps.

\begin{figure}[tb]
\centering
\begin{tikzpicture}[x=1mm,y=1mm,cell/.style={draw,minimum width=19mm,minimum height=5.5mm,inner sep=1pt,font=\small}]
\def\tmrow#1#2#3#4#5#6{%
  \node[anchor=east,font=\small] at (-1,#1) {$#2$};
  \node[cell] at (10,#1) {$#3$};
  \node[cell] at (29,#1) {$#4$};
  \ifnum#5=0 \draw[very thick] (0.5,{#1-2.75}) rectangle (19.5,{#1+2.75});\else \draw[very thick] (19.5,{#1-2.75}) rectangle (38.5,{#1+2.75});\fi
  \node[anchor=west,font=\footnotesize] at (41,#1) {#6};}
\node[font=\footnotesize] at (10,6) {cell $c$};
\node[font=\footnotesize] at (29,6) {cell $c+1$};
\node[font=\footnotesize,anchor=east] at (-1,6) {state};
\tmrow{0}{\mathrm{R}_0}{[i{\uparrow},s]}{u}{0}{the state of $\hat T$ is $q_i$}
\tmrow{-7}{\mathrm{R}_0}{[j{\downarrow},t]}{u}{1}{write $t$ and the number $j$ of $q_j$, move right}
\tmrow{-14}{\mathrm{L}}{[j{\downarrow},t]}{[1{\uparrow},u]}{0}{place a counter with value 1, move left}
\tmrow{-21}{\mathrm{R}_1}{[(j-1){\downarrow},t]}{[1{\uparrow},u]}{1}{decrement, move right}
\tmrow{-28}{\mathrm{L}}{[(j-1){\downarrow},t]}{[2{\uparrow},u]}{0}{increment, move left}
\node at (19.5,-33.5) {$\vdots$};
\node[anchor=west,font=\footnotesize] at (41,-33.5) {the last two steps $j-1$ times in all};
\tmrow{-40}{\mathrm{L}}{[1{\downarrow},t]}{[j{\uparrow},u]}{0}{}
\tmrow{-47}{\mathrm{R}_0}{t}{[j{\uparrow},u]}{1}{remove the mark, move right; the state of $\hat T$ is $q_j$}
\end{tikzpicture}
\caption{Morita's three-state machine executing a quintuple $[q_i,s,t,\mathrm{R},q_j]$ of $\hat T$}
\label{fig:three-carry}
\end{figure}
\begin{figure}[tb]
\centering
\begin{tikzpicture}[x=1mm,y=1mm,font=\footnotesize]
\node[flbox,text width=40mm] (L) at (0,0) {State $\rmL$ (entered by left moves)\\ also the returning state of $\hat T$; reads only down-marked counters; $\delta^\dagger_8$ moves left over non-blank symbols};
\node[flbox,text width=40mm] (R0) at (106,0) {State $\rmR_0$\\ reads only up-marked counters; $\delta^\dagger_2$ executes a right-moving quintuple of $\hat T$};
\node[flbox,text width=40mm] (R1) at (0,-40) {State $\rmR_1$\\ reads only up-marked counters};
\draw[flarr] ([yshift=3mm]L.east) -- node[above,text width=56mm,align=center] {$\delta^\dagger_1$ mark the blank at the left end with $j{\downarrow}$ (a sweep starts)\\ $\delta^\dagger_6$ turn $[1{\downarrow},t]$ back into $t$ (a carry ends)} ([yshift=3mm]R0.west);
\draw[flarr] ([yshift=-3mm]R0.west) -- node[below,text width=56mm,align=center] {$\delta^\dagger_3$ turn a plain $u$ into $[1{\uparrow},u]$ (a carry starts)\\ $\delta^\dagger_7$ remove the counter on the blank (the sweep turns back)} ([yshift=-3mm]L.east);
\draw[flarr] ([xshift=-8mm]L.south) -- node[left] {$\delta^\dagger_4$ (decrement)} ([xshift=-8mm]R1.north);
\draw[flarr] ([xshift=8mm]R1.north) -- node[right] {$\delta^\dagger_5$ (increment)} ([xshift=8mm]L.south);
\end{tikzpicture}
\caption{Flowchart of all rules of the three-state machine of Morita's Lemma~5.10~\cite{Morita17}}
\label{fig:dagger-flow}
\end{figure}
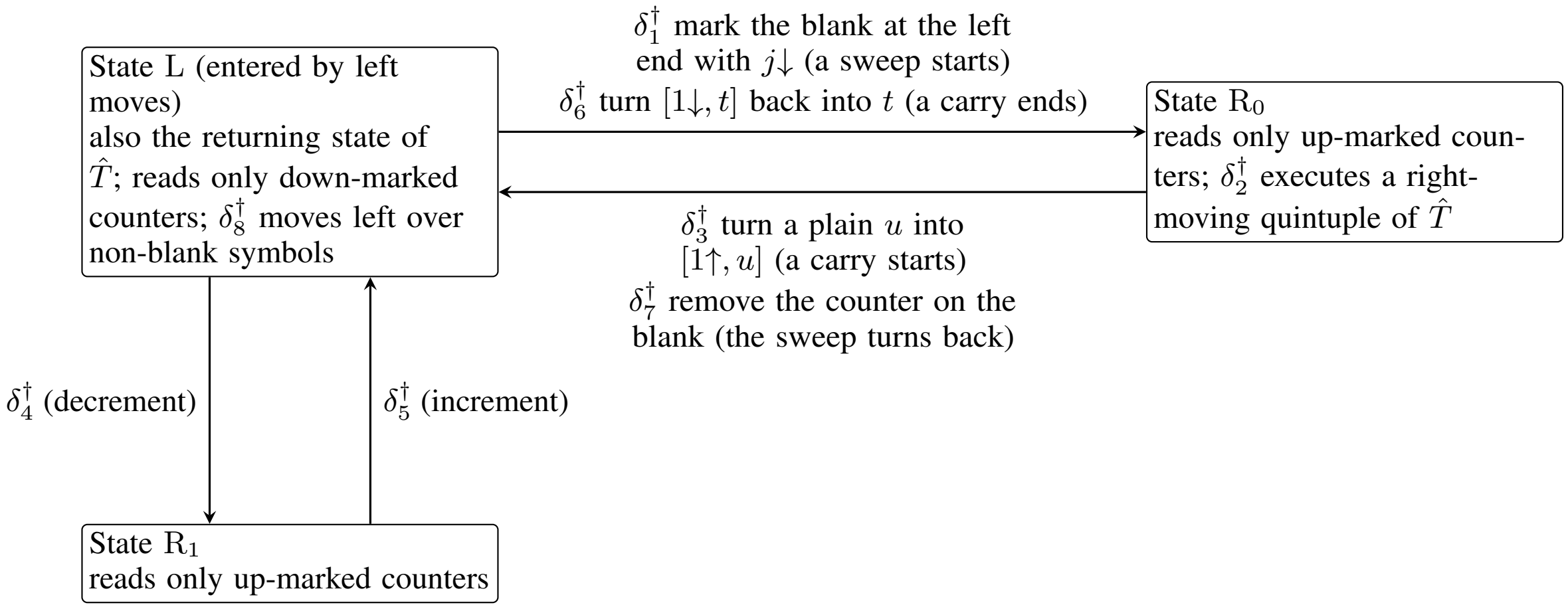

\paragraph{Shared counters.}
In a one-way-updating machine all carries go to the right, so each of the three states always reads counters on a fixed side: $\rmL$ only on the source side, $\rmR_0$ and $\rmR_1$ only on the target side.
In Fig.~\ref{fig:three-carry}, too, $\rmL$ reads only down-marked counters, and $\rmR_0$ and $\rmR_1$ read only up-marked ones.
The same can be seen in Fig.~\ref{fig:dagger-flow}, which shows all rules of Morita's Lemma~5.10 as a flowchart. The labels $\delta^\dagger_1,\dots,\delta^\dagger_8$ are the names that Morita gives to these rules in the proof of Lemma~5.10~\cite{Morita17}; we use them only to relate the figure to that proof.
Hence the up and down marks can be identified without losing determinism, and reversibility is preserved, since the written symbols of the rules entering each state remain distinct.
The count drops from $N(2M-1)$ to $NM$; for $(32,3)$, where $N=298$ and $M=62$, it drops from $N(2M-1)=36654$ to $NM=298\cdot 62=18476$.

\paragraph{Pruned counters.}
A counter $[k,x]$ is needed only if some state with number at least $k$ can read $x$ or is entered by writing $x$.
Keeping only these counters gives $N+\sum_x\mathrm{need}(x)$ symbols, where $\mathrm{need}(x)$ is the largest such number.
The sum depends on the numbering of the states of $\hat T$; states with the same set of relevant symbols can be kept adjacent; there are four such groups, and the best of all $4!=24$ orders of them gives 1,820 symbols for $(32,3)$.
The machines with 18,476 and 1,820 symbols measure the effect of these two steps. For the one with 18,476 symbols we prove universality in the sense of Definition~\ref{def:univ}; for the one with 1,820 symbols we only count its symbols and prove neither that it simulates $T$ nor that the kept counters suffice along every run.
The machines of the next two steps also share and prune counters; their universality follows from general theorems on the conversion (determinism, reversibility, and simulation of $T$), whose hypotheses are checked for the concrete tables.

\paragraph{One symbol per cell.}
The remaining cost lies in the $n^2$ pairs and the $mn^2$ marked pairs.
Instead of packing two cells, we keep one symbol per cell and represent the blank of $T$ by a separate symbol $B$ inside the used region; we write $\bar T$ for the one-way-updating machine obtained in this way.
Since $\bar T$ also only sweeps to the right, the sweep would already have passed the target cell of a left move of $T$. As in Morita's construction, each sweep therefore shifts the whole content of $T$ one cell to the right: a left move of $T$ then keeps the head where it is (the content of the target cell arrives at the current cell), and a right move advances it by two cells.
The current state $p$ of $T$ is recorded as a mark on one cell: on the head cell if $p\in Q_-$, and on the cell to the left of the head if $p\in Q_+$. Since the states of $Q_+$ are entered only by right moves, a sweep that executes a right move can leave the written symbol and the target state together in the mark of one cell. Here, unlike in the four-state conversions and as in Morita's construction, a state that is not entered by any quintuple is put into $Q_-$.
Figure~\ref{fig:onewu-shift} shows the tape before and after one sweep: the state name above a cell is its mark, the thick frame marks the head of $T$, and the dashed arrows show the shift by one cell to the right. If $p\in Q_+$, the mark before the sweep is on the cell to the left of the head instead, and the result of the sweep is the same.
$\bar T$ has $n+3+|R|+|G|$ states, where $R$ is the set of pairs $(q,t)$ such that $T$ has a right-moving quintuple into $q$ writing $t$, and $G=R\cup(\{p_{\mathrm{init}}\}\times S)$ if the state $p_{\mathrm{init}}$ in which the run of $T$ starts is in $Q_+$ and $G=R$ otherwise ($S$ is the set of symbols of $T$).
After the shared and pruned counters, the three-state machine $T'$ has
\[
  1+n+(m_-n+|G|)(n+1)+n(n+2+|R|+|G|)+(n+2)
\]
symbols. Of these, $1+n+(m_-n+|G|)$ are the symbols of $\bar T$ (the blank, the plain symbols, and the state marks), and the rest are counter symbols: $n$ for each mark, $n+2+|R|+|G|$ for each plain symbol, and $n+2$ for the blank.
Since $|R|\le\min(|T|,m_+n)$, where $|T|$ is the number of quintuples of $T$, this is $O(n(mn+|T|))\subseteq O(mn^2)$, compared with $O(m\,m_+n^3)\subseteq O(m^2n^3)$ for Morita's construction.
For $(32,3)$ it gives 659 symbols.

\begin{figure}[tb]
\centering
\begin{tikzpicture}[x=1mm,y=1mm,>=stealth,
  cell/.style={draw,minimum width=14mm,minimum height=6mm,inner sep=1pt,font=\small},
  head/.style={cell,very thick},
  stmark/.style={draw,rounded corners=1pt,fill=white,inner sep=0.8pt,font=\scriptsize},
  shiftarrow/.style={->,densely dashed,gray},
  note/.style={anchor=west,font=\footnotesize,align=left}]
\def\before#1{%
  \foreach \x/\lbl in {0/i-1,16/i,32/i+1,48/i+2} \node[font=\footnotesize] at (\x,{#1+7}) {$\lbl$};
  \node[anchor=east,font=\footnotesize] at (-9,#1) {before the sweep};
  \node[cell] at (0,#1) {$s$};
  \node[head] at (16,#1) {$u$};
  \node[stmark] at (20.5,{#1+3}) {$p$};
  \node[cell] at (32,#1) {$x$};
  \node[note] at (58,#1) {$T$ is in state $p\in Q_-$ and reads $u$};
  \draw[shiftarrow] (1,{#1-3}) -- (15,{#1-11});
  \draw[shiftarrow] (17,{#1-3}) -- (31,{#1-11});
  \draw[shiftarrow] (33,{#1-3}) -- (47,{#1-11});
  \node[anchor=east,font=\footnotesize] at (-9,{#1-14}) {after the sweep};}
\node[anchor=west] at (-34,20) {(a) left-moving quintuple $[p,u,t,\mathrm{L},q]$};
\before{5}
\node[head] at (16,-9) {$s$};
\node[stmark] at (20.5,-6) {$q$};
\node[cell] at (32,-9) {$t$};
\node[cell] at (48,-9) {$x$};
\node[note] at (58,-9) {the head stays\\the mark $q\in Q_-$ is on the head cell};
\node[anchor=west] at (-34,-22) {(b) right-moving quintuple $[p,u,t,\mathrm{R},q]$};
\before{-37}
\node[cell] at (16,-51) {$s$};
\node[cell] at (32,-51) {$t$};
\node[stmark] at (36.5,-48) {$q$};
\node[head] at (48,-51) {$x$};
\node[note] at (58,-51) {the head advances two cells\\the mark $q\in Q_+$ is left of the head};
\end{tikzpicture}
\caption{Execution of a quintuple of $T$ by one sweep of $\bar T$}
\label{fig:onewu-shift}
\end{figure}
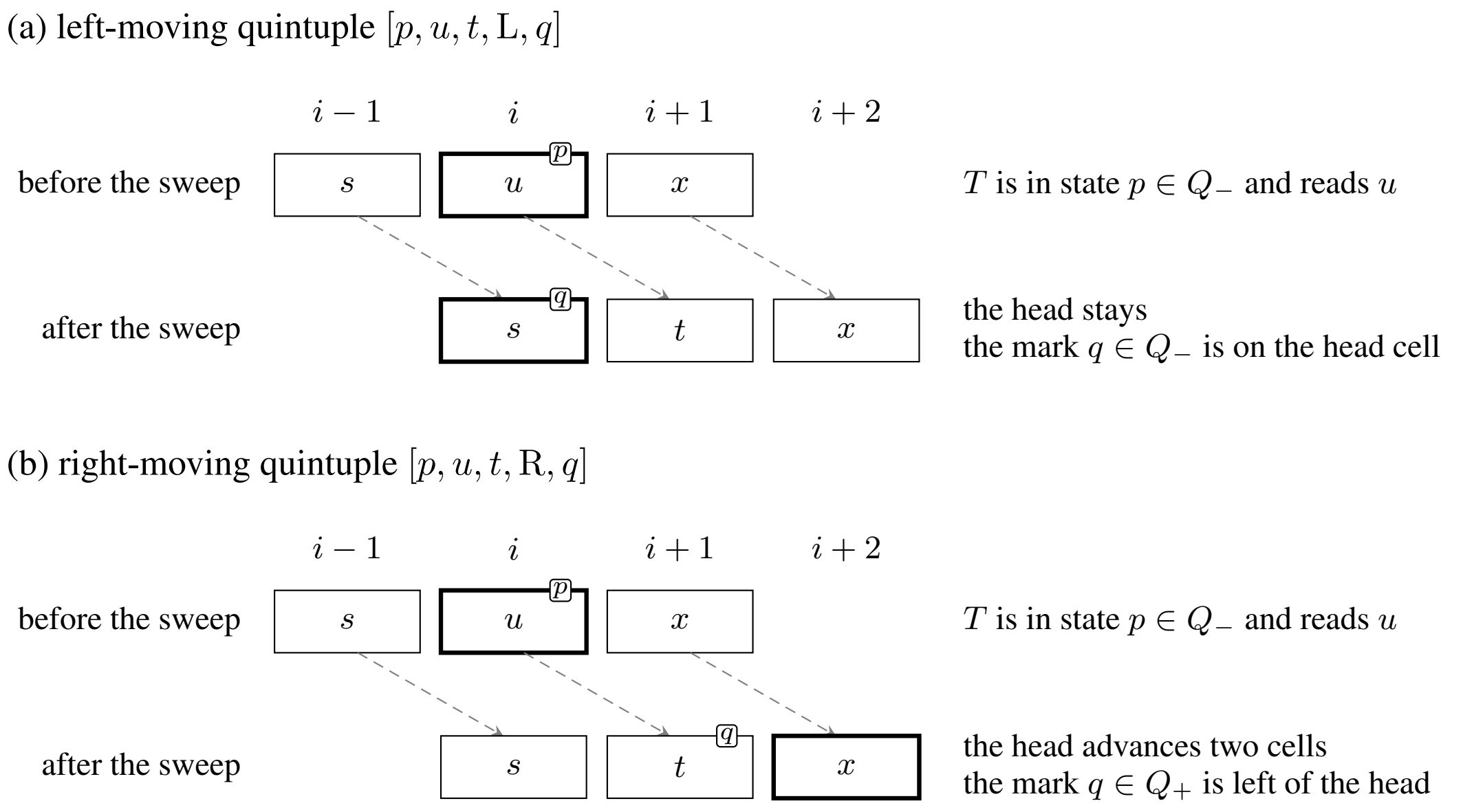

\begin{figure}[tb]
\centering
\begin{tikzpicture}[x=1mm,y=1mm,font=\footnotesize]
\node[anchor=west] at (-4,9) {(a) The steps of reflection};
\node[flbox] (A) at (4,0) {$T$};
\node[flbox] (B) at (40,0) {$\mathrm{mir}(T)$};
\node[flbox,align=center] (C) at (84,0) {three-state machine\\simulating $\mathrm{mir}(T)$};
\node[flbox,align=center] (D) at (130,0) {three-state machine\\simulating $T$};
\draw[flarr] (A) -- node[above] {L $\leftrightarrow$ R} (B);
\draw[flarr] (B) -- node[above] {conversion} (C);
\draw[flarr] (C) -- node[above] {L $\leftrightarrow$ R} (D);
\node[anchor=west] at (-4,-13) {(b) Quintuples and the side of their targets};
\node[anchor=west] at (2,-20) {quintuples of $T$};
\node[anchor=west] at (70,-20) {quintuples of $\mathrm{mir}(T)$};
\node[anchor=west] (r1) at (2,-27) {$[p,s,t,\mathrm{R},q]$: $q\in Q_+$, $(q,t)\in R$};
\node[anchor=west] (m1) at (70,-27) {$[p,s,t,\mathrm{L},q]$: $q\in Q_-$, $(q,t)\notin R$};
\node[anchor=west] (r2) at (2,-34) {$[p',s',t',\mathrm{L},q']$: $q'\in Q_-$};
\node[anchor=west] (m2) at (70,-34) {$[p',s',t',\mathrm{R},q']$: $q'\in Q_+$, $(q',t')\in R$};
\draw[flarr] (55,-27) -- (69,-27);
\draw[flarr] (55,-34) -- (69,-34);
\node[anchor=west] at (-4,-45) {(c) The values for $(32,3)$};
\node[anchor=north west] at (2,-49) {\begin{tabular}{lrrrrr}
\hline
 & $m_+$ & $m_-$ & $|R|$ & $|G|$ & symbols\\
\hline
$T$ & 18 & 14 & 46 & 47 & 659\\
$\mathrm{mir}(T)$ & 14 & 18 & 36 & 36 & 600\\
\hline
\end{tabular}};
\end{tikzpicture}
\caption{Reflection in the three-state conversion}
\label{fig:mirror}
\end{figure}

\paragraph{Reflection.}
Applying the conversion to the mirror image $\mathrm{mir}(T)$ of $T$ (directions L and R exchanged) and exchanging L and R back in the result gives a three-state machine that simulates $T$ (Fig.~\ref{fig:mirror}(a)).
In the mirror image, a right-moving quintuple $[p,s,t,\mathrm{R},q]$ becomes the left-moving quintuple $[p,s,t,\mathrm{L},q]$, so $q$ moves from $Q_+$ to $Q_-$ and the pair $(q,t)$ leaves $R$. Conversely, a left-moving quintuple becomes right-moving, and its target enters $Q_+$ and its pair enters $R$ (Fig.~\ref{fig:mirror}(b)).
Thus the roles of $m_+$ and $m_-$ and of the right-moving and left-moving targets are exchanged.
In the symbol count, the coefficient of $m_-$ is $n(n+1)$, that of $|G|$ is $2n+1$, and that of $|R|$ is $n$, so the exchange changes the count.
For $(32,3)$, $T$ itself has $m_-=14$ and $|R|=46$, and $|G|=47$ because the state in which $T$ starts is in $Q_+$; this gives 659 symbols. The mirror image has $m_-=18$ and $|R|=|G|=36$ (Fig.~\ref{fig:mirror}(c)). The increase of $m_-$ by 4 adds $4\cdot 12=48$ symbols, and the decreases of $|G|$ by 11 and of $|R|$ by 10 remove $77+30=107$, which gives 600 symbols, and hence URTM$(3,600)$. Its table has 1,572 quintuples; instead of printing it, we provide it as a text file in the same repository (Section~\ref{sec:merge}).
Applied to the other five original machines, the improved conversion gives between 802 and 1,209 symbols, taking the better of the machine and its mirror image, so $(32,3)$ is the best starting point.

\subsection{Two symbols}
\label{sec:two}

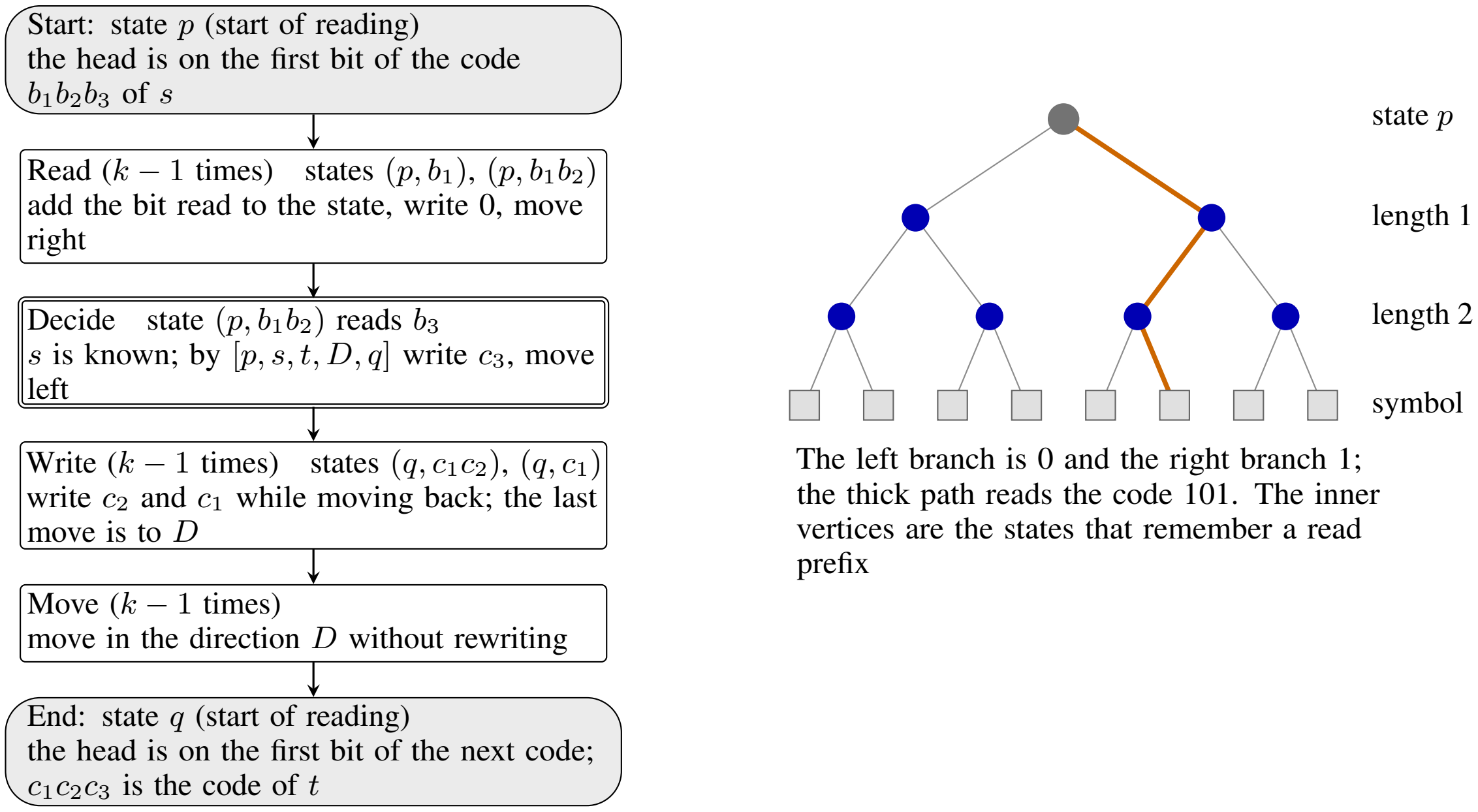
\begin{figure}[tb]
\centering
\begin{tikzpicture}[x=1mm,y=1mm,font=\footnotesize,node distance=3.5mm]
\node[flterm,text width=58mm] (B1) at (0,0) {Start: state $p$ (start of reading)\\ the head is on the first bit of the code $b_1b_2b_3$ of $s$};
\node[flbox,text width=58mm] (B2) [below=of B1] {Read ($k-1$ times)\quad states $(p,b_1)$, $(p,b_1b_2)$\\ add the bit read to the state, write 0, move right};
\node[fldec,text width=58mm] (B3) [below=of B2] {Decide\quad state $(p,b_1b_2)$ reads $b_3$\\ $s$ is known; by $[p,s,t,D,q]$ write $c_3$, move left};
\node[flbox,text width=58mm] (B4) [below=of B3] {Write ($k-1$ times)\quad states $(q,c_1c_2)$, $(q,c_1)$\\ write $c_2$ and $c_1$ while moving back; the last move is to $D$};
\node[flbox,text width=58mm] (B5) [below=of B4] {Move ($k-1$ times)\\ move in the direction $D$ without rewriting};
\node[flterm,text width=58mm] (B6) [below=of B5] {End: state $q$ (start of reading)\\ the head is on the first bit of the next code; $c_1c_2c_3$ is the code of $t$};
\draw[flarr] (B1) -- (B2); \draw[flarr] (B2) -- (B3); \draw[flarr] (B3) -- (B4); \draw[flarr] (B4) -- (B5); \draw[flarr] (B5) -- (B6);
\begin{scope}[shift={(76,-6)}]
\draw[black!45] (0,0) -- (-15,-10);
\draw[line width=1.4pt,orange!80!black] (0,0) -- (15,-10);
\draw[black!45] (-15,-10) -- (-22.5,-20);
\draw[black!45] (-15,-10) -- (-7.5,-20);
\draw[line width=1.4pt,orange!80!black] (15,-10) -- (7.5,-20);
\draw[black!45] (15,-10) -- (22.5,-20);
\draw[black!45] (-22.5,-20) -- (-26.25,-29);
\draw[black!45] (-22.5,-20) -- (-18.75,-29);
\draw[black!45] (-7.5,-20) -- (-11.25,-29);
\draw[black!45] (-7.5,-20) -- (-3.75,-29);
\draw[black!45] (7.5,-20) -- (3.75,-29);
\draw[line width=1.4pt,orange!80!black] (7.5,-20) -- (11.25,-29);
\draw[black!45] (22.5,-20) -- (18.75,-29);
\draw[black!45] (22.5,-20) -- (26.25,-29);
\fill[black!55] (0,0) circle (1.6);
\fill[blue!70!black] (-15,-10) circle (1.4);
\fill[blue!70!black] (15,-10) circle (1.4);
\fill[blue!70!black] (-22.5,-20) circle (1.4);
\fill[blue!70!black] (-7.5,-20) circle (1.4);
\fill[blue!70!black] (7.5,-20) circle (1.4);
\fill[blue!70!black] (22.5,-20) circle (1.4);
\filldraw[fill=black!12,draw=black!60] (-27.75,-30.5) rectangle (-24.75,-27.5);
\filldraw[fill=black!12,draw=black!60] (-20.25,-30.5) rectangle (-17.25,-27.5);
\filldraw[fill=black!12,draw=black!60] (-12.75,-30.5) rectangle (-9.75,-27.5);
\filldraw[fill=black!12,draw=black!60] (-5.25,-30.5) rectangle (-2.25,-27.5);
\filldraw[fill=black!12,draw=black!60] (2.25,-30.5) rectangle (5.25,-27.5);
\filldraw[fill=black!12,draw=black!60] (9.75,-30.5) rectangle (12.75,-27.5);
\filldraw[fill=black!12,draw=black!60] (17.25,-30.5) rectangle (20.25,-27.5);
\filldraw[fill=black!12,draw=black!60] (24.75,-30.5) rectangle (27.75,-27.5);
\node[anchor=west] at (30,0) {state $p$};
\node[anchor=west] at (30,-10) {length 1};
\node[anchor=west] at (30,-20) {length 2};
\node[anchor=west] at (30,-29) {symbol};
\node[text width=62mm,align=left] at (4,-40) {The left branch is 0 and the right branch 1; the thick path reads the code 101. The inner vertices are the states that remember a read prefix};
\end{scope}
\end{tikzpicture}
\caption{Flowchart of Morita's conversion to two symbols ($k=3$) and the tree of reading states}
\label{fig:two-flow}
\end{figure}

Morita's conversion to two symbols~\cite[Sect.~5.3.3]{Morita17} replaces each symbol by a string of $k$ bits, the blank by $0^k$.
A quintuple of $T$ is simulated by reading $k-1$ bits while erasing them, reading the last bit and writing the last bit of the new symbol, writing the remaining bits while moving back, and moving $k-1$ cells in the direction of the move; the states remember the prefix read or the prefix still to write.
If $P_r$ and $P_w$ denote the numbers of read and written prefixes of lengths 1 to $k-1$, summed over the states, the number of states of $T'$ is at most $mk+P_r+P_w$.
Figure~\ref{fig:two-flow} shows the flowchart for $k=3$. The states of the reading stage are counted in $P_r$ and those of the writing stage in $P_w$, while the start state and the states of the moving stage number $k$ for each state of $T$. The inner vertices of the tree on the right are the states that remember a read prefix.
From URTM$(24,4)$ ($k=2$) this gives URTM$(138,2)$, and from URTM$(13,7)$ ($k=3$) it gives 169 states.
For $(24,4)$, $P_r=P_w=45$, so the number of states is $m'=mk+P_r+P_w=24\cdot 2+45+45=138$; for $(13,7)$, $P_r=P_w=65$, so it is $13\cdot 3+65+65=169$.

The states exist to remember the prefix read so far and the prefix still to be written. Choosing the assignment of codes should let more symbols share prefixes, and skipping bits that need not be rewritten on the way back should remove auxiliary states.
We add four degrees of freedom: the assignment of codes to symbols (the blank fixed to $0^k$), the values written in place of the bits that have been read, a path type for each quintuple that says how many bits need not be rewritten on the way back (a right-moving quintuple can then cross its bit string once instead of three times), and, for each state, the end of the bit string at which the head stands when reading begins.
Reversibility of the result reduces to distinct written values on the rules entering each auxiliary state.
For fixed codes and starting ends, the written values and the path types are chosen by a constraint programming model that minimizes the number of auxiliary states; the codes and starting ends are chosen by exhaustive enumeration and hill climbing.
Finally, states whose rules agree on the entries defined for both are merged; a merged table fills undefined entries of one state with the rules of the other, so it simulates $T$ only along runs that never reach an entry that is undefined in one of the merged states but defined in the other, which is the case for runs from encoded CTSH inputs.
For $(13,7)$, choosing only the codes gives 149 states, choosing also the written values and path types gives 135, and merging states gives 92.
A randomized large-neighborhood search over the choices of the table before merging then gave a table with 137 states, more than 135 before merging but one that merges much further, down to 85 states and 153 quintuples, that is, URTM$(85,2)$. Its table is provided at the same location as that of URTM$(3,600)$.
We have no closed formula for the size of this search result; it is verified for the concrete table.
Applied to the other five original machines, the same method, including the search, gives 92, 92, 108, 115, and 115 states for $(10,8)$, $(24,4)$, $(15,6)$, $(17,5)$, and $(32,3)$.
These five tables are also proved universal in the sense of Definition~\ref{def:univ}, and all of them are dominated by URTM$(85,2)$.

\section{The new URTMs and their universality}
\label{sec:univ}

\begin{theorem}[Existence]
\label{thm:exists}
For each of the 13 points
\begin{multline*}
(m,n)\in\{(3,600),(4,38),(5,36),(6,28),(7,26),(8,21),(9,8),(10,7),\\(15,6),(17,5),(24,4),(32,3),(85,2)\},
\end{multline*}
there is a table $U$ that is deterministic and reversible, satisfies $\mathrm{NStates}(U)=m$ and $\mathrm{NSyms}(U)=n$, and satisfies $\mathrm{IsUTMStrict}(U)$.
\end{theorem}

These points are shown in Figure~\ref{fig:compare} (Section~\ref{sec:intro}).
The four points $(15,6)$, $(17,5)$, $(24,4)$, and $(32,3)$ are Morita's original machines, which remain non-dominated; Morita's URTM$(10,8)$ and URTM$(13,7)$ are dominated by URTM$(10,7)$. The other nine points are new.
The witnesses are concrete rule tables: Morita's distributed tables~\cite{MoritaFiles} and the tables produced from them by the conversions of Section~\ref{sec:constr}.

Table~\ref{tab:results} lists the numbers of symbols and states obtained by applying the conversions of Section~\ref{sec:constr} to each of Morita's six machines and to URTM$(85,2)$. The columns of the conversions that reduce states give numbers of symbols, and those of the conversions that reduce symbols give numbers of states. For each conversion, Morita's construction and ours are listed side by side. The column of Morita's four-state conversion gives the value of the formula $(2m+1)n$; Morita gives the table only for $(10,8)$. All other entries count the symbols or states of constructed tables. In the three-state column of our construction, the two values separated by ``/'' are for the machine itself and for its mirror image. Bold entries are points of Theorem~\ref{thm:exists}, and the machines marked $\ast$ are proved universal in the sense of Definition~\ref{def:univ}. A dash marks a conversion that does not apply (a two-symbol conversion of a two-symbol machine). For $(85,2)$, the reverse-counting entry 96 equals the bound $(\max(m_+,m_-)+1)n$ and is not shown to be the least for this conversion; at least 94 symbols are needed. The results of the block decomposition and of symbol merging are in Tables~\ref{tab:blocks} and~\ref{tab:merge}.

\begin{table}[tb]
\centering
\caption{Computed results of the conversions applied to each URTM}
\label{tab:results}
\small
\resizebox{\textwidth}{!}{%
\begin{tabular}{lrrrrrrr}
\toprule
 & \multicolumn{3}{c}{Four states (symbols)} & \multicolumn{2}{c}{Three states (symbols)} & \multicolumn{2}{c}{Two symbols (states)} \\
\cmidrule(lr){2-4}\cmidrule(lr){5-6}\cmidrule(lr){7-8}
From & Morita & disjoint ranges & reverse counting & Morita & one symbol per cell & Morita & ours \\
\midrule
$(10,8)$ & 168 & 88$^\ast$ & 48$^\ast$ & 81075 & 1268 / 1209 & 139 & 92$^\ast$ \\
$(13,7)$ & 189 & 98 & 61 & 98241 & 1073 / 1144 & 169 & \textbf{85}$^\ast$ \\
$(15,6)$ & 186 & 96 & 53 & 77895 & 1038 / 935 & 174 & 108$^\ast$ \\
$(17,5)$ & 175 & 90 & 55 & 55473 & 856 / 844 & 193 & 115$^\ast$ \\
$(24,4)$ & 196 & 100 & 56 & 49323 & 832 / 802 & 138 & 92$^\ast$ \\
$(32,3)$ & 195 & 99 & 57 & 36654 & 659$^\ast$ / \textbf{600}$^\ast$ & 182 & 115$^\ast$ \\
$(85,2)$ & 342 & 172 & 96 & 68655 & 803 / 808 & -- & -- \\
\bottomrule
\end{tabular}}
\end{table}

\subsection{Proof of universality}

Each witness other than the five machines of Table~\ref{tab:merge} is first shown to simulate an arbitrary CTSH: for every CTSH $\hat C$ and initial string $v$, the machine started on the code of $(\hat C,v)$ halts if and only if $\hat C$ halts from $(v,0)$, and its halting ID is the code of the halting ID of $\hat C$.
For Morita's machines this is a direct proof about the given tables, and for URTM$(10,7)$ and URTM$(9,8)$ it is the proof of Section~\ref{sec:tags}, again about the concrete table.
For the converted machines it follows from general simulation theorems for the conversions (preservation of determinism, reversibility, one-step simulation, and the correspondence of halting, halting IDs, and non-halting), instantiated with the concrete tables.
The five merged machines obtained from block tables inherit universality from their tables before merging by Theorem~\ref{thm:merge}.

Universality in the sense of Definition~\ref{def:univ} then requires computable maps between arbitrary TMs and CTSHs.
We formalize the chain
\[
  \mathrm{tm}(i)\ \longrightarrow\ \text{two-register machine}\ \longrightarrow\ \text{2-tag system}\ \longrightarrow\ \text{CTSH}\ \longrightarrow\ U,
\]
which follows the classical route of Cocke and Minsky~\cite{CockeMinsky64} and Morita's Theorem~7.2~\cite{Morita17}.
The first step stores the tape of $\mathrm{tm}(i)$ in two registers and normalizes the symbols of $\mathrm{tm}(i)$, which are arbitrary natural numbers, into binary; this replaces a separate conversion of the TM to two symbols.
A general theorem then states that a machine that simulates an arbitrary CTSH and has a primitive recursive input encoding and a primitive recursive readout of the final string, independent of trailing blanks, satisfies $\mathrm{IsUTMStrict}$.
Each machine supplies only its assignment of natural numbers to states and symbols, its encoding of CTSHs, and its readout of final strings.
URTM$(10,7)$ uses the encoding of URTM$(13,7)$, and its decoding differs from that of URTM$(13,7)$ only in counting the separators $\ast$ and $\mathtt{Y}$ together, minus one in the state $\mathsf{halt}$ (Section~\ref{sec:tags}). URTM$(9,8)$ uses the encoding and the decoding of URTM$(10,7)$.

It follows directly from Definition~\ref{def:univ} that any two of the 13 witnesses of Theorem~\ref{thm:exists} agree: for every TM $\mathrm{tm}(i)$ and every admitted input $a$, both halt on their encodings of $(i,a)$ or neither does, and if both halt, the decoded halting IDs denote the same ID; the encoded inputs, the runs, and the numbers of steps differ between them.

\paragraph{Halting quintuples.}
The distributed tables of Morita's four points represent the halting entries of the tables in the book by two zero-shift quintuples into halting states that are not counted, and the tables of the five points obtained by symbol merging and of $(9,8)$ and $(10,7)$ inherit these two quintuples through the conversions; the tables of $(85,2)$ and $(3,600)$ have no halting quintuples.
We prove Theorem~\ref{thm:exists} for these 11 points also for the tables without the two quintuples (for Morita's four points, these are the tables printed in the book).
Removing them changes neither the number of states nor the number of symbols. A table without them halts one step before the original table enters a halting state, so the decoder first applies the removed quintuple once to the halting ID.

\subsection{Counting states and symbols}
\label{sec:counting}

Morita does not count halting states in the $m$ of URTM$(m,n)$.
The book states that the set of final states is not specified in the problem of reducing the number of states, the state sets of the URTMs in Chapter~7 contain no halting states, and in Chapter~11 it notes that the halting states are not included in URTM$(17,5)$~\cite{Morita17}.
The distributed files also state explicitly that the halting states are not counted.

Our $\mathrm{NStates}$ counts source states and $\mathrm{NSyms}$ counts the symbols occurring in the table together with the blank.
The two conventions could differ for states that occur only as targets, for halting states, and for symbols that do not occur in the table.
For the witness tables of Theorem~\ref{thm:exists} we prove that every target is a source state; for $(85,2)$ and $(3,600)$ this holds already for the tables in the form of the distributed files, and for the other 11 points it holds for the tables without the two halting quintuples.
Hence the number of states remains $m$ even under a convention that also counts targets.
We also prove that the blank occurs in each witness and that every symbol is both read and written, and that the witnesses have exactly $(m,n)$ under Morita's way of counting (targets counted, only symbols occurring in the table).
An independent count outside the proof assistant agrees (Table~\ref{tab:counts}).
For the 11 tables with the additional halting quintuples, counting targets gives $m+2$ states.

\begin{table}[tb]
\centering
\caption{Independent count of states and symbols in the witness tables}
\label{tab:counts}
\small
\begin{tabular}{lrcccc}
\toprule
Table & Quint. & Sources & Target-only & Symbols & Zero-shifts \\
\midrule
Morita's 4 points, distributed & 64--84 & $m$ & 2 & $n$ & 2 \\
same, tables in the book & 62--82 & $m$ & 0 & $n$ & 0 \\
5 symbol-merging points, with 2 halting quintuples & 139--152 & $m$ & 2 & $n$ & 2 \\
same, without them (published tables) & 137--150 & $m$ & 0 & $n$ & 0 \\
$(9,8)$ & 54 & 9 & 2 & 8 & 2 \\
same, without 2 quintuples & 52 & 9 & 0 & 8 & 0 \\
$(10,7)$ & 57 & 10 & 2 & 7 & 2 \\
same, without 2 quintuples & 55 & 10 & 0 & 7 & 0 \\
$(85,2)$ & 153 & 85 & 0 & 2 & 0 \\
$(3,600)$ & 1,572 & 3 & 0 & 600 & 0 \\
\bottomrule
\end{tabular}
\end{table}

\section{Nonexistence of URTMs with at most two states}
\label{sec:twostate}

\begin{theorem}[Halting of two-state RTMs]
\label{thm:halting}
There is a computable Boolean function $f$ such that, for all distinct states $q_0,q_1,q_f$, every deterministic and reversible table $T$ whose source states are among $q_0,q_1$ and whose targets are among $q_0,q_1,q_f$, every position $p\in\mathbb{Z}$, and every code $z\in\Code$,
\[
  f(T,p,z)=\mathrm{true}\iff\mathrm{Halts}_T(\mathrm{toID}(p,z)).
\]
\end{theorem}

\begin{corollary}
\label{cor:twostate}
If $U$ is deterministic and reversible and $\mathrm{NStates}(U)\le 2$, then $\neg\mathrm{IsUTM}(U)$, and hence $\neg\mathrm{IsUTMStrict}(U)$.
Moreover, there is a TM $\mathrm{tm}(i)$ that is simulated, in the sense of conditions~2 and~3 of Definition~\ref{def:univ}, by no deterministic and reversible two-state table.
\end{corollary}

The corollary follows from Theorem~\ref{thm:halting} and the undecidability of the halting problem.
Two reductions bring a general table into the form of the theorem.
A one-state table is a special case of a two-state one.
If a table has several halting targets, removing the quintuples whose target is not a source state does not change whether the machine halts: if the original run halts by executing a removed quintuple, the reduced run halts one step earlier.
Removing quintuples preserves determinism and reversibility.
The corollary therefore holds for any number of symbols, and with zero-shifts and halting states allowed as in Morita's definition.
If this formalization captures Morita's Definition~7.1, it answers the question of the book and of the survey~\cite{Morita23} negatively. Since halting is decidable uniformly in the table (Theorem~\ref{thm:halting}), the same conclusion holds under any definition of universality that reduces the halting problem of Turing machines to halting of the machine by a computable encoding into finite IDs; the second part of the corollary addresses the survey's question, which concerns the class of two-state RTMs.

\paragraph{Proof outline of Theorem~\ref{thm:halting}.}
Let $T$ be as in the theorem, with the quintuples into $q_f$ removed.
Fix an initial finite ID and a window $[a,b]$ containing its non-blank cells and its head; let $w=b-a+1$ and let $n$ be the number of symbols.

\begin{enumerate}
\item \emph{The state is the direction of the last move.}
By the unique shift-direction property, if $T$ has both a left-moving and a right-moving quintuple, then one state is entered only by left moves, the other only by right moves, and there is no zero-shift.
So the state records the direction of the last move; we call the state entered by left moves L and the one entered by right moves R. $T$ is then a partial injection $\pi:\{\rmL,\rmR\}\times S\rightharpoonup S\times\{\rmL,\rmR\}$, where $S$ is the set of symbols of $T$.
(If $T$ moves in only one direction, say to the right, the argument is as follows. When the head first enters a blank cell $b+1,b+2,\dots$ to the right of the window, everything to its right is blank and the head never moves left again, so the run from then on is determined by the state in which it enters. If the head reaches $b+3$, two of the three states in which it enters $b+1$, $b+2$, $b+3$ are equal, since there are only two states (Fig.~\ref{fig:oneway}). Let $c<c'$ be two cells entered in the same state. The run from $c'$ is then the run from $c$ shifted by $c'-c$ cells to the right, so the head keeps entering $c'+(c'-c)$, $c'+2(c'-c)$, \dots{} in the same state and never halts. Hence the head of a halting run stays left of $b+3$, and a halting run is short. The case of moving only to the left is symmetric.)
\item \emph{A half-line reports only that the head came back.}
Fix a boundary and the half-line $\mathcal{H}$ to its right.
The head always enters $\mathcal{H}$ in state R and leaves it in state L, and the behavior inside $\mathcal{H}$ does not depend on the content outside.
Hence the outcome of the $j$-th entry into $\mathcal{H}$ (leave, halt inside, or never return) and the content of $\mathcal{H}$ on leaving depend only on the initial content of $\mathcal{H}$ and on $j$.
In particular, moving back and forth across the boundary carries no information into $\mathcal{H}$ beyond the number of entries, and the outside learns only at which entry the head comes back.
Figure~\ref{fig:halfline} illustrates this.
\item \emph{The head cannot go deep into a blank half-line.}
Let $[k,\infty)$ be initially blank and consider a halting run.
The half-line $[k+1,\infty)$ is a translate of $[k,\infty)$, so by step~2 their contents at the $j$-th entry are translates of each other.
Since, by step~2, the $j$-th entries into the two half-lines have the same outcome, if the number of entries into $[k+1,\infty)$ caught up with that into $[k,\infty)$, the run would repeat itself shifted by one cell and never halt.
So the numbers of entries into $[k,\infty)$, $[k+1,\infty)$, $[k+2,\infty)$, \dots\ strictly decrease, and the head never passes $k$ plus the number of entries into $[k,\infty)$.
\item \emph{Each cell runs its own clock.}
A cell to the right of the initial head is first entered from the left, and afterwards the head can return only from the side to which it left.
So the pair (direction, symbol) at the $i$-th visit of the cell is obtained by iterating a fixed map $\kappa(d,s)=(\overline{d'},v)$, where $\pi(d,s)=(v,d')$, on a set of size $2n$.
All blank cells run the same clock, so a run that halts inside a blank half-line enters it at most $2n+1$ times.
\item \emph{Inside the window the machine is finite-state.}
Replacing each step that leaves the window by a turn at its end gives a machine with at most $K=3w\,n^{w}$ configurations, whose orbit the real run follows inside the window by step~2.
A run that halts inside the window does so before the orbit becomes periodic; a run that halts in a half-line enters it in every period, hence by step~4 in boundedly many periods.
In both cases each half-line is entered at most $K(2n+3)+1$ times.
\item \emph{A computable bound.}
By steps~3 and~5 the head of a halting run stays in a computable interval around the window.
A halting run of a deterministic machine never repeats an ID, so it halts within a number of steps bounded by a computable function of $n$ and $w$.
Running $T$ for that many steps decides halting.
\end{enumerate}
\begin{figure}[tb]
\centering
\begin{tikzpicture}[x=1mm,y=1mm,font=\small,>=stealth,cell/.style={draw,minimum width=10mm,minimum height=6mm,inner sep=0pt}]
\foreach \x in {0,10,20} \node[cell,fill=black!10] at (\x,0) {};
\foreach \x/\lbl in {30/b+1,40/b+2,50/b+3,60/b+4,70/b+5} {\node[cell] at (\x,0) {}; \node[font=\footnotesize] at (\x,-6) {$\lbl$};}
\node at (81,0) {$\cdots$};
\draw[decorate,decoration={brace,mirror,amplitude=3pt}] (-5,-4) -- (25,-4) node[midway,below=3pt,font=\footnotesize] {window $[a,b]$};
\node[anchor=east,font=\footnotesize] at (-6,7) {state on entry};
\foreach \x/\q in {30/q_0,40/q_1,50/q_0,60/q_1,70/q_0} \node[font=\footnotesize] at (\x,7) {$\q$};
\draw[->,bend left=40] (31,10) to node[above,font=\footnotesize] {same state} (49,10);
\draw[->,bend left=40,densely dashed] (51,10) to node[above,font=\footnotesize] {repeats} (69,10);
\end{tikzpicture}
\caption{An example of the states in which a right-moving two-state table enters the blank cells to the right of the window}
\label{fig:oneway}
\end{figure}
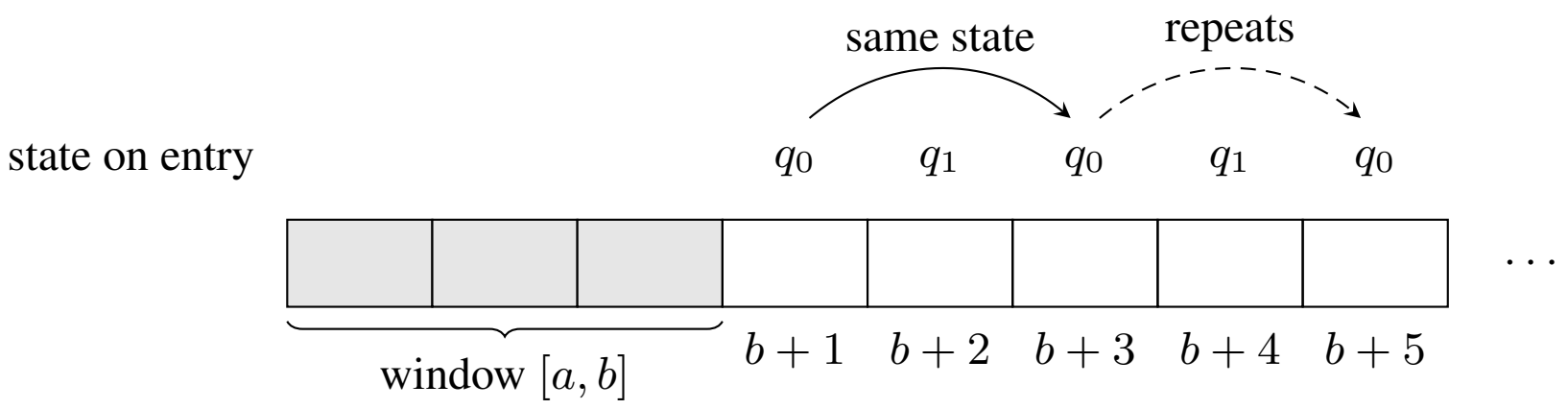
\begin{figure}[tb]
\centering
\begin{tikzpicture}[x=1mm,y=1mm,font=\small]
\fill[black!8] (30,2) rectangle (62,-46);
\draw[thick] (30,4) -- (30,-46);
\node[anchor=south east] at (29.5,4) {boundary};
\node[anchor=south west] at (30.5,4) {half-line $\mathcal{H}$};
\draw[->] (-2,2) -- (-2,-46) node[midway,left] {time};
\draw[->] (0,12) -- (62,12) node[right] {cells};
\draw[thick,blue!70!black] (14,0) -- (22,-6) -- (17,-10) -- (34,-16) -- (42,-20) -- (36,-24) -- (24,-28) -- (20,-31) -- (40,-36) -- (48,-40) -- (44,-42);
\node[anchor=south west,font=\footnotesize] at (30.5,-13.8) {enters in R};
\node[anchor=south west,font=\footnotesize] at (30.5,-32.8) {enters in R};
\node[anchor=south east,font=\footnotesize] at (29.5,-25.2) {leaves in L};
\fill (14,0) circle (0.8);
\node[anchor=east,font=\footnotesize] at (13,0) {start};
\fill (44,-42) circle (0.8);
\node[anchor=west,font=\footnotesize] at (45,-42) {halts inside $\mathcal{H}$};
\end{tikzpicture}
\caption{The head of a two-state RTM $T$ crossing the boundary of a half-line}
\label{fig:halfline}
\end{figure}
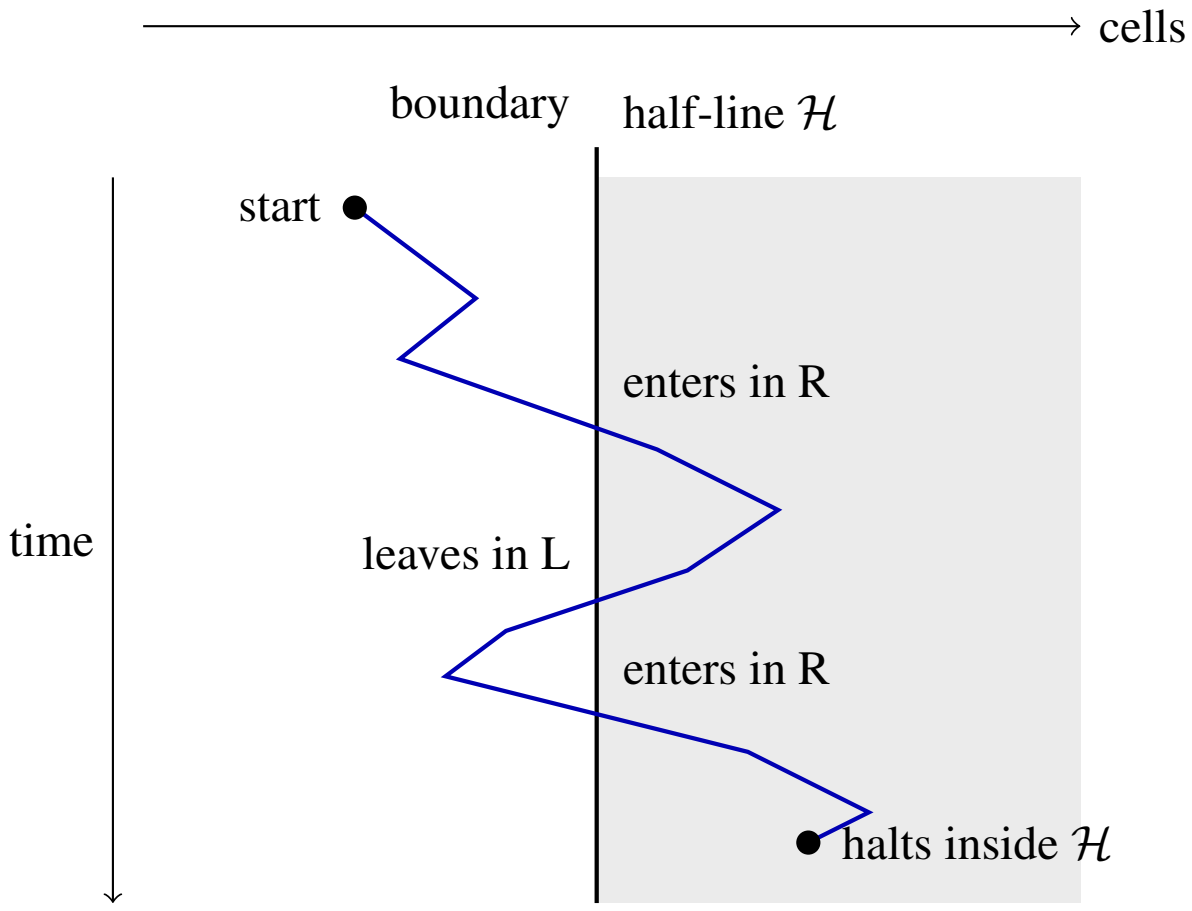
Only step~1 uses reversibility; steps~2 to~6 use only determinism and the fact that the state equals the direction of the last move.
In an irreversible two-state TM the same state can be entered from both sides, so a half-line can send more than one bit of information, and indeed such machines can be universal.
With three states, two states can be entered by right moves (as $\rmR_0$ and $\rmR_1$ in Morita's conversion to three states); the state in which the head enters a half-line then carries information, and step~2 fails.

\paragraph{A further nonexistence result.}
\begin{theorem}
\label{thm:oneL}
Let $U$ be deterministic and reversible with at most one left-moving quintuple $\ell$, and assume that no zero-shift quintuple has the source state of $\ell$ as its target.
Then $\neg\mathrm{IsUTM}(U)$.
The same holds with right and left exchanged.
\end{theorem}
This is a reversible counterpart of Pavlotskaya's theorem that the halting problem is decidable for two-symbol TMs with a single left-moving instruction~\cite{Pavlotskaya73}; here the number of states and symbols is arbitrary.
\paragraph{Proof outline.}
By reversibility, the only quintuple entering the target of $\ell$ is $\ell$ itself.
If $\ell$ enters its own source state $p$, then $p$ is entered only by $\ell$, so the head moves left only in one sweep at the start of the run and never afterwards.
Otherwise the quintuples entering $p$ all move in one direction, which is not L ($\ell$ is the only left move and it enters another state) and not Z (by the assumption), so every execution of $\ell$ immediately follows a right move into $p$; hence the head never goes more than one cell to the left of the rightmost cell it has reached.
In both cases, if the head reaches two new cells to the right of the initial window in the same state and with the same symbol to their left, the run repeats itself shifted and never halts, so a halting run stays within a computable distance of the window and halting is decidable.
Without the assumption on zero-shifts this fails: with $[p,1,1,\rmL,q]$ and $[q,1,1,\rmZ,p]$ the head moves left without bound.

\section{Nonexistence of three-state two-symbol URTMs}
\label{sec:threetwo}

\begin{theorem}
\label{thm:threetwo}
Let $U$ be deterministic and reversible, and suppose there are distinct states $q_0,q_1,q_2$ and a symbol $\sigma\ne 0$ such that every quintuple of $U$ has its source state among $q_0,q_1,q_2$ and reads and writes only $0$ and $\sigma$.
Then $\neg\mathrm{IsUTMStrict}(U)$.
\end{theorem}

The formalized proof first removes repeated quintuples from the table. This changes neither the steps nor halting, so it does not change $\mathrm{IsUTMStrict}$ either.
Targets are arbitrary, so halting states and zero-shifts are allowed as in Morita's definition.
We first prove the theorem for Morita's formulation on string IDs, with the additional requirement that the encoded inputs use only the symbols $0$ and $\sigma$.
For deterministic two-symbol tables, $\mathrm{IsUTMStrict}$ implies this formulation, because the only symbols that occur are the blank and one non-blank symbol; composing the two gives Theorem~\ref{thm:threetwo}.
As a check that the transfer is not vacuous, applying it to the witness for $(85,2)$ shows that at least one machine satisfies the string-ID form.

\begin{figure}[tb]
\centering
\begin{tikzpicture}[x=1mm,y=1mm,font=\footnotesize,node distance=3.5mm]
\node[flterm,text width=84mm] (S) at (0,0) {Start: a deterministic reversible table using only $q_0,q_1,q_2$ and $0,\sigma$};
\node[flbox,text width=84mm] (N) [below=of S] {Reduce to a normal form by renaming states and symbols and removing quintuples\\ A normal form is determined by its key; there are 359,829 keys};
\node[fldec,text width=84mm] (D) [below=of N] {Test\quad does the decision procedure checked by the kernel produce a certificate?};
\node[flbox,text width=84mm] (G) [below=8mm of D] {178 classes of exceptional keys; general arguments close 110 of them:\\ a state has no entering quintuple (6 classes)\\ every quintuple entering a state comes from that state (48 classes)\\ a state is entered by a zero-shift (56 classes)};
\node[flbox,text width=84mm] (R) [below=8mm of G] {Close them by arguments for each type of behavior\\ The last group of 10 classes is closed by showing directly that the head of a halting run stays within a computable distance of the initial window};
\node[flterm,text width=84mm] (X) [below=of R] {End: the head of a halting run stays within a computable distance of the initial window, so halting is decidable and the table is not universal};
\draw[flarr] (S) -- (N); \draw[flarr] (N) -- (D);
\draw[flarr] (D) -- node[right] {no} (G);
\draw[flarr] (G) -- node[right] {remaining 68 classes} (R);
\draw[flarr] (R) -- (X);
\draw[flarr] (D.east) -- node[above] {yes} ++(8,0) |- (X.east);
\draw[semithick] (G.east) -- ++(8,0);
\end{tikzpicture}
\caption{Flow of the proof that three-state two-symbol tables are not universal}
\label{fig:three-flow}
\end{figure}
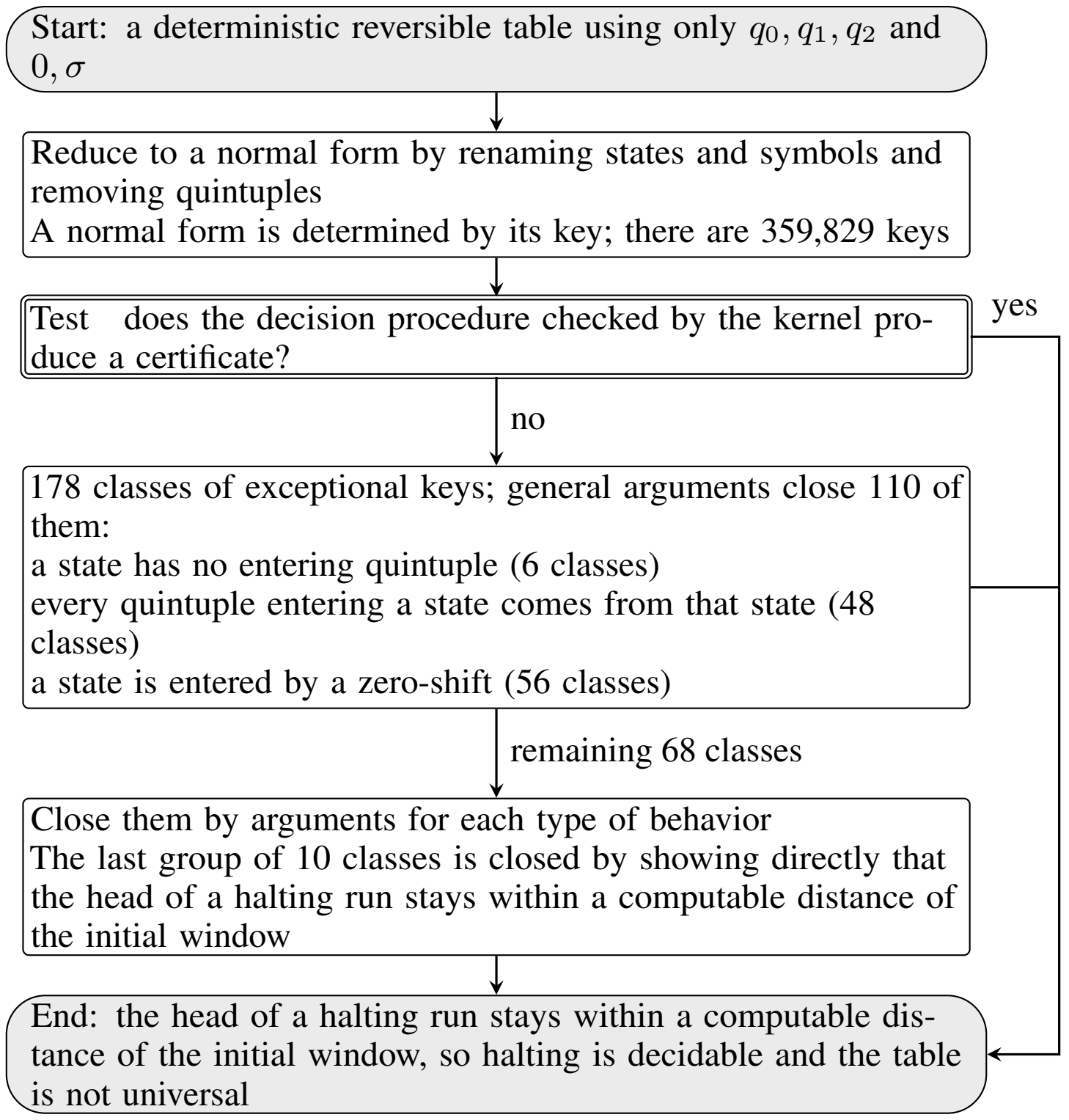

\paragraph{Proof outline.}
In every case we show that the head of a halting run stays within a distance of the initial window that is computable from the table and the window; then, since there are only finitely many configurations within that distance, a halting run halts within a computable number of steps and halting is decidable.
Every table is first reduced to a normal form by renaming states and symbols and by removing the quintuples that enter a state other than $q_0,q_1,q_2$, which changes the halting time by at most one step.
A normal form is determined by assigning to each of the six entries (three states, two symbols) either no quintuple or one of the six pairs of a written symbol and a target state, together with the direction in which each of the three states is entered; we call this a \emph{key} (by reversibility, the assignments of the entries with quintuples are distinct). There are 359,829 keys: choosing the $k$ entries that have quintuples and assigning them distinct pairs gives $\sum_{k=0}^{6}\binom{6}{k}^2k!=13{,}327$ possibilities, and each of the three states is entered in one of the three directions, so the number is $3^3\cdot 13{,}327=359{,}829$.
Figure~\ref{fig:key} shows an example of a key. The rows are the states and the columns the read symbols; each entry gives the written symbol and the target state, and -- means that there is no quintuple.
The direction of each quintuple is the direction in which its target state is entered. Hence this key represents the table of the five quintuples $[q_0,0,\sigma,\rmR,q_1]$, $[q_0,\sigma,0,\rmR,q_2]$, $[q_1,0,\sigma,\rmL,q_0]$, $[q_2,0,\sigma,\rmR,q_2]$ and $[q_2,\sigma,0,\rmL,q_0]$.
For each of them, a decision procedure checked by the kernel either produces a certificate or marks the table as exceptional.
A certificate is a finite set of patterns of visits outside a window that is closed under the transitions of the table.
It implies, by the pigeonhole principle, that a halting run halts within a computable number of steps, so that halting is decidable and the table is not universal.
The exceptional keys are handled by general arguments:
\begin{itemize}
\item if a state has no entering quintuple, then after the first step the table is a two-state table, and Theorem~\ref{thm:halting} applies;
\item if every quintuple entering some state comes from that state itself, then a halting run leaves that state within the width of the window plus one steps and never returns;
\item if some state is entered by a zero-shift, then that quintuple can be composed with the next one, eliminating the state and again reducing to a two-state machine;
\end{itemize}
and the remaining keys by arguments for each type of behavior; the last of them are closed by showing this computable distance directly.
Counted outside the proof assistant up to renaming of states and left-right reflection, the exceptional keys form 178 classes of 2,280 tables, and the last group consists of 10 classes.
Figure~\ref{fig:three-flow} shows this flow and the number of classes at each stage.
\begin{figure}[tb]
\centering
\begin{tabular}{c|cc|c}
 & $0$ & $\sigma$ & entered by \\ \hline
$q_0$ & $\sigma,q_1$ & $0,q_2$ & $\rmL$ \\
$q_1$ & $\sigma,q_0$ & -- & $\rmR$ \\
$q_2$ & $\sigma,q_2$ & $0,q_0$ & $\rmR$ \\
\end{tabular}
\caption{An example of a key}
\label{fig:key}
\end{figure}

\section{Formalization}
\label{sec:formal}

All definitions of Section~\ref{sec:prelim} and all theorems of Sections~\ref{sec:constr} to~\ref{sec:threetwo} are formalized in Lean~4 with Mathlib; computability in Definition~\ref{def:univ} is Mathlib's notion of computable functions.
Every theorem is accepted by the Lean kernel, contains no \texttt{sorry}, and depends only on the standard axioms; all computations used in proofs are checked by the kernel.
The formalized machine model reproduces the halting step counts of the test runs that Morita distributes with his nine URTMs~\cite{MoritaFiles}, with one exception: a test for URTM$(13,7)$ states 1,073 steps, whereas recomputation gives 1,173, as does the same test for URTM$(10,8)$, which the book says takes the same time.

What remains to be trusted is that the definitions and theorem statements express the intended claims, in particular that Definition~\ref{def:univ} captures Morita's Definition~7.1; we state them in full so that the reader can compare them with the original.
Since the existence and the nonexistence results use the same definition, a definition that were trivially true or trivially false would contradict one of them.
The formalization and a full version with further details are available in the same repository (Section~\ref{sec:merge}).

\section{Conclusion}

We have constructed nine new URTMs that, together with four of Morita's machines, give 13 non-dominated points, and we have shown that no URTM exists with at most two states or with three states and two symbols.
The new points come from making Morita's conversions economical: disjoint ranges by direction, counting in reverse, splitting state numbers into blocks, sharing and pruning counters, merging symbols, and merging two states by tags on the cells.
The two-state result rests on the observation that, in a reversible machine with two states, the state is determined by the direction of the last move.

Several questions remain open.
None of the new points is known to be minimal, and the lower and upper bounds are far apart: no URTM with at most two states exists, while the smallest known three-state URTM has 600 symbols.
The numbers of symbols of the machines obtained by symbol merging come from a search over a limited set of block decompositions, starting tables, and annealing runs; trying further combinations may lower them.
It is open whether some three-state, three-symbol or four-state, two-symbol RTM is universal.
For irreversible UTMs, the smallest known machines have far fewer symbols for the same number of states, and it would be interesting to know whether reversibility forces this difference.

\section*{Acknowledgments}
We are grateful to the late Kenichi Morita, who introduced us to this problem.
We thank Toya Makino for discussions.
This paper was written in Japanese and translated into English with a Generative AI tool.
Generative AI tools were also used for computer search and for writing Lean code; all Lean code was checked by the Lean kernel.
This work was supported by JST CREST Grant Number JPMJCR24I4, Japan.

\end{document}